\documentclass[trackchanges,twocolumn]{aastex7}

\usepackage{amsmath}
\usepackage{gensymb}
\usepackage{siunitx} 
\usepackage{hyperref}
\usepackage[super]{nth}

\defcitealias{bruzualStellarPopulationSynthesis2003}{BC03}
\defcitealias{fritz06agnmodel}{F06}
\defcitealias{dsilvaSelfConsistentJWSTCensus2025}{Paper I}

\newcommand{\fescstars}{$14^{+9}_{-7}\%$}

\newcommand{\fescstarsbest}{$12\%$}
\newcommand{\fescagn}{$60^{+28}_{-32}\%$}

\newcommand{\fescagnbest}{$63\%$}

\begin{document}

\title{Dissecting Reionisation with the Cosmic Star Formation and Active Galactic Nuclei Luminosity History}

\correspondingauthor{Jordan C. J. D'Silva}
\author[0000-0002-9816-1931]{Jordan C. J. D'Silva}
\affiliation{International Centre for Radio Astronomy Research (ICRAR) and the
International Space Centre (ISC), The University of Western Australia, M468,
35 Stirling Highway, Crawley, WA 6009, Australia}
\email[show]{jordan.dsilva@research.uwa.edu.au}

\author[0000-0001-9491-7327]{Simon P. Driver}
\affiliation{International Centre for Radio Astronomy Research (ICRAR) and the
International Space Centre (ISC), The University of Western Australia, M468,
35 Stirling Highway, Crawley, WA 6009, Australia}
\email{simon.driver@uwa.edu.au}

\author[0000-0003-3021-8564]{Claudia D. P. Lagos}
\affiliation{International Centre for Radio Astronomy Research (ICRAR) and the
International Space Centre (ISC), The University of Western Australia, M468,
35 Stirling Highway, Crawley, WA 6009, Australia}
\affiliation{ARC Centre of Excellence for All Sky Astrophysics in 3 Dimensions (ASTRO 3D), Australia}
\email{claudia.lagos@uwa.edu.au}

\author[0000-0003-0429-3579]{Aaron S. G. Robotham}
\affiliation{International Centre for Radio Astronomy Research (ICRAR) and the
International Space Centre (ISC), The University of Western Australia, M468,
35 Stirling Highway, Crawley, WA 6009, Australia}
\email{aaron.robotham@uwa.edu.au}

\author[0000-0003-4875-6272]{Nathan J. Adams}
\affiliation{Jodrell Bank Centre for Astrophysics, Alan Turing Building,
University of Manchester, Oxford Road, Manchester M13 9PL, UK}
\email{nathan.adams@manchester.ac.uk}

\author[0000-0003-1949-7638]{Christopher J. Conselice} 
\affiliation{Jodrell Bank Centre for Astrophysics, Alan Turing Building,
University of Manchester, Oxford Road, Manchester M13 9PL, UK}
\email{conselice@manchester.ac.uk}

\author[0000-0003-1625-8009]{Brenda Frye} 
\affiliation{Department of Astronomy/Steward Observatory, University of Arizona, 933 N Cherry Ave,
Tucson, AZ, 85721-0009, USA}
\email{bfrye@arizona.edu}

\author[0000-0001-6145-5090]{Nimish P. Hathi}
\affiliation{Space Telescope Science Institute, 3700 San Martin Drive, Baltimore, MD 21218, USA}
\email{nhathi@stsci.edu}

\author[0000-0002-4130-636X]{Thomas Harvey} 
\affiliation{Jodrell Bank Centre for Astrophysics, Alan Turing Building,
University of Manchester, Oxford Road, Manchester M13 9PL, UK}
\email{thomas.harvey-3@postgrad.manchester.ac.uk}

\author[0000-0002-6610-2048]{Anton M. Koekemoer} 
\affiliation{Space Telescope Science Institute,
3700 San Martin Drive, Baltimore, MD 21218, USA}
\email{koekemoer@stsci.edu}

\author[0000-0002-6150-833X]{Rafael {Ortiz~III}} 
\affiliation{School of Earth and Space Exploration, Arizona State University,
Tempe, AZ 85287-1404, USA}
\email{rortizii@asu.edu}

\author[0000-0003-4223-7324]{Massimo Ricotti}
\affiliation{Department of Astronomy, University of Maryland, College Park, 20742, USA}
\email{ricotti@astro.umd.edu}

\author[0000-0002-5404-1372]{Clayton Robertson}
\affiliation{Department of Physics and Astronomy, University of Louisville, Natural Science Building 102, Louisville, KY 40292, USA}
\email{cdrobe05@louisville.edu}

\author[0000-0001-6564-0517]{Ross M. Silver}
\affiliation{Astrophysics Science Division, NASA Goddard Space Flight Center, Greenbelt, MD 20771, USA}
\email{ross.m.silver@nasa.gov}

\author[0000-0003-3903-6935]{Stephen M. Wilkins}
\affiliation{Astronomy Centre, University of Sussex, Falmer, Brighton BN1 9QH, UK}
\email{s.wilkins@sussex.ac.uk}

\author[0000-0001-9262-9997]{Christopher N. A. Willmer} 
\affiliation{Steward Observatory, University of Arizona,
933 N Cherry Ave, Tucson, AZ, 85721-0009, USA}
\email{cnaw@as.arizona.edu}

\author[0000-0001-8156-6281]{Rogier A. Windhorst}
\affiliation{School of Earth and Space Exploration, Arizona State University,
Tempe, AZ 85287-1404, USA}
\email{Rogier.Windhorst@asu.edu}

\author[0000-0003-3329-1337]{Seth H. Cohen} 
\affiliation{School of Earth and Space Exploration, Arizona State University,
Tempe, AZ 85287-1404, USA}
\email{seth.cohen@asu.edu}

\author[0000-0003-1268-5230]{Rolf A. Jansen} 
\affiliation{School of Earth and Space Exploration, Arizona State University,
Tempe, AZ 85287-1404, USA}
\email{rolfjansen.work@gmail.com}

\author[0000-0002-7265-7920]{Jake Summers} 
\affiliation{School of Earth and Space Exploration, Arizona State University,
Tempe, AZ 85287-1404, USA}
\email{jssumme1@asu.edu}

\author[0000-0001-7410-7669]{Dan Coe} 
\affiliation{Space Telescope Science Institute, 3700 San Martin Drive, Baltimore, MD 21218, USA}
\affiliation{Association of Universities for Research in Astronomy (AURA) for the European Space Agency (ESA), STScI, Baltimore, MD 21218, USA}
\affiliation{Center for Astrophysical Sciences, Department of Physics and Astronomy, The Johns Hopkins University, 3400 N Charles St. Baltimore, MD 21218, USA}
\email{dcoe@stsci.edu}

\author[0000-0001-9440-8872]{Norman A. Grogin} 
\affiliation{Space Telescope Science Institute,
3700 San Martin Drive, Baltimore, MD 21218, USA}
\email{nagrogin@stsci.edu}

\author[0000-0001-6434-7845]{Madeline A. Marshall} 
\affiliation{Los Alamos National Laboratory, Los Alamos, NM 87545, USA}
\email{madeline_marshall@outlook.com}

\author[0000-0003-3382-5941]{Nor Pirzkal} 
\affiliation{Space Telescope Science Institute,
3700 San Martin Drive, Baltimore, MD 21218, USA}
\email{npirzkal@stsci.edu}

\author[0000-0003-0894-1588]{Russell E. Ryan, Jr.} 
\affiliation{Space Telescope Science Institute,
3700 San Martin Drive, Baltimore, MD 21218, USA}
\email{rryan.asu@stsci.edu}

\author[0000-0001-7592-7714]{Haojing Yan} 
\affiliation{Department of Physics and Astronomy, University of Missouri,
Columbia, MO 65211, USA}
\email{yanhaojing@gmail.com}


\begin{abstract}
The combination of the $z=0-13.5$ cosmic star formation history and active galactic nuclei (AGN) luminosity history as inferred by the \textit{James Webb Space Telescope} is connected to the cosmic spectral energy distribution (CSED) to explore the sources of reionisation. We compute the redshift evolution of the corresponding cosmic ionising photon emissivity, the neutral fraction and the cosmic microwave background optical depth. We use the generative SED modelling code \textsc{ProSpect} to bracket the ionising emissivity between escape fractions of $f_{\mathrm{esc}} = 1 - 100\%$ for both the stars and AGN. Stars alone could have achieved reionisation by $z\approx 6$ with $f_{\mathrm{esc}} \gtrsim 30\%$ for solar metallicity ($Z=0.02$) stars or $f_{\mathrm{esc}} \gtrsim 10\%$ for metal-poor ($Z=10^{-4}$) stars. On the other hand, AGN by themselves would have struggled to produce sufficiently many ionising photons even with $f_{\mathrm{esc}} = 100\%$. A hybrid model containing both stars and AGN is explored where we find best fit (median$\pm 1\sigma$) $f_{\mathrm{esc}}=$ \fescstarsbest{} (\fescstars{}) for the stars and $f_{\mathrm{esc}}=$ \fescagnbest{} (\fescagn{}) for the AGN, maintained at all redshifts. In essence, the joint growth of stellar mass and supermassive black holes produces neither more nor fewer ionising photons than needed to reionise $\gtrsim 99\%$ of the intergalactic medium by $z\approx 6$.
\end{abstract}



\section{Introduction} \label{sec:intro}
Following both Big Bang nucleosynthesis and subsequent recombination, the overwhelming baryonic content of the intergalactic medium (IGM) consisted of neutral hydrogen (HI) and helium (HeI). Because the cosmic microwave background (CMB) photons could no longer keep the hydrogen ionised at $z\approx 1100$, the neutral contents only became ionised again with the emergence of the first sources of light from stars and supermassive black holes \citep[SMBHs, e.g.,][]{barkanaBeginningFirstSources2001a}. 

A key piece of evidence of reionisation comes in the form of the Gunn-Peterson trough in the spectra of distant quasars \citep{gunnDensityNeutralHydrogen1965}. The trough appears as total absorption at wavelengths shorter than the Lyman-$\alpha$ line ($\lambda < \SI{1216}{\angstrom}$) indicating a Universe completely filled with HI. Quasars at $z\lesssim 6$, on the other hand, show the Lyman-$\alpha$ forest instead of the Gunn-Peterson trough indicating sparser spatial coverage of HI; hence, reionisation is thought to have completed -- when most of the Universe became uniformly ionised -- by $z\approx 6$ \citep[e.g.,][]{fanSurvey58Quasarsin2001,beckerEvidenceReionizationAtz2001}. Additionally, CMB photons experience Thomson scattering off free electrons produced during reionisation. Thus, the optical depth of the CMB as constrained by, for example the \textit{Wilkinson Microwave Anisotropy Probe} \citep[\textit{WMAP,}][]{bennettFirstYearWilkinsonMicrowave2003} or the \textit{Planck} telescope \citep{adamPlanckIntermediateResults2016,aghanimPlanck2018Results2020b}, provide further constraints on the timing of reionisation, having been $\approx 50\%$ completed by $z\approx 7-8$. 

Because of the almost complete absorption of $\lambda \leq \SI{912}{\angstrom}$ photons due to the intergalactic medium (IGM), nature has impeded our ability to study the ionising radiation of the ionising sources. Hence, often low redshift, $z=0-3$, galaxies are instead investigated to establish relationships between the intrinsic Lyman continuum and intrinsic ultraviolet (UV) luminosity  \citep[$\mathrm{L^{intrinsic}_{UV}}$, e.g.,][]{steidelKeckLymanContinuum2018,begleyVANDELSSurveyMeasurement2022,chisholmFarultravioletContinuumSlope2022,saldana-lopezLowRedshiftLymanContinuum2022,pahlConnectionEscapeIonizing2023}. Obtaining the intrinsic luminosity involves accounting for the proportion of photons that enter the line of sight compared to the total photon production, otherwise known as the escape fraction, $f_{\mathrm{esc}}$. Hence, the number of ionising photons produced in a galaxy and transmitted into the IGM is calculated as,
\begin{equation} \label{eq:nions_galaxy}
    \dot{N}_{\mathrm{ion}} = \xi_{\mathrm{ion}}\mathrm{L^{intrinsic}_{UV}} f_{\mathrm{esc}}.
\end{equation}
The proportionality constant in Equation~\ref{eq:nions_galaxy} is the ionising efficiency, $\xi_{\mathrm{ion}} = \mathrm{\dot{N}_{ion}^{intrinsic}} / \mathrm{L^{intrinsic}_{UV}}$, that connects the UV luminosity to the intrinsic production rate of Lyman continuum photons from astrophysical processes occurring in galaxies. The canonical value of $\log_{10}(\xi_{\mathrm{ion}}/\mathrm{Hz \, erg^{-1}}) = 25.2$, appropriate for $z\gtrsim 6$ galaxies as per the \citet{bruzualStellarPopulationSynthesis2003} stellar population synthesis library, is often used in the literature \citep[e.g.,][]{robertsonNEWCONSTRAINTSCOSMIC2013,kuhlenConcordanceModelsReionization2012}. The use of model spectra from the \citet{bruzualStellarPopulationSynthesis2003} library is critical because $\mathrm{\dot{N}_{ion}^{intrinsic}}$ of the ionising sources is virtually unobservable in practice.   

Because of the connection between $\dot{N}_{\mathrm{ion}}$ and $\mathrm{L^{intrinsic}_{UV}}$, it is reasoned that the predominant source of ionising photons originates in star forming galaxies \citep[e.g.,][]{robertsonNEWCONSTRAINTSCOSMIC2013,robertsonCosmicReionizationEarly2015a,finkelsteinConditionsReionizingUniverse2019}. Obtaining a volume-averaged census of ionising photons involves computing the quantities in Equation~\ref{eq:nions_galaxy} for statistical samples. The UV luminosity function steeply rises toward fainter luminosities without a significant turnover at least to $\mathrm{M_{UV}}\approx -15.5$ at $z\approx 9$ \citep{atekExtremeFaintEnd2018,bouwens29GalaxiesMagnified2022}. As such, the UV luminosity density \citep[$\rho_{\mathrm{UV}}(z)$,][]{lillyCanadaFranceRedshiftSurvey1996a} as inferred by the \textit{Hubble Space Telescope} (\textit{HST}) showed a steady increase between $z\approx 8-10$ \citep[e.g.,][]{bouwensUVLuminosityFunctions2015a,oeschDearth10Galaxies2018,harikaneGOLDRUSHIVLuminosity2022}. Newer inferences of $\rho_{\mathrm{UV}}(z)$ with the \textit{James Webb Space Telescope} (\textit{JWST}) show an even more rapid increase at $z\gtrsim 10$ in comparison because of its ability to better probe the rest-frame UV-optical at these extreme redshifts \citep[e.g.,][]{harikaneComprehensiveStudyGalaxies2023,harikaneJWSTNIRSpecFirst2023b,donnanJWSTPRIMERNew2024,adamsEPOCHSIIUltraviolet2024}. Therefore, $\rho_{\mathrm{UV}}(z)$ indicates vigorous star formation at $z\approx 5-16$ as the UV luminosity originates predominantly from young, massive stars \citep{madaudickinson2014}.

At the same time, \textit{JWST} has also found that $\xi_{\mathrm{ion}}$ of $z\gtrsim 9$ galaxies may be up to $\approx 1$~dex higher than the assumed canonical value \citep[e.g.,][]{curtis-lakeSpectroscopicConfirmationFour2023a,atekMostPhotonsThat2024,simmondsLowmassBurstyGalaxies2024}. This in combination with bright UV luminosities points to a strong evolution in the budget of ionising photons at $z\gtrsim 6$. \citet{munozReionizationJWSTPhoton2024}, using these new \textit{JWST} results, found that the timeline of reionisation would be inconsistent with the constraints inferred from the CMB, with reionisation completing far too early at $z\approx 9$. In addition to UV bright star forming galaxies, \textit{JWST} has identified many active galactic nuclei (AGN) at $z\gtrsim6$ \citep[e.g.,][]{kocevskiHiddenLittleMonsters2023a,juodzbalisEPOCHSVIIDiscovery2023,larsonCEERSDiscoveryAccreting2023a,greeneUNCOVERSpectroscopyConfirms2024,mattheeLittleRedDots2024,kocevskiRiseFaintRed2024a} that could be accreting well beyond their Eddington rates \citep[e.g.,][]{schneiderAreWeSurprised2023b}. Thus, the contribution of ionising photons from accreting SMBHs could be significant \citep{madauCosmicReionizationPlanck2015,harikaneJWSTNIRSpecFirst2023b,madauCosmicReionizationJWST2024}, potentially adding to the surplus of ionising photons seemingly already produced through star formation.

In light of this, it is timely to investigate the stellar and AGN contributions to reionisation. We approach this by connecting the cosmic stellar mass assembly and SMBH growth to an equivalent cosmic spectral energy distribution (CSED). Cosmic stellar mass assembly is quantified with the cosmic star formation history (CSFH) that is the redshift evolution of the sum of all star formation rates (SFRs) per unit volume. Likewise, cosmic SMBH growth is quantified with the cosmic AGN luminosity history (CAGNH) that is the redshift evolution of the sum of all bolometric AGN luminosities per unit volume. We use both the $z=0-13.5$ CSFH and CAGNH, as reported in \citet{dsilvaSelfConsistentJWSTCensus2025}, combined with the flexible generative spectral energy distribution (SED) model \textsc{ProSpect} \citep{robothamProSpectGeneratingSpectral2020}. We explore the redshift evolution of the cosmic ionising emissivity, the neutral fraction of the IGM, the CMB optical depth and the ionising efficiency, striving to do this in a completely self-consistent manner. 

Section~\ref{sec:data} describes the CSFH and CAGNH that we use. In Section~\ref{sec:modeling}, we outline both our reionisation and CSED modelling. We present our results and discussion in Sections~\ref{sec:results} and \ref{sec:discussion}. Finally, Section~\ref{sec:summary} summarizes our work. We use standard concordance $\Lambda$CDM cosmology with $H_{0} = 70 \, \mathrm{km \, s^{-1} \, Mpc^{-1}}$, $\Omega_{\Lambda} = 0.7$ and $\Omega_{M} = 0.3$. We use the AB magnitude system \citet{okeSecondaryStandardStars1983} and the \citet{chabrierGalacticStellarSubstellar2003} IMF. Throughout, we denote the normal distribution with mean, $\mu$, and variance, $\sigma^{2}$, as $\mathcal{N}(\mu, \sigma^{2})$.
 
\section{Data} \label{sec:data}
We used the CSFH and CAGNH from \citet{dsilvaSelfConsistentJWSTCensus2025}, hereafter referred to as \citetalias{dsilvaSelfConsistentJWSTCensus2025}. Briefly, the work in \citetalias{dsilvaSelfConsistentJWSTCensus2025} consisted of using a combination of \textit{HST} and \textit{JWST} to produce $0.2-4.4\,\mu$m photometric catalogues over $400 \, \mathrm{amin}^{2}$ of extragalactic fields. Specifically, data from the following programs were used: the Prime Extragalactic Areas of Reionisation and Lensing Science (PEARLS, PIs: R. Windhorst \& H. Hammel, PIDs: 1176 \& 2738) survey \citep{windhorstJWSTPEARLSPrime2023}, the JWST Advanced Deep Extragalactic Survey \citep[JADES, PIDs: 1180 PI: D. Eisenstein, 1210 PI: N. Luetzgendorf \& 3215, PI: D. Eisenstein \& R. Maiolino, ][]{eisensteinOverviewJWSTAdvanced2023a,riekeJADESInitialData2023}, the Next Generation Deep Extragalactic Exploratory Public (NGDEEP, PID: 2079, PIs: S. Finkelstein, C. Papovich, N. Pirzkal) survey, \citep{bagleyNextGenerationDeep2024}, the Cosmic Evolution Early Release Science \citep[CEERS, PID: 1345, PI: S. Finkelstein,][]{bagleyCEERSEpochNIRCam2023a,finkelsteinCEERSKeyPaper2023a} and, finally, the Public Release IMaging for Extragalactic Research (PRIMER, PID:1837, PI: J. Dunlop) survey. 

Using a combination of the SED fitting software EAZY \citep{brammerEAZYFastPublic2008} and \textsc{ProSpect} \citep{robothamProSpectGeneratingSpectral2020}, in \citetalias{dsilvaSelfConsistentJWSTCensus2025} they obtained a sample of $3751$ galaxies, and their associated stellar masses, SFRs and AGN bolometric luminosities, at $z_{\mathrm{phot}} \approx 5.5-13.5$  with a $95\%$ confidence. A brief overview of the SED fitting methodology is provided below. Full details can be found in \citetalias{dsilvaSelfConsistentJWSTCensus2025}.

An initial sample of $4585$ $z>5$ candidates was obtained with EAZY using the SED templates of \citet{larsonSpectralTemplatesOptimal2023} that were optimised for $z>5$ galaxy searches. They first ran EAZY by restricting the range of possible photometric redshifts to $0\leq z \leq 5$, and then performed a second run by broadening the range to $0\leq z \leq 25$. By comparing the $\chi^{2}$ of the two EAZY fits, robust $z>5$ galaxies were obtained by rejecting the restricted, $0\leq z \leq 5$, fit with a $95\%$ confidence. The full photometric redshift probability distribution from EAZY was then passed into \textsc{ProSpect} to improve the photometric redshift that could have been biased by the limited set of template SEDs in EAZY. This resulted in $3751$ robust $z>5$ galaxies. 

\textsc{ProSpect} is a flexible and fully generative Bayesian SED fitting code built from an assortment of models. To fit the SEDs of the $3751$ galaxies, \textsc{ProSpect} employs the \citet{charlotSimpleModelAbsorption2000} model to implement dust attenuation, the \citet[][hereafter BC03]{bruzualStellarPopulationSynthesis2003} stellar population synthesis libraries combined with the \citet{chabrierGalacticStellarSubstellar2003} IMF to model starlight and the \citet[][hereafter F06]{fritz06agnmodel} model to account for AGN. The star formation histories (SFHs) were implemented as smooth four-parameter skewed normal distributions, referred to as \texttt{massfunc\_snorm\_trunc} functions. Finally, the mass fraction of metals in stars, the metallicity ($Z$), was implemented as a linear evolution that is assumed to be $Z=10^{-4}$ (the lowest metallicity template in \citetalias{bruzualStellarPopulationSynthesis2003}) at the genesis of star formation and scales with the buildup of stellar mass. 

The inclusion of an AGN component in the SEDs introduces more parameters than stellar-only models. They did two \textsc{ProSpect} runs, with and without an AGN, and used the deviance information criterion to decide whether the extra complexity of the AGN included model significantly improved the fit. If the inclusion of an AGN did not improve the fit, it either meant that the galaxy does not host an accreting SMBH or that it could not be observed with the NIRCam data on hand. A maximum possible AGN contribution for those galaxies was inferred by calculating the AGN luminosity of a completely unobscured \citetalias{fritz06agnmodel} template that contributed $20\%$ of the flux in the NIRCam F277W, F356W and F444W bands. Essentially, the AGN luminosity could be bound between $0$ and this maximum possible value. 

The SFR distribution function (SFRF) and the AGN luminosity function (AGNLF) at $z\gtrsim 5$ were then computed, and integrating those they further obtained the CSFH and CAGNH. The SFR density distributions produced convergent integrals, with contributions to the CSFH from the faint ends of the SFRF ever diminishing. Only lower and upper limits on the CAGNH could be obtained from the minimum and maximum AGN contributions as per the SED fitting. Then, using the $z\approx 0-5$ constraints of the CSFH and CAGNH from \citet{dsilvaGAMADEVILSCosmic2023} that were obtained in almost an identical manner (i.e., the same photometric extraction, SED models and SFRF and AGNLF integrals), they fitted the $z\approx0-13.5$ CSFH and CAGNH with smooth, four-parameter functions. Thus, they produced a self-consistent census of star formation and AGN activity over $\gtrsim 13$~Gyr of cosmic history, allowing us to account for the two prevailing sources of (re)ionising radiation. 

\begin{figure}
    \centering
    \includegraphics[width=\linewidth]{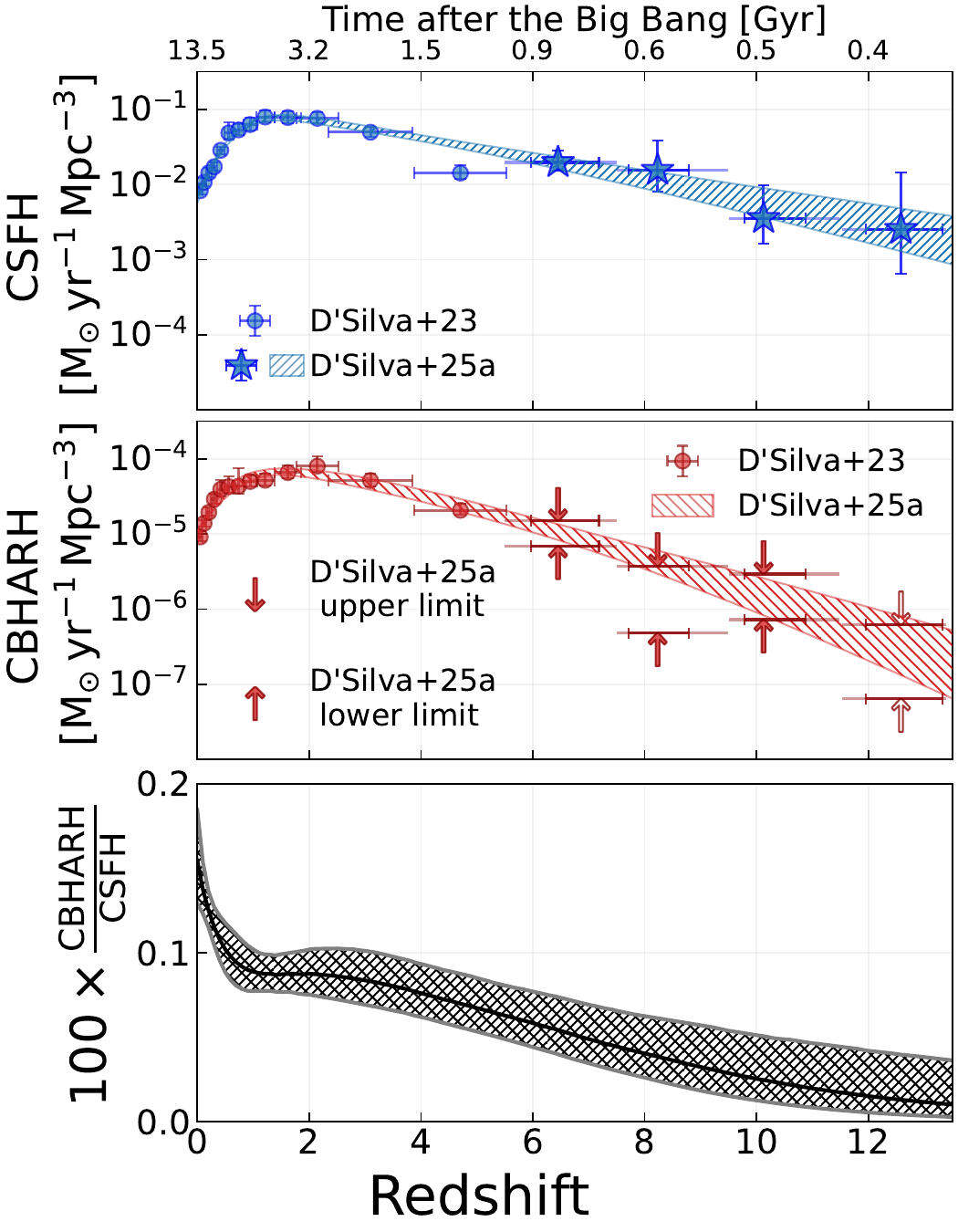}
    \caption{\textit{Top:} CSFH as a function of redshift. The circles with error bars show the results and $1\sigma$ uncertainties from \citet{dsilvaGAMADEVILSCosmic2023} at $z<5$. The stars with error bars show the results and $1\sigma$ uncertainties at $z \gtrsim 5$ from \citetalias{dsilvaSelfConsistentJWSTCensus2025}. The horizontal error bars show both the $1\sigma$ spread of the redshift distribution in each bin and the width of the bin. The hatched region shows the $16-\nth{84}$ percentile spread of the model fit to the data points from \citetalias{dsilvaSelfConsistentJWSTCensus2025}. \textit{Middle:} CBHARH as a function of redshift. Points and hatches have the same meaning as the top panel. The upward (downward) facing arrows show the lower (upper) limits at $z\gtrsim5.5$ from \citetalias{dsilvaSelfConsistentJWSTCensus2025}. The last points of the CBHARH are highlighted with white to signify that they are uncertain and were not used in the model fitting. \textit{Bottom:} percentage ratio of the model fits to the CBHARH and CSFH. The solid line shows the median and the hatched regions show the $16-\nth{84}$ percentiles propagated from the fits.}
    \label{fig:cbharhvscsfh}
\end{figure}

Figure~\ref{fig:cbharhvscsfh} shows the results of the CSFH and cosmic black hole accretion rate history (CBHARH) that \citet{dsilvaSelfConsistentJWSTCensus2025} obtained by integrating the SFRF and AGNLF. 

The CBHARH is computed as \begin{equation}
    \mathrm{CBHARH = CAGNH / \epsilon c^{2}},
\end{equation} where $\epsilon = 0.1$ is the radiative efficiency and $c$ is the speed of light. Taking the ratio of the CBHARH to CSFH equals the dimensionless baryon conversion ratio. It can be seen that the dimensionless ratio of CBHARH to CSFH tentatively rises by a factor of $\approx 3$ between $z\approx 11 \to 6$, reflecting the relative increase in the SMBH growth compared to star formation in the early Universe \citep[e.g.,][]{maiolinoJADESDiversePopulation2023a,harikaneJWSTNIRSpecFirst2023b}.

The constraints on the CSFH and the CAGNH at $z\gtrsim 5$ from \citet{dsilvaSelfConsistentJWSTCensus2025} were inferred from the rest-frame UV-optical. As discussed in \citet{dsilvaSelfConsistentJWSTCensus2025}, being limited to these rest-frame wavelengths may miss significant AGN contributions to the SEDs of galaxies at $z \gtrsim 12$, and so the results in the highest redshift bin were not included in their fitting of the smooth four-parameter function presented in Figure~\ref{fig:cbharhvscsfh}. The CAGNH could be consistently flat at $z\approx 12$ instead of rapidly rising as portrayed by the smooth four-parameter function. Moreover, any dust-enshrouded obscured component of either the CSFH or CAGNH at $z\gtrsim 5$ will be missing, potentially affecting inferences on $f_{\mathrm{esc}}$, and so this is an important caveat to keep in mind. 

\section{Modelling cosmic reionisation} \label{sec:modeling}
The rapid rise of the CSFH and CBHARH hints toward an equivalently rapid production of ionising photons that reionised the Universe. 

\subsection{The reionisation equation}
The timeline of reionisation can be modelled with 
\begin{equation}
    \frac{dQ(z)}{dz} = \frac{\dot{n}_{\mathrm{ion}}(z)/n_{\mathrm{H}}(z) - Q(z)/\mathrm{t_{rec}}(z)}{(1+z)\times H(z)},
\label{eq:madaureionize}
\end{equation}
which essentially tracks the balance of ionising photon production to the rate at which free electrons recombine with nucleons \citep{madauRadiativeTransferClumpy1999,gnedinModelingCosmicReionization2022}. $Q(z)$ is the fraction of $\mathrm{HII}$, where $Q(z)=1$ indicates complete reionisation and $1-Q(z)$ is the neutral fraction. $H(z)$ is the Hubble parameter.

\begin{equation}\label{eq:nH}
    \mathrm{n_{H}}(z) = X_{\mathrm{H}}\Omega_{b}\rho_{0}(1+z)^{3} / m_{\mathrm{H}}
\end{equation}
is the number density of hydrogen, where $X_{\mathrm{H}} \approx 0.76$ is the hydrogen fraction from Big Bang nucleosynthesis, $\Omega_{b} \approx 0.04$ is the cosmic baryon fraction, $\rho_{0}$ is the critical density and $m_{\mathrm{H}}$ is the proton mass. 

\begin{equation}\label{eq:trec}
    \mathrm{t_{rec}} = [\mathcal{C} \, \alpha_{\mathrm{B}} \, n_{e}(z)]^{-1}
\end{equation}
is the recombination timescale, where $\mathcal{C}\approx1-6$ is the clumping factor \citep[e.g.,][]{kaurovCOSMICREIONIZATIONCOMPUTERS2015}, which we fiducially set to $\mathcal{C}=3$, and $\alpha_{B} = 2.6\times 10^{-13} \, \mathrm{cm^3 \, s^{-1}}$ is the case-B recombination coefficient for gas temperatures $\approx 10^{4}$~K. The model has nucleons recombining with electrons inside of ionised regions of the IGM. Electrons may be obtained from hydrogen and helium. Hence,
\begin{equation}
\label{eq:ne}
    \begin{split}
            n_{e}(z) &= n_{\mathrm{H}}(z) + \eta \, n_{\mathrm{He}}(z) \\
            &= n_{\mathrm{H}}(z)\left[1 +  \eta \, \frac{n_{\mathrm{He}}(z)}{n_{\mathrm{H}}(z)}\right]\\
            &= n_{\mathrm{H}}(z)\left[1 +  \eta \, \frac{(1-X_{\mathrm{H}})\Omega_{b}\rho_{0}(1+z)^{3} / 4 \, m_{\mathrm{H}}}{X_{\mathrm{H}}\Omega_{b}\rho_{0}(1+z)^{3} / m_{\mathrm{H}}}\right]\\
            &= n_{\mathrm{H}}(z)\left[1 +  \eta \, \frac{1-X_{\mathrm{H}}}{4 \, X_{\mathrm{H}}}\right]\\
            & \approx n_{\mathrm{H}}(z) \left[ 1 + 0.08\eta \right],
    \end{split}
\end{equation}
where $\eta = 1$ at $z>4$ to account for HeII and $\eta = 2$ at $z \leq 4$ to account for HeIII \citep[e.g.,][]{kuhlenConcordanceModelsReionization2012, bouwensREIONIZATIONPLANCKDERIVED2015,masonModelindependentConstraintsHydrogenionizing2019,munozReionizationJWSTPhoton2024}. Notice that Equations~\ref{eq:nH},\ref{eq:trec},\ref{eq:ne} give a recombination term in the numerator of Equation~\ref{eq:madaureionize} proportional to $Q(z)n_{\mathrm{H}}(z)$ to account for the free electrons produced as the neutral contents of the IGM are ionised. 

\subsection{The cosmic ionising emissivity}
$\dot{n}_{\mathrm{ion}}(z)$ is the rate at which the total number of ionising photons are produced in the cosmic volume, otherwise referred to as the cosmic ionising emissivity, and calculated in this work as
\begin{equation}
\label{eq:nions}
    \dot{n}_{\mathrm{ion}}(z) = (1+z)^{3}\times\int^{\SI{912}{\angstrom}}_{0} \frac{L(\lambda,z)}{hc/\lambda}d\lambda,
\end{equation}
where $L(\lambda,z)$ is the rest-frame CSED at each redshift, $h$ is Planck's constant and $c$ is the speed of light. The CSED is the sum of all galaxy SEDs and therefore represents the population- and volume-averaged radiation budget.

To quantify the contribution to reionisation from stars and AGN, we computed a model CSED redshift evolution with \textsc{ProSpect}. The process of `SED fitting' with \textsc{ProSpect} entails using Bayesian analysis to infer the stellar and AGN properties of galaxies given their observed spectroscopic/photometric measurements \citep[e.g.,][]{conroyModelingPanchromaticSpectral2013a}. Because \textsc{ProSpect} is fully Bayesian, it can also work in `SED generation' mode where, instead, the (model) SED is generated for a given set of stellar and AGN properties. For example, \textsc{ProSpect} has been used to generate synthetic SEDs from the SFHs of simulated galaxies when operating in the generative mode \citep{lagosFarultravioletFarinfraredGalaxy2019}. Now, just as \textsc{ProSpect} can generate a SED in units of $\mathrm{L_{\odot} \, \SI{}{\angstrom}^{-1}}$ for the given SFH and AGN properties of a single galaxy, it can likewise produce the redshift evolution of the CSED in units of $\mathrm{L_{\odot} \, \SI{}{\angstrom}^{-1} \, Mpc^{-3}}$ for the given CSFH and CAGNH. We highlight that this method of `one canonical galaxy' as per the CSFH and CAGNH has been used to infer the extragalactic background light using generative SED models (\citealt{andrewsModellingCosmicSpectral2018}; \citealt{fermi-latcollaborationGammarayDeterminationUniverses2018}; \citealt{koushanGAMADEVILSConstraining2021}; S. Driver et al. in prep; S. Tompkins et al. in prep.). 

\subsection{The \textsc{Prospect} model}
The same \textsc{ProSpect} set up used in \citetalias{dsilvaSelfConsistentJWSTCensus2025} and discussed in Section~\ref{sec:data} is also used here. For further details on \textsc{ProSpect}, the reader is directed elsewhere \citep[i.e.,][]{robothamProSpectGeneratingSpectral2020,bellstedtGalaxyMassAssembly2020,thorneDeepExtragalacticVIsible2021,thorneDeepExtragalacticVIsible2022,dsilvaGAMADEVILSCosmic2023}. 

The Lyman continuum and UV flux, $91 \leq \lambda \, [\text{\AA}] \leq 1500$, is characterized by young O- and B-type stars in the \citetalias{bruzualStellarPopulationSynthesis2003} stellar population synthesis library, while in the \citetalias{fritz06agnmodel} model for the AGN the spectrum is a combination of power laws, 
\begin{equation}
L(\lambda) \propto
    \begin{cases} 
      \lambda^{1} & 10 \leq \lambda \, [\text{\AA}] \leq 500 \\
      \lambda^{-0.2} & 500 < \lambda \, [\text{\AA}] \leq 1250 \\
      \lambda^{-1.5} & 1250 < \lambda \, [\text{\AA}] \leq 10^{4}.
   \end{cases}
\end{equation}

We modelled the transmission of ionising photons into the IGM with the \texttt{escape-frac} parameter for the stars and the viewing angle, \texttt{AGNan}, parameter in the \citetalias{fritz06agnmodel} model for the AGN inside of our canonical galaxy. \texttt{escape-frac} in \textsc{ProSpect} scales the proportion of Lyman continuum from stars that can leave the galaxy. \texttt{AGNan} is connected to an equivalent escape fraction for the AGN as \SI{90}{\degree} corresponds to face-on, completely unobscured, type I AGN while \SI{0}{\degree} corresponds to edge-on, obscured, type II AGN. The combination of these two parameters allowed us to conveniently explore $f_{\mathrm{esc}}$ for the whole canonical galaxy. 

In relation to Equation~\ref{eq:nions}, 
\begin{equation}
    \label{eq:fesc}
    f_{\mathrm{esc}} = \frac{\dot{n}_{\mathrm{ion}}}{\dot{n}^{\mathrm{intrinsic}}_{\mathrm{ion}}} \in [0,1],
\end{equation}
where $\dot{n}^{\mathrm{intrinsic}}_{\mathrm{ion}}$ is the cosmic ionising emissivity for a completely unobscured $L(\lambda,z)$. In this work, $f_{\mathrm{esc}}$ is essentially a wavelength and redshift independent scaling of the flux below the Lyman limit. In other words, $f_{\mathrm{esc}} = 1 (0)$ means $100\%$$(0\%)$ transmission of all $\lambda \leq \SI{912}{\angstrom}$ photons into the IGM. 

Using the above setup meant that we necessarily had to set some of the other parameters of the \textsc{ProSpect} model. We set the V-band/$\SI{5500}{\angstrom}$ optical depth normalizing coefficients of the birth clouds and dust screen in the interstellar medium for the \citet{charlotSimpleModelAbsorption2000} model to a low value of $\tau^{\mathrm{BC/ISM}}_{V} = 10^{-3}$. Setting $\tau^{\mathrm{BC/ISM}}_{V}$ to this low value means that the net attenuation and resulting transmission of Lyman continuum photons into the IGM is solely encapsulated by the \texttt{escape-frac} parameter. The $9.7\mu$m optical depth normalizing coefficient for the \citetalias{fritz06agnmodel} model was also set to $\tau_{9.7} = 10^{-3}$. The \citetalias{fritz06agnmodel} model uses additional parameters to control the opening angle of the AGN (\texttt{AGNct}), the ratio of the inner and outer torus radii as per the flared disk (\texttt{AGNrm}) and the density distribution of the dust (\texttt{AGNbe}, \texttt{AGNal}). All these extra parameters were fixed to the same values used for SED fitting in \citet[][see their Tab. 1]{thorneDeepExtragalacticVIsible2022} and \citetalias{dsilvaSelfConsistentJWSTCensus2025}. 

The justification for adopting the fixed optical depth normalizing coefficients, $\tau^{\mathrm{BC/ISM}}_{V} = 10^{-3}$, in the \citet{charlotSimpleModelAbsorption2000} attenuation model is that they would be degenerate with $f_{\mathrm{esc}}$. In terms of the AGN, a key assumption in the \citetalias{fritz06agnmodel} model is that the spatial geometry of the torus is smooth, which is important for the established connection between the viewing angle and $f_{\mathrm{esc}}$. \citet{feltreSmoothClumpyDust2012a} compared the same \citetalias{fritz06agnmodel} smooth torus models with the clumpy models of \citet{nenkovaAGNDustyTori2008} where they found distinct differences with especially the $9.7\mu$m silicate feature and the infrared dust peak.

More latitude with the extra parameters could be allowed to establish a more complex relationship between the model parameters and the escaping Lyman continuum, beyond the parameters \texttt{escape-frac} and \texttt{AGNan}. Even a covering fraction parameter could be included to account for nonuniform dust geometry \citep[e.g.,][]{reddyCONNECTIONREDDENINGGAS2016}. However, increasing the degrees of freedom in this manner would still return values of $f_{\mathrm{esc}}\in[0,1]$. Hence, it suffices to consider $f_{\mathrm{esc}}$ with only these two parameters for the stars and AGN in the toy model of the canonical galaxy. 

Using \textsc{ProSpect} and the implicit \citetalias{bruzualStellarPopulationSynthesis2003} stellar population synthesis library and \citetalias{fritz06agnmodel} model for the AGN, we obtained $\dot{n}_{\mathrm{ion}}(z)$ and solved the differential equation in Equation~\ref{eq:madaureionize} for $Q(z)$ with the initial condition that $Q(z=15.5)=0$. The model does not take Lyman limit systems into account, which may hint at patchy reionisation where the neutral fraction is $\sim 10^{-5}$ at $z\approx 5.3$ \citep{bosmanHydrogenReionizationEnds2022}. For this reason, we manually restricted $Q(z)$ from exceeding unity \citep[see e.g.,][]{madauCosmicReionizationPlanck2017}. 

\section{Results} \label{sec:results}
We investigated three scenarios of reionisation. The first two are, case I: where only stars contribute ionising photons and case II: where only AGN contribute. In each of these cases, we tested a range of $f_{\mathrm{esc}}$ between $1\%$ and $100\%$ to explore essentially minimum and maximum escaping photons. The final case III is a hybrid model of reionisation where both stars and AGN contribute ionising photons, inferred with assumptions about $f_{\mathrm{esc}}$ and the stellar metallicity. 

\subsection{Case I: Stars only}\label{subsec:starlim}
\begin{figure*}
    \centering
    \includegraphics[width=\linewidth]{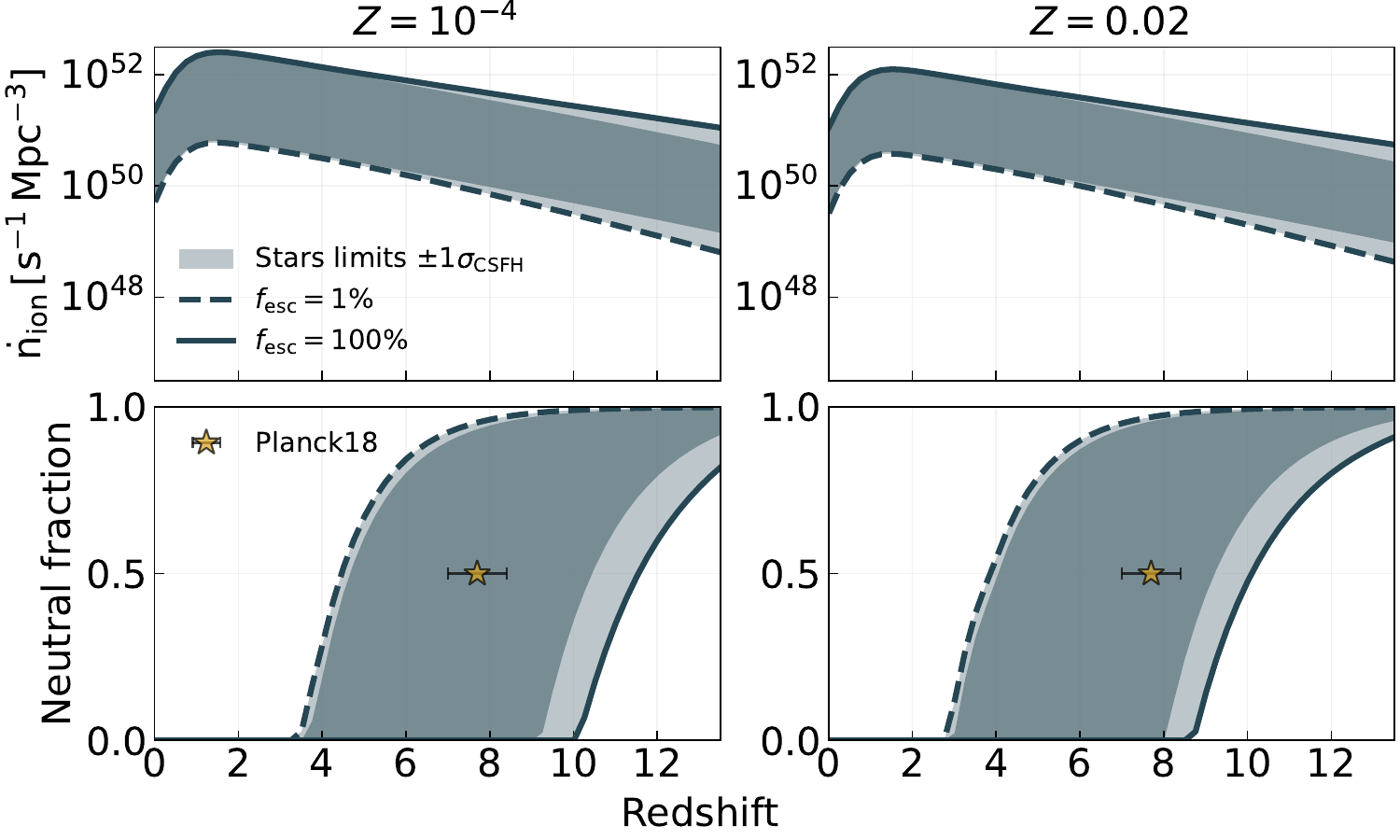}
    \caption{\textit{Top left:} cosmic ionising emissivity as a function of redshift, $\dot{n}_{\mathrm{ion}}(z)$, from stars only and assuming primordial metallicity, $Z=10^{-4}$. The blue filled region shows minimum/maximum bounds by considering $f_{\mathrm{esc}}=1\%$/$100\%$ with the lighter blue showing the $1\sigma$ uncertainty of the CSFH fit propagated through. \textit{Top right:} the same as the top left only for solar metallicity, $Z=0.02$. \textit{Bottom left:} the neutral fraction as per Equation~\ref{eq:madaureionize} for primordial metallicity. The filled regions have the same meaning as the top panels. The gold star with the horizontal error bar shows the mid-point and $1\sigma$ uncertainty of reionisation inferred from \citet{aghanimPlanck2018Results2020b}. \textit{Bottom right:} the same as the bottom left but for solar metallicity.}
    \label{fig:reionisation_stars}
\end{figure*}

Figure~\ref{fig:reionisation_stars} shows both $\dot{n}_{\mathrm{ion}}(z)$ and the neutral fraction inferred from the CSFH and assuming ionising photons only originate from stars. Two constant metallicities, $Z=10^{-4}$ and $Z=0.02$, are shown to bound the extremes of, essentially, primordial and solar chemistry for the stars. 

As expected, more ionising photons are produced in the primordial metallicity case compared to solar with the normalisation of $\dot{n}_{\mathrm{ion}}(z)$ shifting up by $\approx 0.5$~dex. The solar model can however mimic the primordial metallicity case when $f_{\mathrm{esc}}$ is sufficiently high. It is clear that stars alone could have reionised the IGM by $z\approx6$ with $f_{\mathrm{esc}} \gtrsim 10\%$ for primordial metallicity and $f_{\mathrm{esc}} \gtrsim 30\%$ for solar metallicity. 

\subsection{Case II: AGN only}\label{subsec:AGNlim}
\begin{figure*}
    \centering
    \includegraphics[width=\linewidth]{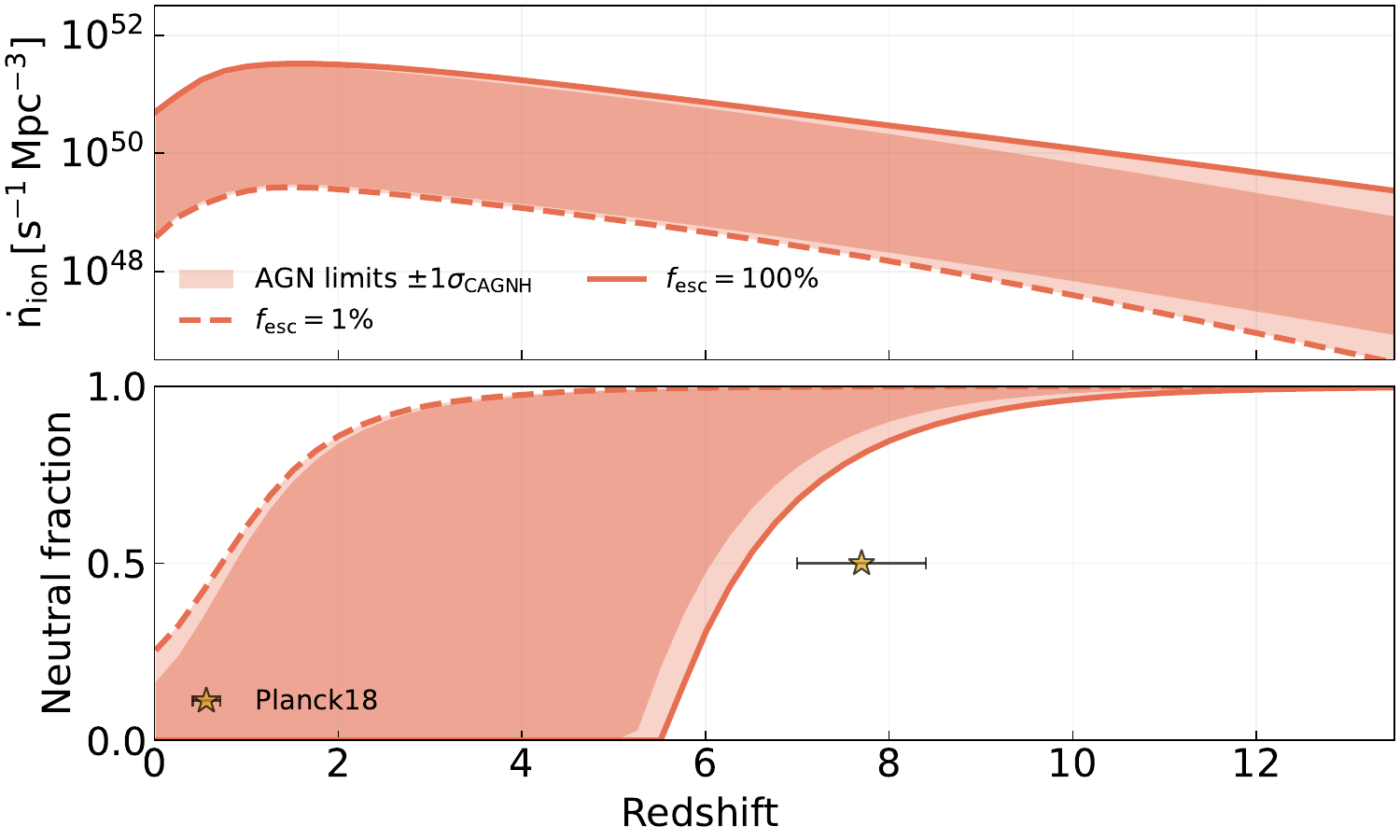}
    \caption{\textit{Top:} cosmic ionising emissivity as a function of redshift, $\dot{n}_{\mathrm{ion}}(z)$, from AGN only. The red filled region minimum/maximum bounds by considering $f_{\mathrm{esc}}=1\%$/$100 \%$ with the lighter red showing the $1 \sigma$ uncertainty of the CAGNH fit propagated through. \textit{Bottom:} the neutral fraction as per Equation~\ref{eq:madaureionize}. The filled regions have the same meaning as the top panels. The gold star with the horizontal error bar shows the mid-point and $1\sigma$ uncertainty of reionisation inferred from \citet{aghanimPlanck2018Results2020b}.}
    \label{fig:reionisation_AGN}
\end{figure*}

Figure~\ref{fig:reionisation_AGN} shows both $\dot{n}_{\mathrm{ion}}(z)$ and the neutral fraction inferred from the CAGNH and assuming only ionising photons originate from AGN. Even with $f_{\mathrm{esc}}=100\%$, AGN alone would have struggled, if not failed, to have achieved reionisation by $z\approx 6$ \citep[though see, for example,][]{madauCosmicReionizationPlanck2015,madauCosmicReionizationJWST2024}. Special attention is drawn to the CMB constraint of the neutral fraction. It can be seen that even completely unobscured AGN alone do not converge on this result for even $f_{\mathrm{esc}}=100\%$, implying that AGN do not quite produce the required cosmic ionising emissivity. 

To further investigate this we calculated the optical depth of the CMB photons: 
\begin{multline}
    \tau_{\mathrm{CMB}}(<z) = \\ 
    \int^{z}_{0}\sigma_{T} n_{e}^{\mathrm{com}}(z') \times Q(z') \times (1+z')^{2} \times\frac{c}{H(z')}  dz' ,
        \label{eq:tauCMB}
\end{multline}
where $\sigma_{T}$ is the Thomson scattering cross section and $n_{e}^{\mathrm{com}}(z)$ is the comoving electron number density. In detail,
\begin{equation}
    n_{e}^{\mathrm{com}}(z) = (1+z)^{-3} \times n_{e}(z).
\end{equation}
The \textit{Planck} result is $\tau_{\mathrm{CMB}} = 0.054 \pm 0.007$ that is an integrated value obtained over the entire path length of the photons since recombination \citep{aghanimPlanck2018Results2020a,aghanimPlanck2018Results2020b}. Equation~\ref{eq:tauCMB} is however a cumulative measure of CMB scattering from free electrons produced up to $z=0$. Hence, $\lim\limits_{z\to0 }\tau_{\mathrm{CMB}}(<z) =0$ because $\lim\limits_{z\to0}Q(z)=1$ and there were negligible new free electrons that were not already produced at earlier times. 

\begin{figure}
    \centering
    \includegraphics[width=\columnwidth]{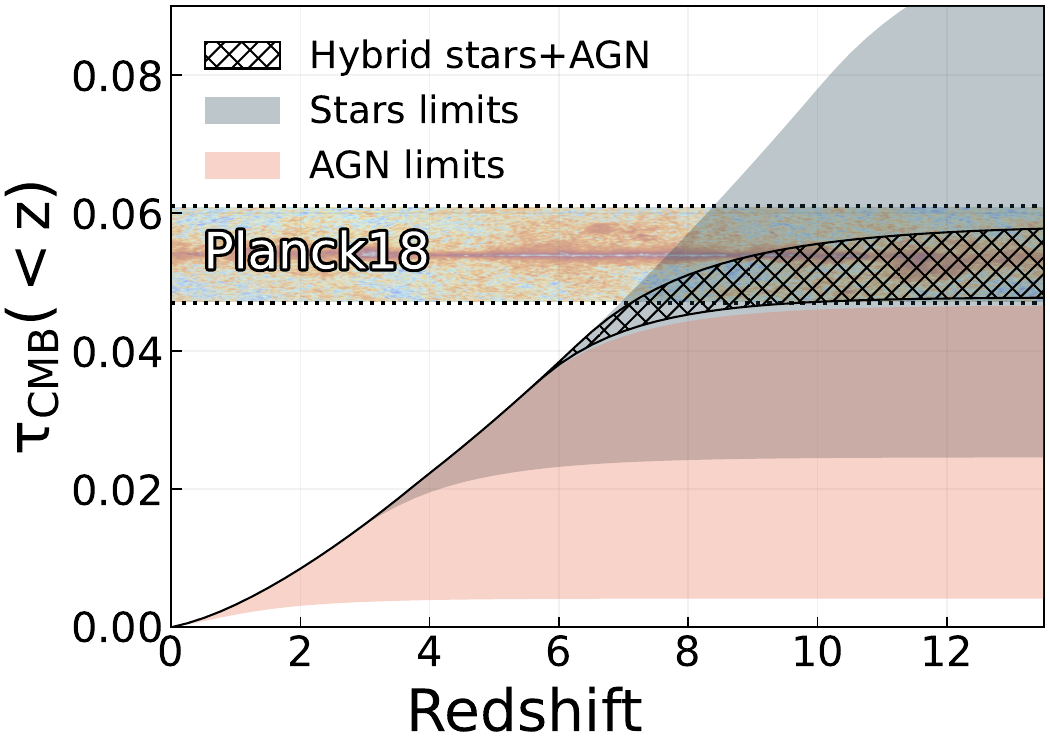}
    \caption{Optical depth of CMB photons as a function of redshift, $\tau_{\mathrm{CMB}}(<z)$. The black hatched region shows the results for the hybrid model as in Figure~\ref{fig:reionisation_hybrid}. The red filled region shows the same AGN limits as in Figure~\ref{fig:reionisation_AGN}. The blue filled region shows the same stellar limits as in Figure~\ref{fig:reionisation_stars} except that the upper/lower bounds correspond to primordial/solar metallicity to encompass the entire range of stellar chemistry. The filled horizontal band shows the $1\sigma$ range obtained from the \textit{Planck} analysis of the CMB \citep{aghanimPlanck2018Results2020a,aghanimPlanck2018Results2020b}.}
    \label{fig:tauCMB}
\end{figure}
Figure~\ref{fig:tauCMB} shows the results for $\tau_{\mathrm{CMB}}(<z)$ using our inferred $Q(z)$. At the very least, for $100\%$ escaping photons, the CAGNH would need to be $\approx 0.5$~dex greater in normalisation to solely achieve reionisation by $z\approx6$. Hence, it is clear that AGN alone do not produce sufficiently many ionising photons and free electrons to be consistent with the CMB optical depth, meaning that there must have been some contribution from the stars, even despite AGN being abundant at $z\gtrsim5$. 

\subsection{Case III: Hybrid stars and AGN}\label{subsec:hybrid}
While the analysis in Section~\ref{subsec:starlim} has shown that stars could have been the sole contributors to reionisation, there would most likely have been a mixture of ionising photons from both stars and AGN since the CSFH and CBHARH are both ramping up, as demonstrated in Figure~\ref{fig:cbharhvscsfh}. So, we explored this with a hybrid reionisation model. For this model of a single canonical galaxy, we assumed the closed box chemical enrichment scheme in \textsc{ProSpect} where the metallicity is tied to the (cosmic) SFH and hence evolves with redshift. The yield was set as $y=0.03$, appropriate for the \citet{chabrierGalacticStellarSubstellar2003} IMF \citep{madaudickinson2014}. In this formalism, the metallicity is that of the final generation of stars at the given epoch that is, in fact, the instantaneous gas phase metallicity of the canonical galaxy in the closed box model \citep{thorneDEVILSCosmicEvolution2022}. We assume that the metallicity is $Z_{\mathrm{start}}=10^{-3}$ at $z=15.5$ when the integration begins to be consistent with the metallicities of galaxies at $z\gtrsim 10$ \citep[e.g.,][]{nakajimaJWSTCensusMassMetallicity2023a,heintzDilutionChemicalEnrichment2023}. The metallicity at $z=0$ is assumed to be $Z_{\mathrm{final}}=0.02$ that was found to be appropriate to recover the optical extragalactic background light with a similar \textsc{ProSpect} setup \citep{koushanGAMADEVILSConstraining2021}. 

To infer the redshift-invariant, volume-averaged $f_{\mathrm{esc}}$ for both the AGN and stars of our canonical galaxy, we fit for the $z\approx 4.75$ constraint of $\dot{n}_{\mathrm{ion}}$ from \citet{beckerNewMeasurementsIonizing2013}, the CMB optical depth at $z>15$ and the $z = 3$ ratio of $\dot{n}_{\mathrm{ion}}$ between AGN and stars that is $\approx 1$. The last constraint is from \citet{smithLymanContinuumEscape2020} who used \textit{HST} WFC3-UVIS to show that the $1\sigma$ half Gaussian upper limits of Lyman continuum signal between star forming and AGN dominated galaxies are approximately unity at $z = 3$. To translate these upper limits to a ratio, we did a Monte-Carlo exercise where we sampled the half Gaussian upper limits and computed the resulting ratio of Lyman continuum signal from AGN and star formation. The resulting distribution of the ratios is log-normal and we empirically found that the standard deviation is $\approx 0.68$. Hence, we implemented the ratio of $\dot{n}_{\mathrm{ion}}(z = 3)$ from AGN to stars constraint as a log-normal distribution with $\mu = 0$ and $\sigma = 0.68$. Importantly, we did not use observations of the neutral fraction to constrain our model so that we could have a true prediction of the reionisation timeline. 

\begin{figure}
    \centering
    \includegraphics[width=\columnwidth]{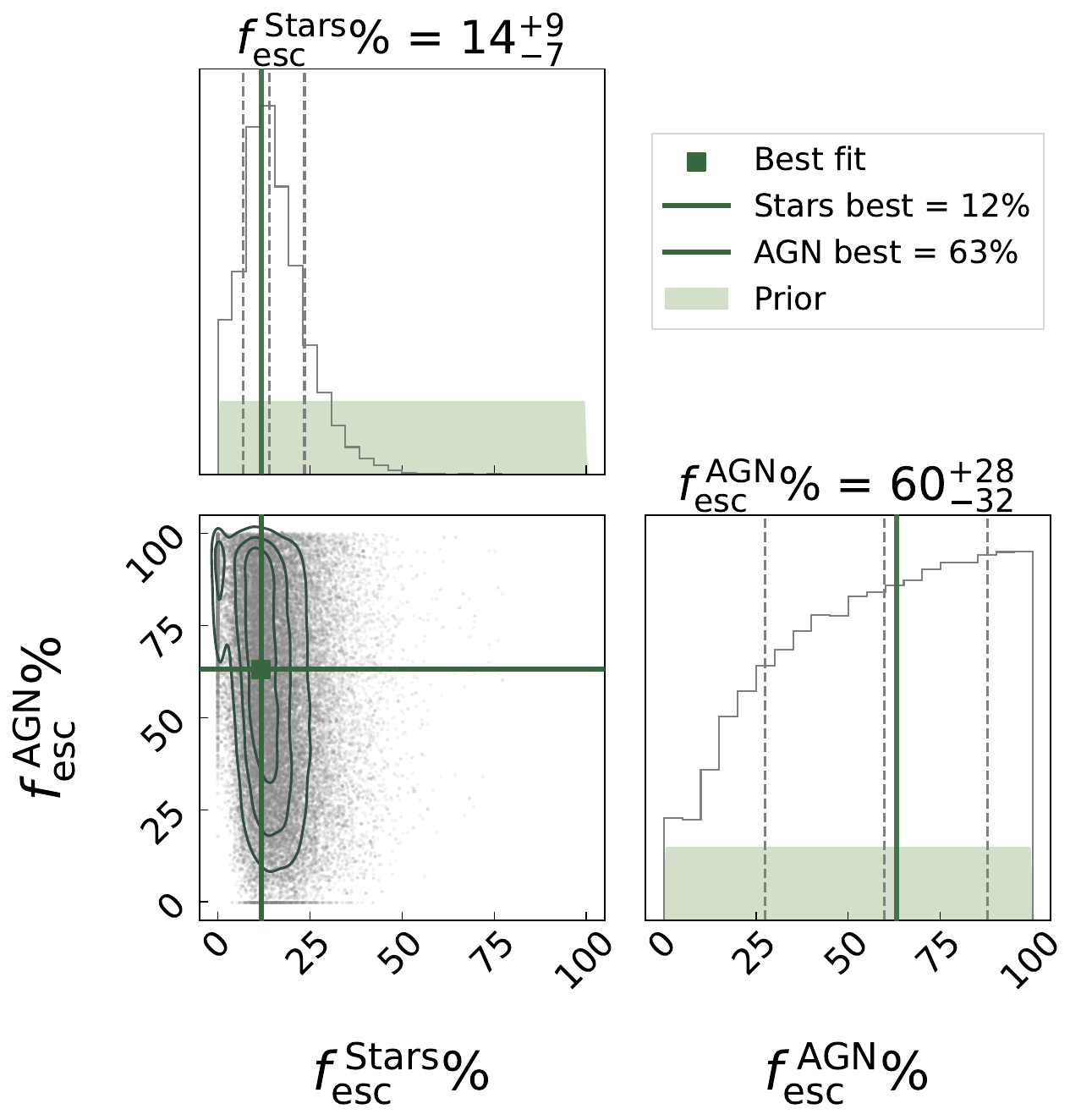}
    \caption{Corner plot of the posterior distribution of $f_{\mathrm{esc}}$ for both stars (top left) and AGN (lower right) when fitting to the \citet{beckerNewMeasurementsIonizing2013} $\dot{n}_{\mathrm{ion}}$ at $z\approx4.75$, the CMB optical depth at $z>15$ and the unity ratio of $\dot{n}_{\mathrm{ion}}(z=3)$ from AGN to stars found by \citet{smithLymanContinuumEscape2020}. The green shaded regions show the uniform prior distributions on $f_{\mathrm{esc}}$. Vertical dashed lines for the 1D histograms are the median and $16-\nth{84}$ percentiles. The median and $16-\nth{84}$ are quoted on top of the histograms, while the best fit values are quoted in the legend. Black contours (bottom left) show the \nth{16}, \nth{50} and \nth{84} percentile levels of the posterior samples. The green square in the bottom left panel and the green lines in all panels show the maximum likelihood/best fit values.}
    \label{fig:cornerplot}
\end{figure}

We used maximum likelihood analysis combined with the fitting software \textsc{Highlander}\footnote{\url{https://github.com/asgr/Highlander}} to explore the posterior distribution. The escape fraction cannot physically be less than $0$ or greater than $1$. As such, when fitting $f_{\mathrm{esc}}$ for both stars and AGN we used a uniform prior between those bounds to disfavour unphysical solutions. Figure~\ref{fig:cornerplot} shows the corner plot of the posterior distribution of $f_{\mathrm{esc}}$ for both stars and AGN. The prior distributions are shown as the green shaded regions. The median and $16-\nth{84}$ percentiles from the posteriors are $f_{\mathrm{esc}}=$ \fescstars{} for the stars and $f_{\mathrm{esc}}=$ \fescagn{} for the AGN. The spread in these escape fractions mostly reflects the uncertainty in the ratio of $\dot{n}_{\mathrm{ion}}$ from AGN to stars, motivating the need to further investigate Lyman continuum in both stellar and AGN dominated galaxies throughout redshift and over many lines of sight. The maximum likelihood/best fit values that we found are $f_{\mathrm{esc}}\approx$ \fescstarsbest{} for the stars and $f_{\mathrm{esc}}\approx$ \fescagnbest{} for the AGN. The hybrid model that is discussed in the remainder of the text uses the maximum likelihood/best fit values for $f_{\mathrm{esc}}$. 

\begin{figure*}
    \centering
    \includegraphics[width=\linewidth]{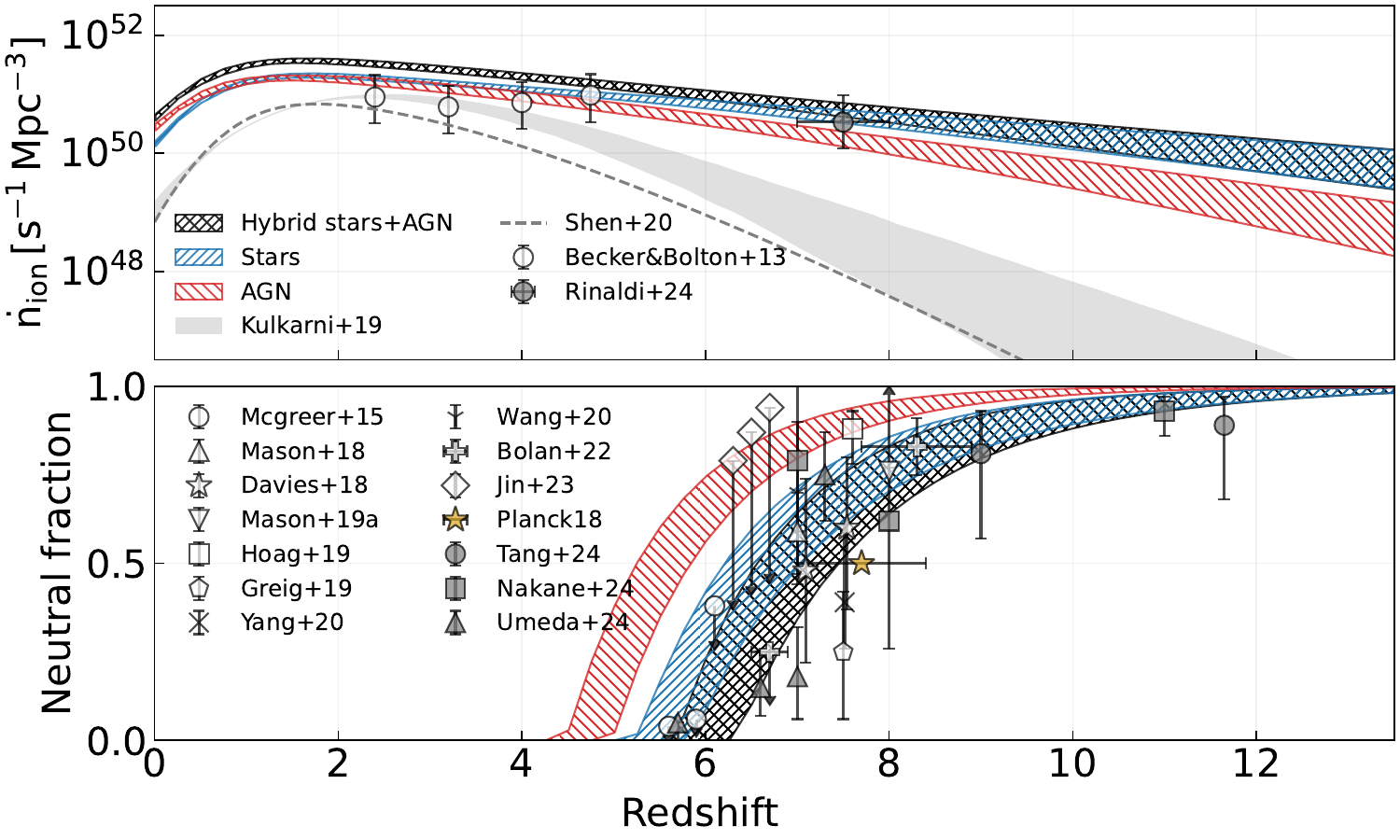}
    \caption{\textit{Top:} cosmic ionising emissivity as a function of redshift, $\dot{n}_{\mathrm{ion}}(z)$, for the hybrid model and inferred $f_{\mathrm{esc}}=$ \fescstarsbest{} for the stars and $f_{\mathrm{esc}}=$ \fescagnbest{} for the AGN. The black hatched corresponds to the combined contribution from stars and AGN, the blue corresponds to the contribution from stars and the red corresponds to the contribution from AGN. The spread of the hatched regions reflects the spread in the fitted CSFH and CAGNH. We show observational constraints and their $1\sigma$ uncertainties from \citet{beckerNewMeasurementsIonizing2013}. \textit{Bottom:} the neutral fraction as a function of redshift as per Equation~\ref{eq:madaureionize}. The hatched regions have the same meaning as the top panels. We show results and $1\sigma$ uncertainties from \citet{masonUniverseReionizing72018,daviesQuantitativeConstraintsReionization2018,hoagConstrainingNeutralFraction2019,greigConstraintsReionization752019, yangPoniuaenaLuminous752020, wangSignificantlyNeutralIntergalactic2020,bolanInferringIntergalacticMedium2022,tangJWSTNIRSpecObservations2024,nakaneLyaEmission7132024a,umedaSILVERRUSHXIVLya2024} as indicated in the legend. Upper/lower limit observations from \citet{mcgreerModelindependentEvidenceFavour2015,masonInferencesTimelineReionization2019, bolanInferringIntergalacticMedium2022, jinNearlyModelindependentConstraints2023} are shown with arrows as indicated. The gold star with the horizontal error bar shows the mid-point and $1\sigma$ uncertainty of reionisation redshift inferred from \citet{aghanimPlanck2018Results2020b}.}
    \label{fig:reionisation_hybrid}
\end{figure*}

Figure~\ref{fig:reionisation_hybrid} shows the key results of $\dot{n}_{\mathrm{ion}}(z)$ and the neutral fraction while Figure~\ref{fig:tauCMB} shows the key result of the CMB optical depth for the hybrid model. There is overall agreement between the hybrid model and these key observables. 

Our model $\dot{n}_{\mathrm{ion}}(z)$ for the AGN is $\approx 1-5$~dex higher at $z\approx 5-13$ than previous results from \citet{kulkarniEvolutionAGNUV2019}, who compute UV luminosity functions for AGN up to $z\approx 7.5$, and \cite{shenBolometricQuasarLuminosity2020}, who compile X-ray, UV and IR AGN luminosity functions up to $z\approx 7$. The main reason for this difference is likely that the $z>5$ AGN number densities from \textit{JWST} are $\approx 1$~dex greater than the pre-\textit{JWST} measurements because of the ability to better reveal early AGN than previously possible, as noted in \citetalias{dsilvaSelfConsistentJWSTCensus2025}.

We draw special attention to the neutral fraction since none of the shown observations were used to constrain the model. While stars drive reionisation in the hybrid model, the average ionising photon contribution from AGN increases by a factor of $\approx 3$ in comparison between $z\approx 11 \to 6$ that reflects the relative rate of change increase of the CAGNH (and therefore CBHARH) compared to the CSFH. Hence, while stars were most likely the dominant sources of reionisation, AGN would have supplied necessary photons to keep the IGM ionised at $z\lesssim 6$. 

\begin{figure*}
    \centering
    \includegraphics[width=\linewidth]{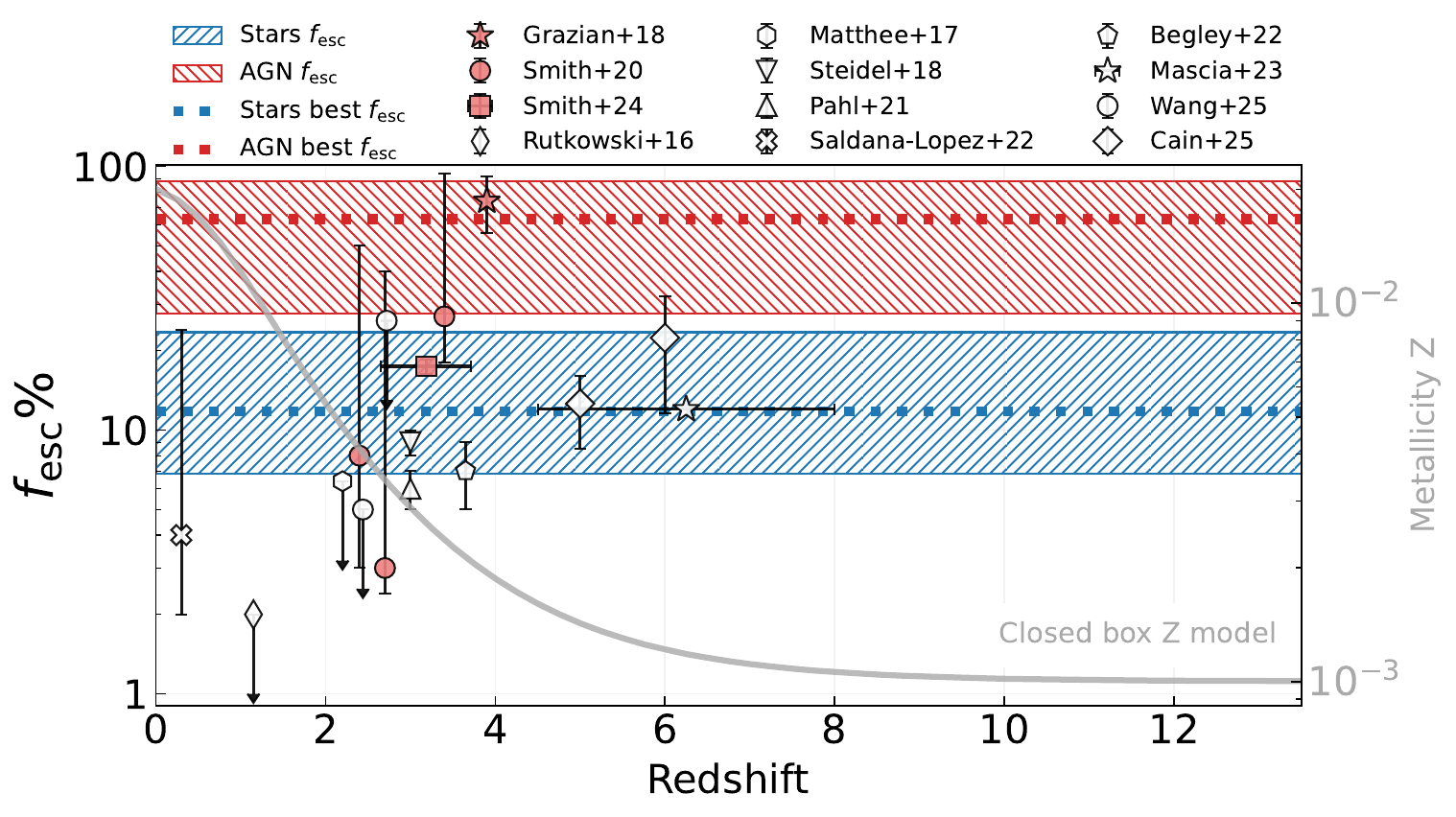}
    \caption{Escape fraction, $f_{\mathrm{esc}}$, as a function of redshift. The blue hatched box shows the $1\sigma$ ranges, $f_{\mathrm{esc}}=$ \fescstars{}, and the dotted line shows the best fit, $f_{\mathrm{esc}}=$ \fescstarsbest{}, for the stars. The red hatched box shows the $1\sigma$ ranges, $f_{\mathrm{esc}}=$ \fescagn{}, and the line shows the best fit, $f_{\mathrm{esc}}=$ \fescagnbest{}, for the AGN. Observational constraints and $1\sigma$ uncertainties of $f_{\mathrm{esc}}$ for star forming galaxies from \citet{steidelKeckLymanContinuum2018,pahlUncontaminatedMeasurementEscaping2021,saldana-lopezLowRedshiftLymanContinuum2022,begleyVANDELSSurveyMeasurement2022,masciaClosingSourcesCosmic2023,cainNewConstraintsGalactic2025} are shown as indicated. $1\sigma $ upper limits of $f_{\mathrm{esc}}$ for star forming galaxies from \citet{rutkowskiLYMANCONTINUUMESCAPE2016,mattheeProductionEscapeLymanContinuum2017,wangLymanContinuumEscape2025} are shown with down-facing arrows. Light red data points with error bars are from \citet{smithLymanContinuumEscape2020, smithLymanContinuumEmission2024} and \citet{grazianContributionFaintAGNs2018} and show results and $1\sigma$ uncertainties of $f_{\mathrm{esc}}$ for AGN. The gray line shows the closed box metallicity history on the right-side vertical axis.}
    \label{fig:fesc}
\end{figure*}

\subsection{Comparison of $f_{\mathrm{esc}}$ from the literature}
It can be seen that the median $\dot{n}_{\mathrm{ion}}(z)$ of our model at $z\lesssim 4$ is in excess by $\approx 0.3-0.7$~dex compared to the median of the observations by \citet{beckerNewMeasurementsIonizing2013}. Our $\dot{n}_{\mathrm{ion}}(z)$ for the AGN is also $\approx 0.5-1.5$~dex higher at $z\lesssim 2$ than \citet{kulkarniEvolutionAGNUV2019} and \citet{shenBolometricQuasarLuminosity2020}. This suggests that $f_{\mathrm{esc}}$ for both stars and AGN are most probably functions of redshift \citep[e.g.,][]{shullCriticalStarFormation2012}. Indeed, $f_{\mathrm{esc}}$ for both AGN and stellar dominated galaxies was found to anticorrelate with, for example, the gas-phase metallicity of the galaxies and the slope, $\beta$, of the UV spectrum that both evolve with redshift \citep{leitetEscapeLymanContinuum2013,chisholmFarultravioletContinuumSlope2022,masciaClosingSourcesCosmic2023,smithLymanContinuumEmission2024}. 

Figure~\ref{fig:fesc} shows our escape fractions compared to results from the literature across redshift. For the AGN, $f_{\mathrm{esc}}=$ \fescagnbest{} is much higher than $f_{\mathrm{esc}}\approx18 \pm 1\%$ at $z\approx 3.6$ reported by \citet{smithLymanContinuumEmission2024}. This result was obtained from WFC3-UVIS observations of 12 broad-line AGN and fitting their SEDs with the software CIGALE \citep{boquienCIGALEPythonCode2019b}, where the average $f_{\mathrm{esc}}$ was calculated as the uncertainty weighted mean of the 12 individual escape fractions. The highest redshift $f_{\mathrm{esc}}$ for AGN at $z\approx 3.4$ from \citet{smithLymanContinuumEscape2020} agrees within $1\sigma$ with our constraint. Our result agrees closer with that from \citet[][$f_{\mathrm{esc}} \approx 74\%$]{grazianContributionFaintAGNs2018} that is the mean value from a sample of 16 AGN. As noted by \citet{grazianContributionFaintAGNs2018} the individual escape fractions of their studied AGN were between $44-100\%$.  In general, observations of the AGN $f_{\mathrm{esc}}$ are uncertain due to both low number statistics and the intrinsic faintness of Lyman continuum. Our inferred $f_{\mathrm{esc}}=$ \fescstarsbest{} for the stars is slightly lower than $f_{\mathrm{esc}}= 20\%$ that is often assumed in reionisation models \citep[e.g.,][]{ouchiLARGEAREASURVEY2009,robertsonNEWCONSTRAINTSCOSMIC2013,robertsonCosmicReionizationEarly2015a,munozReionizationJWSTPhoton2024}. The evolution of the $f_{\mathrm{esc}}$ for the star forming galaxies is clear, going from $\approx 10\%$ at $z\gtrsim 3$ \citep{mattheeProductionEscapeLymanContinuum2017,steidelKeckLymanContinuum2018,pahlUncontaminatedMeasurementEscaping2021,begleyVANDELSSurveyMeasurement2022,masciaClosingSourcesCosmic2023,wangLymanContinuumEscape2025} to $\approx 2 \%$ by $z=0.4-1.2$ \citep{rutkowskiLYMANCONTINUUMESCAPE2016,saldana-lopezLowRedshiftLymanContinuum2022} that is contradictory with our assumed constant $f_{\mathrm{esc}}$. Despite this, the hybrid model agrees with the CMB optical depth (by construction) and the general reionisation history at $z\gtrsim 6$. Within the context of the hybrid model, this means that $f_{\mathrm{esc}}$ does not change much over the first billion years after the Big Bang to significantly affect the reionisation timeline. Without an evolving $f_{\mathrm{esc}}$ the increase of $\dot{n}_{\mathrm{ion}}(z)$ up to $z\approx 6$ for the given CSFH and CAGNH is somewhat controlled by the closed-box metallicity history because metal-rich stars, which become more numerous with time, intrinsically produce fewer ionising photons (as shown in Figure~\ref{fig:reionisation_stars}). There are, however, additional processes not implicit in our \textsc{ProSpect} model that affect $f_{\mathrm{esc}}$. Metal-rich systems tend to be more optically thick to ionising photons compared to metal-poor systems because the relatively lower radiation pressure inhibits the formation of effective photon escape channels \citep{reddyCONNECTIONREDDENINGGAS2016,yooOriginLowEscape2020} and because dust is relatively more prevalent \citep{remy-ruyerGastodustMassRatios2014,vijayanDetailedDustModelling2019,chisholmFarultravioletContinuumSlope2022}. 

\subsection{Comparison of $\mathcal{C}$ from the literature}
Additionally, the clumping factor, $\mathcal{C}$, may also change with redshift. \citet{masonModelindependentConstraintsHydrogenionizing2019} showed that $\dot{n}_{\mathrm{ion}}(z)$ does not change much except for only extreme values of $\mathcal{C} \approx 10-30$, which is ostensibly at odds with numerical simulations of the IGM where $\mathcal{C} \approx 1-6$ at $z\gtrsim 6$ \citep{kaurovCOSMICREIONIZATIONCOMPUTERS2015}. More recently however, \citet{daviesPredicamentAbsorptiondominatedReionization2024} showed that the mean free path of ionising photons and the photoionisation rates at $z\approx 5$ imply that $\mathcal{C} \approx 12$, much higher than our assumed $\mathcal{C}=3$. 

\begin{figure}
    \centering
    \includegraphics[width=\columnwidth]{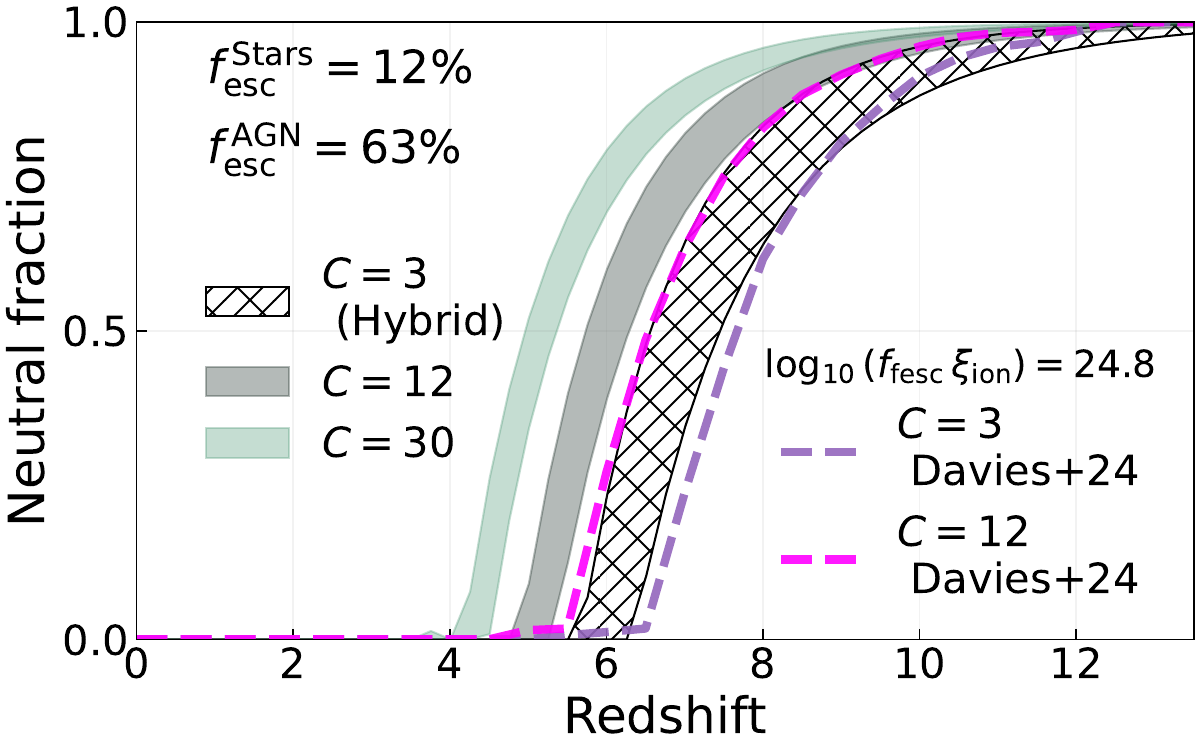}
    \caption{Neutral fraction as a function of redshift using the same $f_{\mathrm{esc}}$ for both stars and AGN found in the previous section. The hatched and filled regions show the results for $\mathcal{C}=3,12,30$ as indicated in the legend. The width of the regions corresponds to the $1\sigma$ uncertainties from the CSFH and CAGNH (as in Figure~\ref{fig:reionisation_hybrid}). The dashed lines are from \citet{daviesPredicamentAbsorptiondominatedReionization2024} corresponding to $\mathcal{C}=3,12$ as indicated. \citet{daviesPredicamentAbsorptiondominatedReionization2024} assumed that $\log_{10}( f_{\mathrm{esc}} \, \xi_{\mathrm{ion}} ) = 24.8$ (see Equation~\ref{eq:nions_galaxy}), which is why their curves are generally shifted to the right in comparison to ours.}
    \label{fig:neutral_clump}
\end{figure}

Figure~\ref{fig:neutral_clump} shows the neutral fraction as a function of redshift for $\mathcal{C}=3,12,30$ using the same $f_{\mathrm{esc}}$ that we found for the hybrid model. As expected, the reionisation timeline gets stretched to lower redshifts as we increase $\mathcal{C}$, with the relative increase in the redshift between $\mathcal{C}=3$ and $\mathcal{C}=12 \, (30)$ being $\Delta z \approx 1.0 \, (1.5)$. This relative difference in the reionisation timeline as a function of $\mathcal{C}$ is in qualitative agreement with the results of \citet{daviesPredicamentAbsorptiondominatedReionization2024}; although, they compute the results for different assumptions about $f_{\mathrm{esc}}$ and $\xi_{\mathrm{ion}}$ compared to our approach. 

\begin{figure}
    \centering
    \includegraphics[width=\columnwidth]{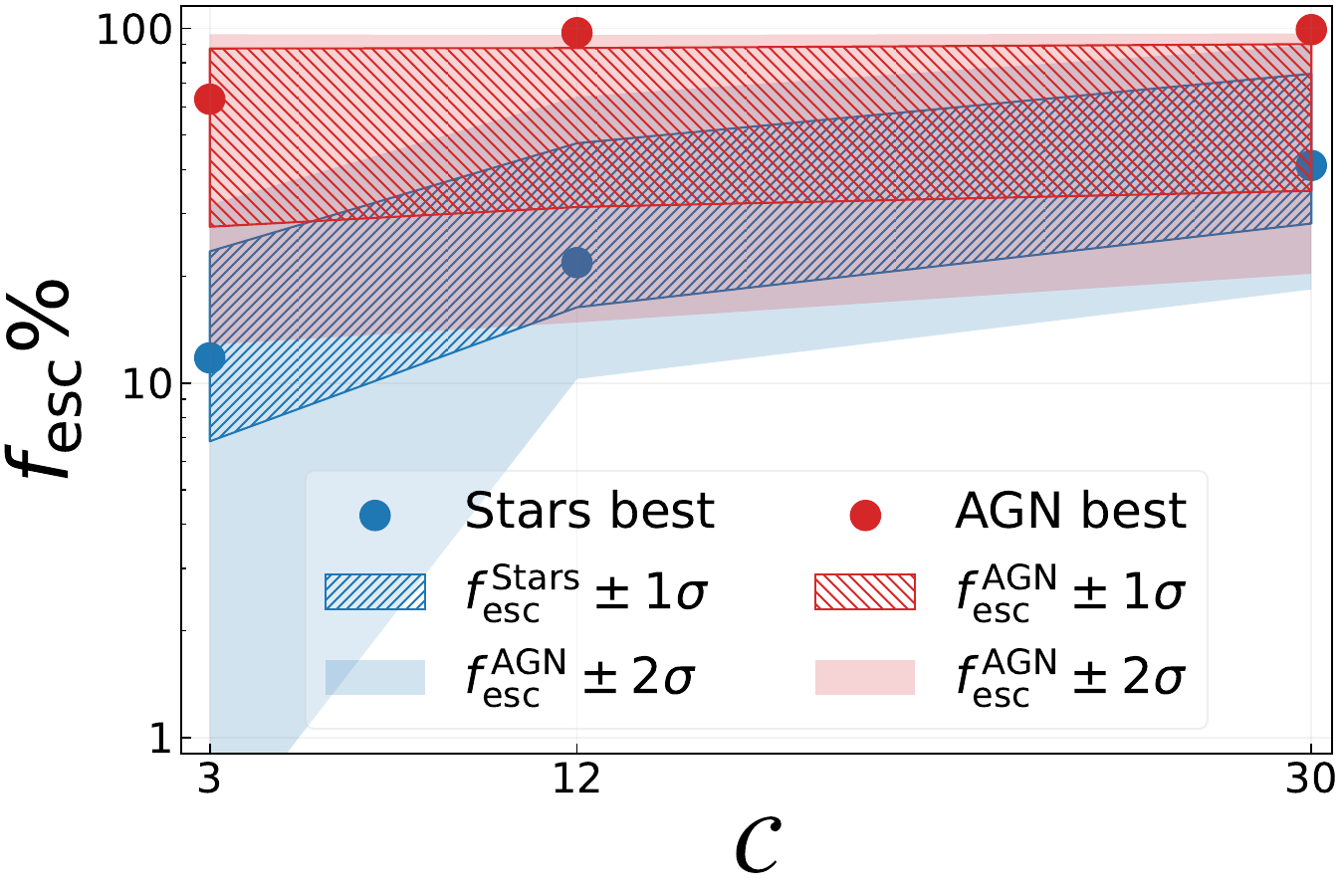}
    \caption{$f_{\mathrm{esc}}$ as function of $\mathcal{C}$. The blue hatched region shows the $1\sigma$ ranges and the circle shows the best fit for the stars. The red hatched region and circle shows the $1\sigma$ ranges and best fit for the AGN. The filled shaded regions show the $2\sigma$ ranges with the same colour scheme as above.}
    \label{fig:fesc_clump}
\end{figure}

Figure~\ref{fig:fesc_clump} shows how the assumed value of $\mathcal{C}$ effects the fitted $f_{\mathrm{esc}}$ for the stars and AGN. It can be seen that $f_{\mathrm{esc}}$ for the AGN is unaffected because the model already settles on a high value of $f_{\mathrm{esc}}$ even for $\mathcal{C}=3$. The stars however show a tentative increase in $f_{\mathrm{esc}}$ with an increase in $\mathcal{C}$. In terms of the best fit values, $f_{\mathrm{esc}}$ would need to increase by a factor of $\approx 1.9 \, (3.5)$ for $\mathcal{C}=12 \, (30)$ to be consistent with the measurements of $\dot{n}_{\mathrm{ion}}(z)$ and the CMB optical depth at $z>15$ used to constrain the model and described in Section~\ref{subsec:hybrid}. In any case, as we change $\mathcal{C}$, the resultant $f_{\mathrm{esc}}$ for both stars and AGN are still within their $1\sigma$ uncertainties. As such, we reason that the main limitation on the predictive power of the model is more so due to the uncertainty on $f_{\mathrm{esc}}$, with second order effects due to the uncertainty on $\mathcal{C}$. 

\subsection{Comparison to other models} \label{subsec:othermodels}
\begin{figure}
    \centering
    \includegraphics[width=\columnwidth]{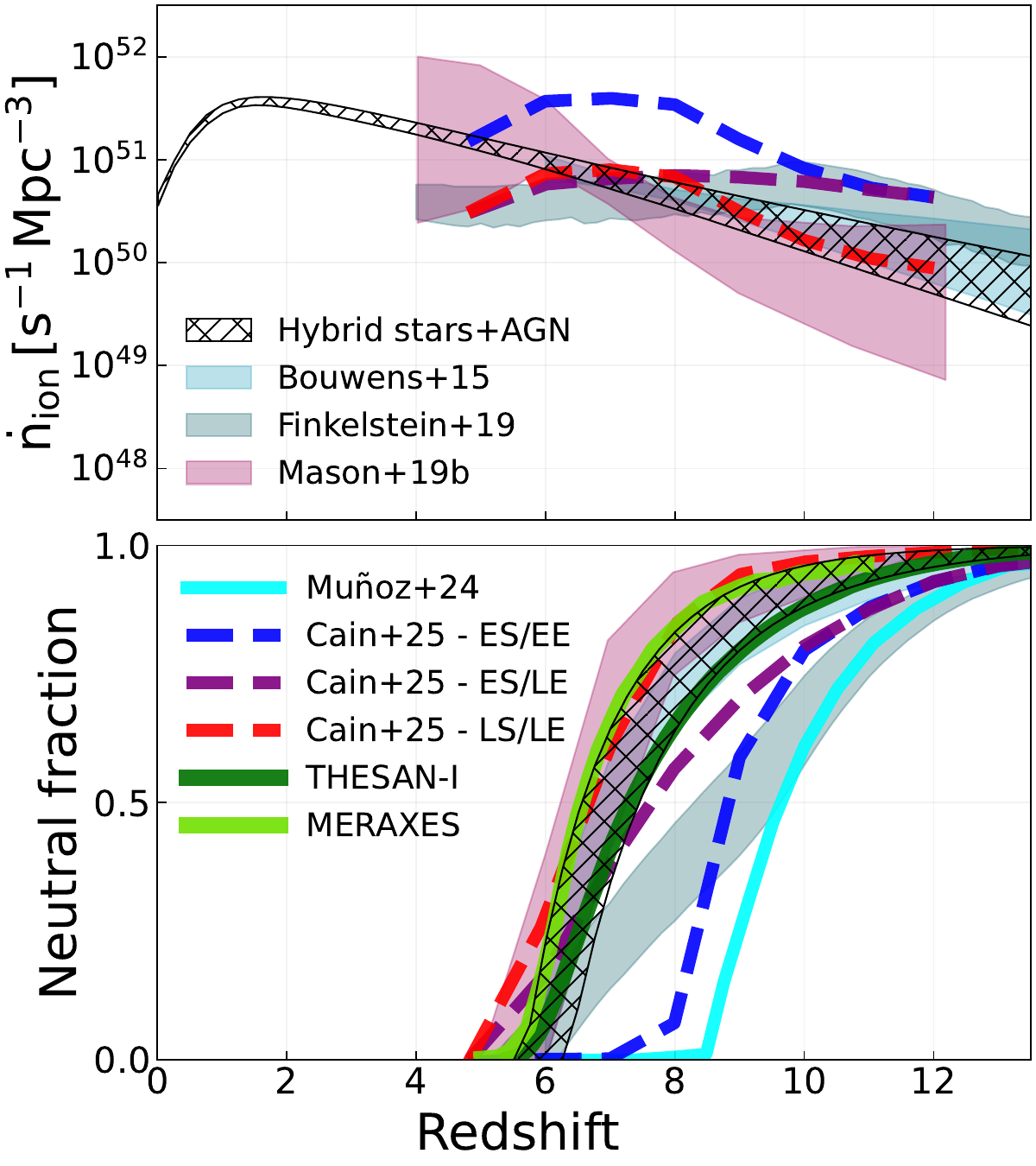}
    \caption{\textit{Top:} $\dot{n}_{\mathrm{ion}}(z)$ as a function of redshift. The black hatched region is the result for the hybrid model as in Figure~\ref{fig:reionisation_hybrid}. We show with filled regions the $1\sigma$ spread of the models from \citet{masonModelindependentConstraintsHydrogenionizing2019} in pink,  \citet{finkelsteinConditionsReionizingUniverse2019} in dark blue and \citet{bouwensREIONIZATIONPLANCKDERIVED2015} in light blue. The three different reionisation models: early start/early end (ES/EE), early start/late end (ES/LE) and late start/late end (LS/LE) from \citet{cainChasingBeginningReionization2025} are shown with blue, purple and red dashed lines.
    \textit{Bottom:} neutral fraction as a function of redshift. The filled regions and dashed lines have the same meaning as the top panel. The cyan line is the JWST-$\xi_{\mathrm{ion}}$ model from \citet{munozReionizationJWSTPhoton2024}. We also show with green lines the results from the \textsc{Thesan-I} \citep{kannanIntroducingThesanProject2022} and the \textsc{Meraxes} \citep{baluThermalReionizationHistory2023} simulations.}
    \label{fig:reionisationModels}
\end{figure}

Figure~\ref{fig:reionisationModels} shows a comparison between the hybrid model in this work and other models from the literature. In particular, \citet{bouwensREIONIZATIONPLANCKDERIVED2015} and \citet{masonModelindependentConstraintsHydrogenionizing2019} both inferred $\dot{n}_{\mathrm{ion}}(z)$ by essentially fitting observations of neutral fractions (similar to the ones shown in Figure~\ref{fig:reionisation_hybrid}) and the CMB optical depth \citep{adamPlanckIntermediateResults2016,aghanimPlanck2018Results2020b}, essentially solving for $1-Q(z)$ using Equation~\ref{eq:madaureionize}. Hence, these comparisons are the most similar to our work and there is general agreement between them. 

\citet{finkelsteinConditionsReionizingUniverse2019} combined extrapolations of the UV luminosity function and halo mass-dependent escape fractions inferred from abundance matching $N$ body simulations to compute $\dot{n}_{\mathrm{ion}}(z)$. Compared to our results, the \citet{finkelsteinConditionsReionizingUniverse2019} model shows $\approx 1$~dex greater $\dot{n}_{\mathrm{ion}}(z)$ at $z\gtrsim 10$ and a flatter $\dot{n}_{\mathrm{ion}}(z)$ at $z\lesssim10$, resulting in an earlier genesis of reionisation where the neutral fraction is $0.5$ at $z\approx 9$. The reason for this difference is that in their model low mass galaxies with high $f_{\mathrm{esc}}$ dominate the ionising budget at early times causing an increase up to $z\approx 10$, while higher mass galaxies with lower $f_{\mathrm{esc}}$ dominate at later times causing the $\dot{n}_{\mathrm{ion}}(z)$ to flatten. 

\citet{munozReionizationJWSTPhoton2024} modelled the neutral fraction by combining UV luminosity functions of \citet{bouwensNewDeterminationsUV2021,donnanJWSTPRIMERNew2024} with $\xi_{\mathrm{ion}}$ from \citet{simmondsLowmassBurstyGalaxies2024}, and solving for $1-Q(z)$ using Equation~\ref{eq:madaureionize}. Their reionisation model starts early and finishes early at $z\approx 9$, in tension with our model, on account of the rapid evolution of $\xi_{\mathrm{ion}}$ at $4<z<9$. 

\citet{cainChasingBeginningReionization2025} presented three unique models of reionisation based on hydrodynamical simulations. Our model agrees with the late start/late end (LS/LE) scenario that was tuned to the CMB optical depth and the Ly$\alpha$ forest. The LS/LE scenario is a scaling down of $\dot{n}_{\mathrm{ion}}(z)$ from the early start/early end (ES/EE) scenario, which was itself motivated by results from \citet{munozReionizationJWSTPhoton2024}, to essentially demand that reionisation completed at $z\approx 5$. The early-start/late end scenario from \citet{cainChasingBeginningReionization2025} was constructed to lie within the ranges of the ES/EE and LS/LE models. Although our hybrid model agrees better with the LS/LE model, we could achieve agreement with the other two scenarios, especially the ES/LE, if for example the $f_{\mathrm{esc}}$ changes with redshift, being higher at $z\gtrsim 7$ and lower thereafter. \citet{cainChasingBeginningReionization2025} argue that the ES/EE model is incompatible with the CMB optical depth and the Ly$\alpha$ forest and it is hence safe to rule out that scenario. Between the LS/LE and ES/LE scenarios, no one observation such as the galaxy properties at $z\gtrsim 5$, Lyman alpha emitters, the CMB and QSO absorption spectra, can discern them, but together there is a preference for the LS/LE scenario--likewise favoured by our model. 

Finally, we compared our results of the neutral fraction with the full radiation-magnetohydrodynamic simulation \textsc{Thesan-I} \citep{kannanIntroducingThesanProject2022} and the semi-analytical model \textsc{Meraxes} \citep[simulation L210\_AUG,][]{baluThermalReionizationHistory2023}. In general, we find good agreement with all of these models potentially by construction as they also used the CMB optical depth as one of the constraints to tune their models. 

\section{Discussion} \label{sec:discussion}
The analysis presented in Section~\ref{sec:results} means that the $z=0-13.5$ CSFH and CAGNH are reconcilable with the established reionisation timeline. In this section, we draw our results into the context of Equation~\ref{eq:nions_galaxy} and calculate $\xi_{\mathrm{ion}}$.

\begin{figure*}
    \centering
    \includegraphics[width=\linewidth]{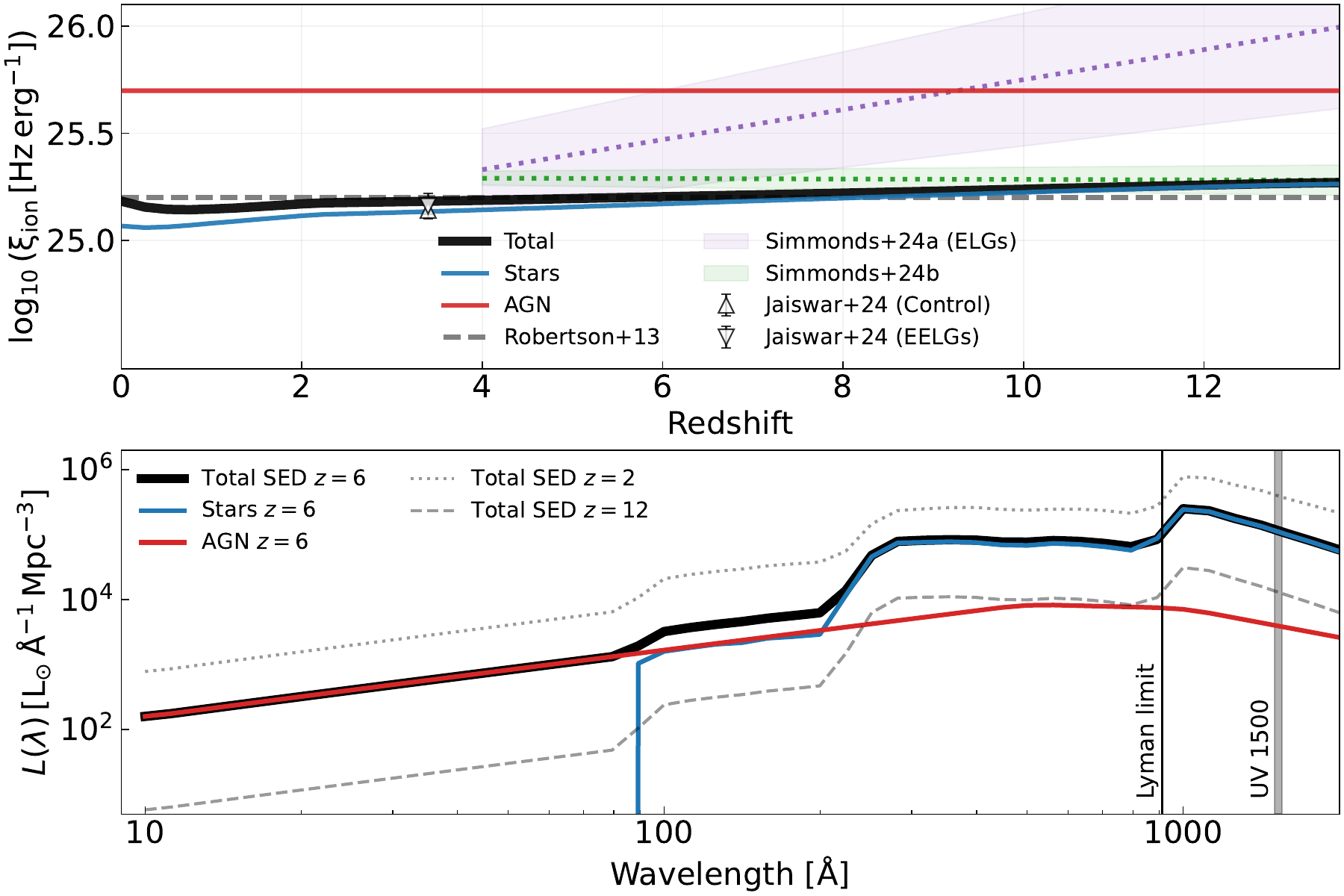}
    \caption{\textit{Top:} ionising efficiency, $\xi_{\mathrm{ion}}$, as a function of redshift. The black line shows the $\xi_{\mathrm{ion}}$ obtained from the median CSFH and CAGNH for the hybrid model, while the blue and red lines show the same for the stellar and AGN contributions. The dashed gray line shows the canonical value, $\log_{10}(\xi_{\mathrm{ion}})= 25.2$, from \citet{robertsonNEWCONSTRAINTSCOSMIC2013}. The dotted purple/green line and filled regions show the linear fits and $1\sigma$ uncertainties from \citet{simmondsLowmassBurstyGalaxies2024,simmondsIonizingPropertiesGalaxies2024}, where we are extrapolating at $z>9$. Triangle points with error bars show the results and $1\sigma$ uncertainties from \citet{jaiswarMOSELSurveyUnwrapping2024}. \textit{Bottom:} CSED from the \textsc{ProSpect} model computed at $z=6$. The black line shows the total SED obtained from the median CSFH and CAGNH, while the blue and red lines show the SED for the stars and AGN. The grey dotted and dashed lines shows the total SEDs at $z=2$ and $z=12$. The vertical line shows the Lyman limit, $\lambda=\SI{912}{\angstrom}$, and the vertical band shows the tophat filter of $\Delta\lambda=\SI{100}{\angstrom}$ centred on $\SI{1500}{\angstrom}$ for the UV.}
    \label{fig:xi_ion}
\end{figure*}

Figure~\ref{fig:xi_ion} shows $\xi_{\mathrm{ion}}$ obtained with \textsc{ProSpect} for the hybrid model. We divided $\dot{n}_{\mathrm{ion}}(z)$ by $\rho_{\mathrm{UV}}(z)$ that was computed from the CSEDs in a tophat filter centred on \SI{1500}{\angstrom} and width $\Delta\lambda=\SI{100}{\angstrom}$. This was computed for both unobscured stars and AGN, i.e., $f_{\mathrm{esc}}=0$, to demonstrate the intrinsic $\xi_{\mathrm{ion}}$ as per the \citetalias{bruzualStellarPopulationSynthesis2003} stellar population synthesis library and \citetalias{fritz06agnmodel} AGN model. In practice, $\xi_{\mathrm{ion}}$ for individual galaxies depends on the details of dust correction to obtain the intrinsic UV and ionising photon fluxes \citep{shivaeiMOSDEFSurveyDirect2018a}.

The hybrid model agrees with the canonical value of $\log_{10}(\xi_{\mathrm{ion}}) = 25.2$. The canonical value is obtained using the \citetalias{bruzualStellarPopulationSynthesis2003} stellar population synthesis library, the same fiducial stellar population synthesis library used in \textsc{ProSpect}. It is the approximate median ionising efficiency for the range of metallicities and constant SFRs that are appropriate for $z\approx 6$ galaxies with UV slopes, $\beta\approx -2$ \citep{robertsonNEWCONSTRAINTSCOSMIC2013,robertsonCosmicReionizationEarly2015a}. In detail, $\xi_{\mathrm{ion}}$ depends on metallicity and the shape of the CSFH. Indeed, the stellar $\xi_{\mathrm{ion}}$ that we find evolves with redshift, tracing both the CSFH and the metallicity history with the closed box model. As can be seen from the CSEDs in Figure~\ref{fig:xi_ion}, the integrated ionising photon flux increases with redshift, likewise increasing $\xi_{\mathrm{ion}}$, as the stars were more metal poor at ever higher redshifts.

Only the normalisation of the AGN SED changes with the CAGNH and so the AGN $\xi_{\mathrm{ion}}$ is constant in this model. $\xi_{\mathrm{ion}}$ for the AGN is $\gtrsim 0.5$~dex higher than the stars because the AGN in the \citetalias{fritz06agnmodel} model allows harder radiation than the stars as per the \citetalias{bruzualStellarPopulationSynthesis2003} stellar population synthesis library. This does not quite translate to a higher ionising photon contribution at $z\gtrsim 6$ since even modestly obscured stars, $f_{\mathrm{esc}} \gtrsim 15\%$, contribute more ionising photons than completely unobscured, $f_{\mathrm{esc}} \approx 100\%$, AGN. Therefore, the difference in the ionising contribution between stars and AGN is mostly due to the $\approx 1000$ times greater space density of star formation in comparison to accreting SMBHs at $z\gtrsim 6$. 

\subsection{Comparison to literature}
\citet{simmondsLowmassBurstyGalaxies2024} calculated $\xi_{\mathrm{ion}}$ with the SED fitting code \textsc{prospector}\footnote{Note that this is an entirely different code to \textsc{ProSpect} that is used in this work.} \citep{johnsonProspectorStellarPopulation2019,johnsonStellarPopulationInference2021} for $z\approx 4-9$ emission line galaxies from photometry in the JADES survey. They found that the median $\xi_{\mathrm{ion}}$ calculated for this sample of galaxies evolves significantly with redshift, being $\gtrsim 0.5$~dex greater than the canonical value, $\log_{10}(\xi_{\mathrm{ion}}) = 25.2$, at $z\gtrsim 6$. Their results for emission line, star forming galaxies lie $\gtrsim 0.5$~dex above what we might expect for the cosmic average (while recognizing that we fix the optical depths of the dust) as per the CSFH. \citet{munozReionizationJWSTPhoton2024}, using the results of $\xi_{\mathrm{ion}}$ from \citet{simmondsLowmassBurstyGalaxies2024}, predicted a reionisation history that would have ended far too early at $z\approx 9$ compared to the \textit{Planck} results. 

In a follow up paper, \citet{simmondsIonizingPropertiesGalaxies2024} used a mass complete sample of $z=4-9$ galaxies from JADES, instead of emission line galaxies, and found a far milder redshift evolution, much closer to the canonical value and our results for the cosmic average. Hence, they concluded that selection effects were the main drivers of the excessive $\xi_{\mathrm{ion}}$ found previously. At $z\approx 4$, $\xi_{\mathrm{ion}}$ between the emission lines galaxies and the mass complete sample are similar. Indeed, \citet{jaiswarMOSELSurveyUnwrapping2024} found very similar $\xi_{\mathrm{ion}}$, close to the canonical value, for both a control sample and a sample of extreme-emission-line-galaxies at $z\approx 3$. 

While we agree with the results of \citet{simmondsIonizingPropertiesGalaxies2024}, it is worth mentioning that the calculation of $\xi_{\mathrm{ion}}$ in comparison to our results is slightly different on account of the different SED fitting codes. A key difference is that \citet{simmondsIonizingPropertiesGalaxies2024} used \textsc{prospector} to fit the metallicity as a free parameter but kept it constant over the entire SFH of the galaxies, while our results employ an evolving metallicity history that we have shown somewhat controls the ionising photon flux in the absence of an evolving $f_{\mathrm{esc}}$. An investigation into model dependencies is beyond the scope of this work, though we caution that inferences on reionisation will likely subsume systematic biases of the SED codes. 

While we may be introducing a prior on our results with the implicit assumption of the \citetalias{bruzualStellarPopulationSynthesis2003} stellar population library with the \citet{chabrierGalacticStellarSubstellar2003} IMF at all redshifts, the \texttt{massfunc\_snorm\_trunc} SFH implementation in \textsc{ProSpect} and AGN model of \citetalias{fritz06agnmodel}, we highlight that the CSFH, CAGNH, their connection to the CSED and, hence, the reionisation timeline have been self-consistently calculated throughout the entire analysis of \citetalias{dsilvaSelfConsistentJWSTCensus2025} and the results presented here. Hence, this work represents a useful baseline to explore additions/amendments to existing models, both from the perspectives of observations \citep[e.g.,][]{robothamProGenyNewSimple2024,bellstedtProGenyIIImpact2024} and simulations. 

\section{Summary} \label{sec:summary}
Closing the chapter on the cosmic dark ages, the epoch of reionisation marks the introduction of the first stars and SMBHs. As such, better characterizing the sources of cosmic reionisation is a necessary exposition to better understand the $\gtrsim 13$~Gyr long odyssey of galaxy formation. In this work, we inferred the sources of reionisation by using the generative SED modelling code \textsc{ProSpect} to connect the computed CSFH and CAGNH from \citetalias{dsilvaSelfConsistentJWSTCensus2025} to the CSED at $z=0-13.5$. The key results are:

\begin{itemize}
    \item The connection between the CSED and the combined CSFH/CAGNH can recover the reionisation timeline that is consistent with, for example, the CMB optical depth and the existence of the Gunn-Peterson trough at $z\gtrsim 6$. Specifically, the cosmic ionising emissivity, neutral fraction and CMB optical depth point are all reconcilable with the shapes of the $z\gtrsim 6$ CSFH and CAGNH.  

    \item By considering scenarios of minimum/maximum $f_{\mathrm{esc}}$ we put physically motivated bounds on both the stellar and AGN contributions to reionisation. AGN alone most likely could not have reionised the IGM at $z\approx 6$ even if $f_{\mathrm{esc}} \approx 100\%$. Hence, star forming galaxies were the dominant contributors to the ionising background. Indeed, stars alone could have achieved reionisation provided that $f_{\mathrm{esc}} \gtrsim 30\%$ for solar metallicity ($Z=0.02$) stars or $f_{\mathrm{esc}} \gtrsim 10\%$ for metal-poor ($Z=0.02$) stars.
    
    Indeed, stars could have achieved reionisation provided that $f_{\mathrm{esc}} \gtrsim 10-30\%$, depending on their metallicity. 

    \item In light of both the CSFH and CBHARH ramping up shortly after the Big Bang, we conceived a hybrid model of reionisation, where both stars and AGN would have contributed ionising photons. The hybrid model indicates that reionisation could have completed at $z\approx 6$ with $f_{\mathrm{esc}}=$ \fescstarsbest{} for the stars and  $f_{\mathrm{esc}}=$ \fescagnbest{} for the AGN, maintained at all redshifts. The median and $1\sigma$ ranges that we found are $f_{\mathrm{esc}}=$ \fescstars{} for the stars and $f_{\mathrm{esc}}=$ \fescagn{} for the AGN. 

    \item Assumptions in SED modelling could be a source of systematic bias in analyses of $z\gtrsim 5$ Lyman continuum. In particular, the assumption of redshift-invariance for $f_{\mathrm{esc}}$ is likely too naive. Better understanding the systematics of our inference tools remains an exciting challenge for the community. It is worth highlighting that this analysis is completely self-consistent throughout. This work is therefore a touchstone study against which alternative methods of SED fitting/modelling and the resultant effects on reionisation may be compared.
\end{itemize}
    
\section*{Data availability}
All the data for the fields used in \citetalias{dsilvaSelfConsistentJWSTCensus2025} can be accessed via \dataset[doi: 10.17909/1c6w-dy61]{https://doi.org/10.17909/1c6w-dy61}. Catalogues will be made available upon reasonable request to the corresponding author.

\section*{Acknowledgements}
We thank the anonymous referee for their helpful comments and suggestions for improving the quality of this work.

J.C.J.D is supported by the Australian Government Research Training Program (RTP) Scholarship. CL is a recipient of the ARC Discovery Project DP210101945. ASGR acknowledges funding by the Australian Research Council (ARC) Future Fellowship scheme (FT200100375). 

This research was supported by the Australian Research Council Centre of Excellence for All Sky Astrophysics in 3 Dimensions (ASTRO 3D), through project number CE170100013.  This work was supported by resources provided by The Pawsey Supercomputing Centre with funding from the Australian Government and the Government of Western Australia.

This work is based on observations made with the NASA/ESA \textit{Hubble Space Telescope} and NASA/ESA/CSA \textit{James Webb Space Telescope} (\textit{JWST}) obtained from the Mikulski Archive for Space Telescopes (MAST) at the Space Telescope Science Institute (STScI), which is operated by the Association of Universities for Research in Astronomy, Inc., under NASA contract NAS 5-03127 for JWST, and NAS 5–26555 for HST. The observations used in this work are associated with JWST programs 1176, 2738, 1345, 2079, 1180, 1210, 3250 and 1837. We acknowledge all those involved in the development of the \textit{JWST} and the execution of these large observing programs. 


\bibliography{ref}{}

@article{grazianContributionFaintAGNs2018,
  title = {The Contribution of Faint {{AGNs}} to the Ionizing Background at z {\textasciitilde} 4},
  author = {Grazian, A. and Giallongo, E. and Boutsia, K. and Cristiani, S. and Vanzella, E. and Scarlata, C. and Santini, P. and Pentericci, L. and Merlin, E. and Menci, N. and Fontanot, F. and Fontana, A. and Fiore, F. and Civano, F. and Castellano, M. and Brusa, M. and Bonchi, A. and Carini, R. and Cusano, F. and Faccini, M. and Garilli, B. and Marchetti, A. and Rossi, A. and Speziali, R.},
  year = {2018},
  month = may,
  journal = {Astronomy \& Astrophysics},
  volume = {613},
  pages = {A44},
  publisher = {EDP Sciences},
  issn = {0004-6361, 1432-0746},
  doi = {10.1051/0004-6361/201732385},
  urldate = {2025-09-17},
  copyright = {{\copyright} ESO 2018},
  langid = {english}
}

@article{kulkarniEvolutionAGNUV2019,
  title = {Evolution of the {{AGN UV}} Luminosity Function from Redshift 7.5},
  author = {Kulkarni, Girish and Worseck, G{\'a}bor and Hennawi, Joseph F.},
  year = {2019},
  month = sep,
  journal = {Monthly Notices of the Royal Astronomical Society},
  volume = {488},
  number = {1},
  pages = {1035--1065},
  issn = {0035-8711},
  doi = {10.1093/mnras/stz1493},
  urldate = {2025-09-19},
  langid = {english}
}

@article{daviesPredicamentAbsorptiondominatedReionization2024,
  title = {The {{Predicament}} of {{Absorption-dominated Reionization II}}: {{Observational Estimate}} of the {{Clumping Factor}} at the {{End}} of {{Reionization}}},
  shorttitle = {The {{Predicament}} of {{Absorption-dominated Reionization II}}},
  author = {Davies, Frederick B. and Bosman, Sarah E. I. and Furlanetto, Steven R.},
  year = {2024},
  month = jun,
  journal = {arXiv e-prints},
  pages = {arXiv:2406.18186},
  doi = {10.48550/arXiv.2406.18186},
  urldate = {2025-04-16},
  langid = {english}
}

@article{nenkovaAGNDustyTori2008,
  title = {{{AGN Dusty Tori}}. {{I}}. {{Handling}} of {{Clumpy Media}}},
  author = {Nenkova, Maia and Sirocky, Matthew M. and Ivezi{\'c}, {\v Z}eljko and Elitzur, Moshe},
  year = {2008},
  month = sep,
  journal = {The Astrophysical Journal},
  volume = {685},
  number = {1},
  pages = {147},
  publisher = {IOP Publishing},
  issn = {0004-637X},
  doi = {10.1086/590482},
  urldate = {2025-04-09},
  langid = {english}
}

@article{feltreSmoothClumpyDust2012a,
  title = {Smooth and Clumpy Dust Distributions in {{AGN}}: A Direct Comparison of Two Commonly Explored Infrared Emission Models},
  shorttitle = {Smooth and Clumpy Dust Distributions in {{AGN}}},
  author = {Feltre, A. and Hatziminaoglou, E. and Fritz, J. and Franceschini, A.},
  year = {2012},
  month = oct,
  journal = {Monthly Notices of the Royal Astronomical Society},
  volume = {426},
  number = {1},
  pages = {120--127},
  issn = {0035-8711},
  doi = {10.1111/j.1365-2966.2012.21695.x},
  urldate = {2025-04-09}
}

@article{cainChasingBeginningReionization2025,
  title = {Chasing the {{Beginning}} of {{Reionization}} in the {{JWST Era}}},
  author = {Cain, Christopher and Lopez, Garett and D'Aloisio, Anson and Mu{\~n}oz, Julian B. and Jansen, Rolf A. and Windhorst, Rogier A. and Gangolli, Nakul},
  year = {2025},
  month = feb,
  journal = {The Astrophysical Journal},
  volume = {980},
  number = {1},
  pages = {83},
  publisher = {The American Astronomical Society},
  issn = {0004-637X},
  doi = {10.3847/1538-4357/ada152},
  urldate = {2025-03-18},
  langid = {english}
}

@article{andrewsModellingCosmicSpectral2018,
  title = {Modelling the Cosmic Spectral Energy Distribution and Extragalactic Background Light over All Time},
  author = {Andrews, S. K. and Driver, S. P. and Davies, L. J. M. and Lagos, C. d. P. and Robotham, A. S. G.},
  year = {2018},
  month = feb,
  journal = {Monthly Notices of the Royal Astronomical Society},
  volume = {474},
  number = {1},
  pages = {898--916},
  issn = {0035-8711},
  doi = {10.1093/mnras/stx2843},
  urldate = {2025-04-07}
}

@ARTICLE{dsilvaSelfConsistentJWSTCensus2025,
       author = {{D'Silva}, Jordan C.~J. and {Driver}, Simon P. and {Lagos}, Claudia D.~P. and {Robotham}, Aaron S.~G. and {Adams}, Nathan J. and {Conselice}, Christopher J. and {Frye}, Brenda and {Hathi}, Nimish P. and {Harvey}, Thomas and {Ortiz}, III, Rafael and {Ricotti}, Massimo and {Robertson}, Clayton and {Silver}, Ross M. and {Wilkins}, Stephen M. and {Willmer}, Christopher N.~A. and {Windhorst}, Rogier A. and {Cohen}, Seth H. and {Jansen}, Rolf A. and {Summers}, Jake and {Koekemoer}, Anton M. and {Coe}, Dan and {Grogin}, Norman A. and {Marshall}, Madeline A. and {Nonino}, Mario and {Pirzkal}, Nor and {Ryan}, Jr., Russell E. and {Yan}, Haojing},
        title = "{Self-Consistent JWST Census of Star Formation and AGN activity at z=5.5-13.5}",
      journal = {arXiv e-prints},
         year = 2025,
        month = mar,
          eid = {arXiv:2503.03431},
        pages = {arXiv:2503.03431},
          doi = {10.48550/arXiv.2503.03431},
archivePrefix = {arXiv},
       eprint = {2503.03431},
 primaryClass = {astro-ph.GA},
       adsurl = {https://ui.adsabs.harvard.edu/abs/2025arXiv250303431D}
}

@article{heintzDilutionChemicalEnrichment2023,
  title = {Dilution of Chemical Enrichment in Galaxies 600 {{Myr}} after the {{Big Bang}}},
  author = {Heintz, Kasper E. and Brammer, Gabriel B. and {Gim{\'e}nez-Arteaga}, Clara and Strait, Victoria B. and {del P. Lagos}, Claudia and Vijayan, Aswin P. and Matthee, Jorryt and Watson, Darach and Mason, Charlotte A. and Hutter, Anne and Toft, Sune and Fynbo, Johan P. U. and Oesch, Pascal A.},
  year = {2023},
  month = dec,
  journal = {Nature Astronomy},
  volume = {7},
  number = {12},
  pages = {1517--1524},
  publisher = {Nature Publishing Group},
  issn = {2397-3366},
  doi = {10.1038/s41550-023-02078-7},
  urldate = {2025-03-20},
  copyright = {2023 The Author(s), under exclusive licence to Springer Nature Limited},
  langid = {english}
}

@article{remy-ruyerGastodustMassRatios2014,
  title = {Gas-to-Dust Mass Ratios in Local Galaxies over a 2 Dex Metallicity Range},
  author = {{R{\'e}my-Ruyer}, A. and Madden, S. C. and Galliano, F. and Galametz, M. and Takeuchi, T. T. and Asano, R. S. and Zhukovska, S. and Lebouteiller, V. and Cormier, D. and Jones, A. and Bocchio, M. and Baes, M. and Bendo, G. J. and Boquien, M. and Boselli, A. and DeLooze, I. and {Doublier-Pritchard}, V. and Hughes, T. and Karczewski, O. {\L} and Spinoglio, L.},
  year = {2014},
  month = mar,
  journal = {Astronomy \& Astrophysics},
  volume = {563},
  pages = {A31},
  publisher = {EDP Sciences},
  issn = {0004-6361, 1432-0746},
  doi = {10.1051/0004-6361/201322803},
  urldate = {2025-03-19},
  copyright = {{\copyright} ESO, 2014},
  langid = {english}
}

@article{vijayanDetailedDustModelling2019,
  title = {Detailed Dust Modelling in the {{L-Galaxies}} Semi-Analytic Model of Galaxy Formation},
  author = {Vijayan, Aswin P and Clay, Scott J and Thomas, Peter A and Yates, Robert M and Wilkins, Stephen M and Henriques, Bruno M},
  year = {2019},
  month = nov,
  journal = {Monthly Notices of the Royal Astronomical Society},
  volume = {489},
  number = {3},
  pages = {4072--4089},
  issn = {0035-8711},
  doi = {10.1093/mnras/stz1948},
  urldate = {2025-03-19}
}

@ARTICLE{cainNewConstraintsGalactic2025,
       author = {{Cain}, Christopher and {D'Aloisio}, Anson and {Munoz}, Julian B.},
        title = "{New constraints on the galactic ionizing efficiency and escape fraction at 2.5 < z < 6 based on quasar absorption spectra}",
      journal = {arXiv e-prints},
         year = 2025,
        month = mar,
          eid = {arXiv:2503.08778},
        pages = {arXiv:2503.08778},
          doi = {10.48550/arXiv.2503.08778},
archivePrefix = {arXiv},
       eprint = {2503.08778},
 primaryClass = {astro-ph.GA},
       adsurl = {https://ui.adsabs.harvard.edu/abs/2025arXiv250308778C}
}

@article{ouchiLARGEAREASURVEY2009,
  title = {{{LARGE AREA SURVEY FOR}} z = 7 {{GALAXIES IN SDF AND GOODS-N}}: {{IMPLICATIONS FOR GALAXY FORMATION AND COSMIC REIONIZATION}}*},
  shorttitle = {{{LARGE AREA SURVEY FOR}} z = 7 {{GALAXIES IN SDF AND GOODS-N}}},
  author = {Ouchi, Masami and Mobasher, Bahram and Shimasaku, Kazuhiro and Ferguson, Henry C. and Fall, S. Michael and Ono, Yoshiaki and Kashikawa, Nobunari and Morokuma, Tomoki and Nakajima, Kimihiko and Okamura, Sadanori and Dickinson, Mark and Giavalisco, Mauro and Ohta, Kouji},
  year = {2009},
  month = nov,
  journal = {The Astrophysical Journal},
  volume = {706},
  number = {2},
  pages = {1136},
  publisher = {The American Astronomical Society},
  issn = {0004-637X},
  doi = {10.1088/0004-637X/706/2/1136},
  urldate = {2025-03-19},
  langid = {english}
}

@article{reddyCONNECTIONREDDENINGGAS2016,
  title = {{{THE CONNECTION BETWEEN REDDENING}}, {{GAS COVERING FRACTION}}, {{AND THE ESCAPE OF IONIZING RADIATION AT HIGH REDSHIFT}}{$\ast$}},
  author = {Reddy, Naveen A. and Steidel, Charles C. and Pettini, Max and Bogosavljevi{\'c}, Milan and Shapley, Alice E.},
  year = {2016},
  month = sep,
  journal = {The Astrophysical Journal},
  volume = {828},
  number = {2},
  pages = {108},
  publisher = {The American Astronomical Society},
  issn = {0004-637X},
  doi = {10.3847/0004-637X/828/2/108},
  urldate = {2025-03-17},
  langid = {english}
}

@article{yooOriginLowEscape2020,
  title = {On the Origin of Low Escape Fractions of Ionizing Radiation from Massive Star-Forming Galaxies at High Redshift},
  author = {Yoo, Taehwa and Kimm, Taysun and Rosdahl, Joakim},
  year = {2020},
  month = nov,
  journal = {Monthly Notices of the Royal Astronomical Society},
  volume = {499},
  number = {4},
  pages = {5175--5193},
  issn = {0035-8711},
  doi = {10.1093/mnras/staa3187},
  urldate = {2025-03-17}
}

@article{bolanInferringIntergalacticMedium2022,
  title = {Inferring the Intergalactic Medium Neutral Fraction at z {$\sim$} 6--8 with Low-Luminosity {{Lyman}} Break Galaxies},
  author = {Bolan, Patricia and Lemaux, Brian C and Mason, Charlotte and Brada{\v c}, Maru{\v s}a and Treu, Tommaso and Strait, Victoria and Pelliccia, Debora and Pentericci, Laura and Malkan, Matthew},
  year = {2022},
  month = dec,
  journal = {Monthly Notices of the Royal Astronomical Society},
  volume = {517},
  number = {3},
  pages = {3263--3274},
  issn = {0035-8711},
  doi = {10.1093/mnras/stac1963},
  urldate = {2025-03-17}
}

@article{jinNearlyModelindependentConstraints2023,
  title = {({{Nearly}}) {{Model-independent Constraints}} on the {{Neutral Hydrogen Fraction}} in the {{Intergalactic Medium}} at z {$\sim$} 5--7 {{Using Dark Pixel Fractions}} in {{Ly$\alpha$}} and {{Ly$\beta$ Forests}}},
  author = {Jin, Xiangyu and Yang, Jinyi and Fan, Xiaohui and Wang, Feige and Ba{\~n}ados, Eduardo and Bian, Fuyan and Davies, Frederick B. and Eilers, Anna-Christina and Farina, Emanuele Paolo and Hennawi, Joseph F. and Pacucci, Fabio and Venemans, Bram and Walter, Fabian},
  year = {2023},
  month = jan,
  journal = {The Astrophysical Journal},
  volume = {942},
  number = {2},
  pages = {59},
  publisher = {The American Astronomical Society},
  issn = {0004-637X},
  doi = {10.3847/1538-4357/aca678},
  urldate = {2025-03-14},
  langid = {english}
}

@article{leitetEscapeLymanContinuum2013,
  title = {Escape of {{Lyman}} Continuum Radiation from Local Galaxies - {{Detection}} of Leakage from the Young Starburst {{Tol}} 1247-232},
  author = {Leitet, E. and Bergvall, N. and Hayes, M. and Linn{\'e}, S. and Zackrisson, E.},
  year = {2013},
  month = may,
  journal = {Astronomy \& Astrophysics},
  volume = {553},
  pages = {A106},
  publisher = {EDP Sciences},
  issn = {0004-6361, 1432-0746},
  doi = {10.1051/0004-6361/201118370},
  urldate = {2025-03-17},
  copyright = {{\copyright} ESO, 2013},
  langid = {english}
}

@article{masciaClosingSourcesCosmic2023,
  title = {Closing in on the Sources of Cosmic Reionization: {{First}} Results from the {{GLASS-JWST}} Program},
  shorttitle = {Closing in on the Sources of Cosmic Reionization},
  author = {Mascia, S. and Pentericci, L. and Calabr{\`o}, A. and Treu, T. and Santini, P. and Yang, L. and Napolitano, L. and {Roberts-Borsani}, G. and Bergamini, P. and Grillo, C. and Rosati, P. and Vulcani, B. and Castellano, M. and Boyett, K. and Fontana, A. and Glazebrook, K. and Henry, A. and Mason, C. and Merlin, E. and Morishita, T. and Nanayakkara, T. and Paris, D. and Roy, N. and Williams, H. and Wang, X. and Brammer, G. and Brada{\v c}, M. and Chen, W. and Kelly, P. L. and Koekemoer, A. M. and Trenti, M. and Windhorst, R. A.},
  year = {2023},
  month = apr,
  journal = {Astronomy \& Astrophysics},
  volume = {672},
  pages = {A155},
  publisher = {EDP Sciences},
  issn = {0004-6361, 1432-0746},
  doi = {10.1051/0004-6361/202345866},
  urldate = {2025-03-14},
  copyright = {{\copyright} The Authors 2023},
  langid = {english}
}

@article{mattheeProductionEscapeLymanContinuum2017,
  title = {The Production and Escape of {{Lyman-Continuum}} Radiation from Star-Forming Galaxies at z {$\sim$} 2 and Their Redshift Evolution},
  author = {Matthee, Jorryt and Sobral, David and Best, Philip and Khostovan, Ali Ahmad and Oteo, Iv{\'a}n and Bouwens, Rychard and R{\"o}ttgering, Huub},
  year = {2017},
  month = mar,
  journal = {Monthly Notices of the Royal Astronomical Society},
  volume = {465},
  number = {3},
  pages = {3637--3655},
  issn = {0035-8711},
  doi = {10.1093/mnras/stw2973},
  urldate = {2025-03-14}
}

@article{pahlUncontaminatedMeasurementEscaping2021,
  title = {An Uncontaminated Measurement of the Escaping {{Lyman}} Continuum at z {$\sim$} 3},
  author = {Pahl, Anthony J and Shapley, Alice and Steidel, Charles C and Chen, Yuguang and Reddy, Naveen A},
  year = {2021},
  month = aug,
  journal = {Monthly Notices of the Royal Astronomical Society},
  volume = {505},
  number = {2},
  pages = {2447--2467},
  issn = {0035-8711},
  doi = {10.1093/mnras/stab1374},
  urldate = {2025-03-14}
}

@article{rutkowskiLYMANCONTINUUMESCAPE2016,
  title = {{{LYMAN CONTINUUM ESCAPE FRACTION OF STAR-FORMING DWARF GALAXIES AT}} z {$\sim$} 1},
  author = {Rutkowski, Michael J. and Scarlata, Claudia and Haardt, Francesco and Siana, Brian and Henry, Alaina and Rafelski, Marc and Hayes, Matthew and Salvato, Mara and Pahl, Anthony J. and Mehta, Vihang and Beck, Melanie and Malkan, Matthew and Teplitz, Harry I.},
  year = {2016},
  month = mar,
  journal = {The Astrophysical Journal},
  volume = {819},
  number = {1},
  pages = {81},
  publisher = {The American Astronomical Society},
  issn = {0004-637X},
  doi = {10.3847/0004-637X/819/1/81},
  urldate = {2025-03-14},
  langid = {english}
}

@article{steidelKeckLymanContinuum2018,
  title = {The {{Keck Lyman Continuum Spectroscopic Survey}} ({{KLCS}}): {{The Emergent Ionizing Spectrum}} of {{Galaxies}} at z {$\sim$} 3{$\ast$}},
  shorttitle = {The {{Keck Lyman Continuum Spectroscopic Survey}} ({{KLCS}})},
  author = {Steidel, Charles C. and Bogosavljevi{\'c}, Milan and Shapley, Alice E. and Reddy, Naveen A. and Rudie, Gwen C. and Pettini, Max and Trainor, Ryan F. and Strom, Allison L.},
  year = {2018},
  month = dec,
  journal = {The Astrophysical Journal},
  volume = {869},
  number = {2},
  pages = {123},
  publisher = {The American Astronomical Society},
  issn = {0004-637X},
  doi = {10.3847/1538-4357/aaed28},
  urldate = {2025-03-14},
  langid = {english}
}

@article{wangLymanContinuumEscape2025,
  title = {The {{Lyman Continuum Escape Fraction}} of {{Star-forming Galaxies}} at 2.4 {$\lessequivlnt$} z {$\lessequivlnt$} 3.0 from {{UVCANDELS}}},
  author = {Wang, Xin and Teplitz, Harry I. and Smith, Brent M. and Windhorst, Rogier A. and Rafelski, Marc and Mehta, Vihang and Alavi, Anahita and Ji, Zhiyuan and Brammer, Gabriel and Colbert, James and Grogin, Norman and Hathi, Nimish P. and Koekemoer, Anton M. and Prichard, Laura and Scarlata, Claudia and Sunnquist, Ben and Arrabal Haro, Pablo and Conselice, Christopher and Gawiser, Eric and Guo, Yicheng and Hayes, Matthew and Jansen, Rolf A. and Lucas, Ray A. and O'Connell, Robert and Robertson, Brant and Rutkowski, Michael and Siana, Brian and Vanzella, Eros and Ashcraft, Teresa and Bagley, Micaela and Baronchelli, Ivano and Barro, Guillermo and Blanche, Alex and Broussard, Adam and Carleton, Timothy and Chartab, Nima and Cheng, Yingjie and Codoreanu, Alex and Cohen, Seth and Dai, Y. Sophia and Darvish, Behnam and Dav{\'e}, Romeel and Degroot, Laura and {de Mello}, Duilia and Dickinson, Mark and Emami, Najmeh and Ferguson, Henry and Ferreira, Leonardo and Finkelstein, Keely and Finkelstein, Steven and Gardner, Jonathan P. and Gburek, Timothy and Giavalisco, Mauro and Grazian, Andrea and Gronwall, Caryl and Hemmati, Shoubaneh and Howell, Justin and Iyer, Kartheik and Kaviraj, Sugata and Kurczynski, Peter and Lazar, Ilin and MacKenty, John and Mantha, Kameswara Bharadwaj and Martin, Alec and Martin, Garreth and McCabe, Tyler and Mobasher, Bahram and Nedkova, Kalina and Olsen, Charlotte and Otteson, Lillian and Ravindranath, Swara and Redshaw, Caleb and Sattari, Zahra and Soto, Emmaris and Yung, L. Y. Aaron and Zabelle, Bonnabelle and {The Uvcandels Team}},
  year = {2025},
  month = feb,
  journal = {The Astrophysical Journal},
  volume = {980},
  pages = {74},
  publisher = {IOP},
  issn = {0004-637X},
  doi = {10.3847/1538-4357/ada4ab},
  urldate = {2025-03-04}
}

@article{wangSignificantlyNeutralIntergalactic2020,
  title = {A {{Significantly Neutral Intergalactic Medium Around}} the {{Luminous}} z = 7 {{Quasar J0252}}--0503},
  author = {Wang, Feige and Davies, Frederick B. and Yang, Jinyi and Hennawi, Joseph F. and Fan, Xiaohui and Barth, Aaron J. and Jiang, Linhua and Wu, Xue-Bing and Mudd, Dale M. and Ba{\~n}ados, Eduardo and Bian, Fuyan and Decarli, Roberto and Eilers, Anna-Christina and Farina, Emanuele Paolo and Venemans, Bram and Walter, Fabian and Yue, Minghao},
  year = {2020},
  month = jun,
  journal = {The Astrophysical Journal},
  volume = {896},
  number = {1},
  pages = {23},
  publisher = {The American Astronomical Society},
  issn = {0004-637X},
  doi = {10.3847/1538-4357/ab8c45},
  urldate = {2025-03-14},
  langid = {english}
}

@article{yangPoniuaenaLuminous752020,
  title = {P{\=o}niu{\=a}`ena: {{A Luminous}} z = 7.5 {{Quasar Hosting}} a 1.5 {{Billion Solar Mass Black Hole}}},
  shorttitle = {P{\=o}niu{\=a}`ena},
  author = {Yang, Jinyi and Wang, Feige and Fan, Xiaohui and Hennawi, Joseph F. and Davies, Frederick B. and Yue, Minghao and Banados, Eduardo and Wu, Xue-Bing and Venemans, Bram and Barth, Aaron J. and Bian, Fuyan and Boutsia, Konstantina and Decarli, Roberto and Farina, Emanuele Paolo and Green, Richard and Jiang, Linhua and Li, Jiang-Tao and Mazzucchelli, Chiara and Walter, Fabian},
  year = {2020},
  month = jul,
  journal = {The Astrophysical Journal Letters},
  volume = {897},
  number = {1},
  pages = {L14},
  publisher = {The American Astronomical Society},
  issn = {2041-8205},
  doi = {10.3847/2041-8213/ab9c26},
  urldate = {2025-03-17},
  langid = {english}
}

@misc{bellstedtProGenyIIImpact2024,
  title = {{{ProGeny II}}: The Impact of Libraries and Model Configurations on Inferred Galaxy Properties in {{SED}} Fitting},
  shorttitle = {{{ProGeny II}}},
  author = {Bellstedt, Sabine and Robotham, Aaron S. G.},
  year = {2024},
  month = oct,
  journal = {arXiv e-prints},
  doi = {10.48550/arXiv.2410.17698},
  urldate = {2025-02-25}
}

@misc{robothamProGenyNewSimple2024,
  title = {{{ProGeny I}}: A New Simple Stellar Population Generator and Impact of Isochrones / Stellar Atmospheres / Initial Mass Functions},
  shorttitle = {{{ProGeny I}}},
  author = {Robotham, A. S. G. and Bellstedt, S.},
  year = {2024},
  month = oct,
  journal = {arXiv e-prints},
  doi = {10.48550/arXiv.2410.17697},
  urldate = {2025-02-25}
}

@article{shivaeiMOSDEFSurveyDirect2018a,
  title = {The {{MOSDEF Survey}}: {{Direct Observational Constraints}} on the {{Ionizing Photon Production Efficiency}}, {$\Xi$}ion, at z {$\sim$} 2},
  shorttitle = {The {{MOSDEF Survey}}},
  author = {Shivaei, Irene and Reddy, Naveen A. and Siana, Brian and Shapley, Alice E. and Kriek, Mariska and Mobasher, Bahram and Freeman, William R. and Sanders, Ryan L. and Coil, Alison L. and Price, Sedona H. and Fetherolf, Tara and Azadi, Mojegan and Leung, Gene and Zick, Tom},
  year = {2018},
  month = mar,
  journal = {The Astrophysical Journal},
  volume = {855},
  number = {1},
  pages = {42},
  publisher = {The American Astronomical Society},
  issn = {0004-637X},
  doi = {10.3847/1538-4357/aaad62},
  urldate = {2025-02-21},
  langid = {english}
}

@article{madauCosmicReionizationPlanck2017,
  title = {Cosmic {{Reionization}} after {{Planck}} and before {{JWST}}: {{An Analytic Approach}}},
  shorttitle = {Cosmic {{Reionization}} after {{Planck}} and before {{JWST}}},
  author = {Madau, Piero},
  year = {2017},
  month = dec,
  journal = {The Astrophysical Journal},
  volume = {851},
  number = {1},
  pages = {50}, 
  publisher = {The American Astronomical Society},
  issn = {0004-637X},
  doi = {10.3847/1538-4357/aa9715},
  urldate = {2025-02-10},
  langid = {english}
}

@article{bosmanHydrogenReionizationEnds2022,
  title = {Hydrogen Reionization Ends by z = 5.3: {{Lyman-$\alpha$}} Optical Depth Measured by the {{XQR-30}} Sample},
  shorttitle = {Hydrogen Reionization Ends by z = 5.3},
  author = {Bosman, Sarah E I and Davies, Frederick B and Becker, George D and Keating, Laura C and Davies, Rebecca L and Zhu, Yongda and Eilers, Anna-Christina and D'Odorico, Valentina and Bian, Fuyan and Bischetti, Manuela and Cristiani, Stefano V and Fan, Xiaohui and Farina, Emanuele P and Haehnelt, Martin G and Hennawi, Joseph F and Kulkarni, Girish and Mesinger, Andrei and Meyer, Romain A and Onoue, Masafusa and Pallottini, Andrea and Qin, Yuxiang and {Ryan-Weber}, Emma and Schindler, Jan-Torge and Walter, Fabian and Wang, Feige and Yang, Jinyi},
  year = {2022},
  month = jul,
  journal = {Monthly Notices of the Royal Astronomical Society},
  volume = {514},
  number = {1},
  pages = {55--76},
  issn = {0035-8711},
  doi = {10.1093/mnras/stac1046},
  urldate = {2025-02-10}
}

@article{simmondsIonizingPropertiesGalaxies2024,
  title = {Ionizing Properties of Galaxies in {{JADES}} for a Stellar Mass Complete Sample: Resolving the Cosmic Ionizing Photon Budget Crisis at the {{Epoch}} of {{Reionization}}},
  shorttitle = {Ionizing Properties of Galaxies in {{JADES}} for a Stellar Mass Complete Sample},
  author = {Simmonds, C and Tacchella, S and Hainline, K and Johnson, B D and Pusk{\'a}s, D and Robertson, B and Baker, W M and Bhatawdekar, R and Boyett, K and Bunker, A J and Cargile, P A and Carniani, S and Chevallard, J and Curti, M and {Curtis-Lake}, E and Ji, Z and Jones, G C and Kumari, N and Laseter, I and Maiolino, R and Maseda, M V and Rinaldi, P and Stoffers, A and {\"U}bler, H and Villanueva, N C and Williams, C C and Willott, C and Witstok, J and Zhu, Y},
  year = {2024},
  month = dec,
  journal = {Monthly Notices of the Royal Astronomical Society},
  volume = {535},
  number = {4},
  pages = {2998--3019},
  issn = {0035-8711},
  doi = {10.1093/mnras/stae2537},
  urldate = {2025-02-10}
}

@article{atekMostPhotonsThat2024,
  title = {Most of the Photons That Reionized the {{Universe}} Came from Dwarf Galaxies},
  author = {Atek, Hakim and Labb{\'e}, Ivo and Furtak, Lukas J. and Chemerynska, Iryna and Fujimoto, Seiji and Setton, David J. and Miller, Tim B. and Oesch, Pascal and Bezanson, Rachel and Price, Sedona H. and Dayal, Pratika and Zitrin, Adi and Kokorev, Vasily and Weaver, John R. and Brammer, Gabriel and van Dokkum, Pieter and Williams, Christina C. and Cutler, Sam E. and Feldmann, Robert and Fudamoto, Yoshinobu and Greene, Jenny E. and Leja, Joel and Maseda, Michael V. and Muzzin, Adam and Pan, Richard and Papovich, Casey and Nelson, Erica J. and Nanayakkara, Themiya and Stark, Daniel P. and Stefanon, Mauro and Suess, Katherine A. and Wang, Bingjie and Whitaker, Katherine E.},
  year = {2024},
  month = feb,
  journal = {Nature},
  volume = {626},
  number = {8001},
  pages = {975--978},
  publisher = {Nature Publishing Group},
  issn = {1476-4687},
  doi = {10.1038/s41586-024-07043-6},
  urldate = {2025-01-30},
  copyright = {2024 The Author(s), under exclusive licence to Springer Nature Limited},
  langid = {english}
}

@article{curtis-lakeSpectroscopicConfirmationFour2023a,
  title = {Spectroscopic Confirmation of Four Metal-Poor Galaxies at z = 10.3-13.2},
  author = {{Curtis-Lake}, Emma and Carniani, Stefano and Cameron, Alex and Charlot, Stephane and Jakobsen, Peter and Maiolino, Roberto and Bunker, Andrew and Witstok, Joris and Smit, Renske and Chevallard, Jacopo and Willott, Chris and Ferruit, Pierre and Arribas, Santiago and Bonaventura, Nina and Curti, Mirko and D'Eugenio, Francesco and Franx, Marijn and Giardino, Giovanna and Looser, Tobias J. and L{\"u}tzgendorf, Nora and Maseda, Michael V. and Rawle, Tim and Rix, Hans-Walter and {Rodr{\'i}guez del Pino}, Bruno and {\"U}bler, Hannah and Sirianni, Marco and Dressler, Alan and Egami, Eiichi and Eisenstein, Daniel J. and Endsley, Ryan and Hainline, Kevin and Hausen, Ryan and Johnson, Benjamin D. and Rieke, Marcia and Robertson, Brant and Shivaei, Irene and Stark, Daniel P. and Tacchella, Sandro and Williams, Christina C. and Willmer, Christopher N. A. and Bhatawdekar, Rachana and Bowler, Rebecca and Boyett, Kristan and Chen, Zuyi and {de Graaff}, Anna and Helton, Jakob M. and Hviding, Raphael E. and Jones, Gareth C. and Kumari, Nimisha and Lyu, Jianwei and Nelson, Erica and Perna, Michele and Sandles, Lester and Saxena, Aayush and Suess, Katherine A. and Sun, Fengwu and Topping, Michael W. and Wallace, Imaan E. B. and Whitler, Lily},
  year = {2023},
  month = may,
  journal = {Nature Astronomy},
  volume = {7},
  pages = {622--632},
  issn = {2397-3366},
  doi = {10.1038/s41550-023-01918-w},
  urldate = {2024-05-14}
}

@article{begleyVANDELSSurveyMeasurement2022,
  title = {The {{VANDELS}} Survey: A Measurement of the Average {{Lyman-continuum}} Escape Fraction of Star-Forming Galaxies at z = 3.5},
  shorttitle = {The {{VANDELS}} Survey},
  author = {Begley, R and Cullen, F and McLure, R J and Dunlop, J S and Hall, A and Carnall, A C and Hamadouche, M L and McLeod, D J and Amor{\'i}n, R and Calabr{\`o}, A and Fontana, A and Fynbo, J P U and Guaita, L and Hathi, N P and Hibon, P and Ji, Z and Llerena, M and Pentericci, L and {Saldana-Lopez}, A and Schaerer, D and Talia, M and Vanzella, E and Zamorani, G},
  year = {2022},
  month = jul,
  journal = {\mnras},
  volume = {513},
  number = {3},
  pages = {3510--3525},
  issn = {0035-8711},
  doi = {10.1093/mnras/stac1067},
  urldate = {2025-01-30}
}

@article{chisholmFarultravioletContinuumSlope2022,
  title = {The Far-Ultraviolet Continuum Slope as a {{Lyman Continuum}} Escape Estimator at High Redshift},
  author = {Chisholm, J and {Saldana-Lopez}, A and Flury, S and Schaerer, D and Jaskot, A and Amor{\'i}n, R and Atek, H and Finkelstein, S L and Fleming, B and Ferguson, H and Fern{\'a}ndez, V and Giavalisco, M and Hayes, M and Heckman, T and Henry, A and Ji, Z and {Marques-Chaves}, R and Mauerhofer, V and McCandliss, S and Oey, M S and {\"O}stlin, G and Rutkowski, M and Scarlata, C and Thuan, T and Trebitsch, M and Wang, B and Worseck, G and Xu, X},
  year = {2022},
  month = dec,
  journal = {\mnras},
  volume = {517},
  number = {4},
  pages = {5104--5120},
  issn = {0035-8711},
  doi = {10.1093/mnras/stac2874},
  urldate = {2025-01-30}
}

@article{pahlConnectionEscapeIonizing2023,
  title = {The Connection between the Escape of Ionizing Radiation and Galaxy Properties at z {$\sim$} 3 in the {{Keck Lyman}} Continuum Spectroscopic Survey},
  author = {Pahl, Anthony J and Shapley, Alice and Steidel, Charles C and Reddy, Naveen A and Chen, Yuguang and Rudie, Gwen C and Strom, Allison L},
  year = {2023},
  month = may,
  journal = {\mnras},
  volume = {521},
  number = {3},
  pages = {3247--3259},
  issn = {0035-8711},
  doi = {10.1093/mnras/stad774},
  urldate = {2025-01-30}
}

@article{saldana-lopezLowRedshiftLymanContinuum2022,
  title = {The {{Low-Redshift Lyman Continuum Survey}} - {{Unveiling}} the {{ISM}} Properties of Low-z {{Lyman-continuum}} Emitters},
  author = {{Saldana-Lopez}, Alberto and Schaerer, Daniel and Chisholm, John and Flury, Sophia R. and Jaskot, Anne E. and Worseck, G{\'a}bor and Makan, Kirill and Gazagnes, Simon and Mauerhofer, Valentin and Verhamme, Anne and Amor{\'i}n, Ricardo O. and Ferguson, Harry C. and Giavalisco, Mauro and Grazian, Andrea and Hayes, Matthew J. and Heckman, Timothy M. and Henry, Alaina and Ji, Zhiyuan and {Marques-Chaves}, Rui and McCandliss, Stephan R. and Oey, M. Sally and {\"O}stlin, G{\"o}ran and Pentericci, Laura and Thuan, Trinh X. and Trebitsch, Maxime and Vanzella, Eros and Xu, Xinfeng},
  year = {2022},
  month = jul,
  journal = {\aap},
  volume = {663},
  pages = {A59},
  publisher = {EDP Sciences},
  issn = {0004-6361, 1432-0746},
  doi = {10.1051/0004-6361/202141864},
  urldate = {2025-01-30},
  copyright = {{\copyright} ESO 2022},
  langid = {english}
}

@article{bennettFirstYearWilkinsonMicrowave2003,
  title = {First-{{Year Wilkinson Microwave Anisotropy Probe}}({{WMAP}})* {{Observations}}:{{Preliminary Maps}} and {{Basic Results}}},
  shorttitle = {First-{{Year Wilkinson Microwave Anisotropy Probe}}({{WMAP}})* {{Observations}}},
  author = {Bennett, C. L. and Halpern, M. and Hinshaw, G. and Jarosik, N. and Kogut, A. and Limon, M. and Meyer, S. S. and Page, L. and Spergel, D. N. and Tucker, G. S. and Wollack, E. and Wright, E. L. and Barnes, C. and Greason, M. R. and Hill, R. S. and Komatsu, E. and Nolta, M. R. and Odegard, N. and Peiris, H. V. and Verde, L. and Weiland, J. L.},
  year = {2003},
  month = sep,
  journal = {The Astrophysical Journal Supplement Series},
  volume = {148},
  number = {1},
  pages = {1},
  publisher = {IOP Publishing},
  issn = {0067-0049},
  doi = {10.1086/377253},
  urldate = {2025-01-30},
  langid = {english}
}

@article{beckerEvidenceReionizationAtz2001,
  title = {Evidence for {{Reionization}} Atz {$\sim$} 6: {{Detection}} of {{aGunn-Peterson Trough}} in Az = 6.{{28Quasar}}* **},
  shorttitle = {Evidence for {{Reionization}} Atz {$\sim$} 6},
  author = {Becker, Robert H. and Fan, Xiaohui and White, Richard L. and Strauss, Michael A. and Narayanan, Vijay K. and Lupton, Robert H. and Gunn, James E. and Annis, James and Bahcall, Neta A. and Brinkmann, J. and Connolly, A. J. and Csabai, Istv{\'a}n and Czarapata, Paul C. and Doi, Mamoru and Heckman, Timothy M. and Hennessy, G. S. and Ivezi{\'c}, {\v Z}eljko and Knapp, G. R. and Lamb, Don Q. and McKay, Timothy A. and Munn, Jeffrey A. and Nash, Thomas and Nichol, Robert and Pier, Jeffrey R. and Richards, Gordon T. and Schneider, Donald P. and Stoughton, Chris and Szalay, Alexander S. and Thakar, Aniruddha R. and York, D. G.},
  year = {2001},
  month = dec,
  journal = {\aj},
  volume = {122},
  number = {6},
  pages = {2850},
  publisher = {IOP Publishing},
  issn = {1538-3881},
  doi = {10.1086/324231},
  urldate = {2025-01-30},
  langid = {english}
}

@article{fanSurvey58Quasarsin2001,
  title = {A {{Survey}} of z {$>$} 5.8 {{Quasarsin}} the {{Sloan Digital Sky Survey}}. {{I}}. {{Discovery}} of {{Three New Quasars}} and {{theSpatial Density}} of {{Luminous Quasars}} Atz {$\sim$} 6* **},
  author = {Fan, Xiaohui and Narayanan, Vijay K. and Lupton, Robert H. and Strauss, Michael A. and Knapp, Gillian R. and Becker, Robert H. and White, Richard L. and Pentericci, Laura and Leggett, S. K. and Haiman, Zolt{\'a}n and Gunn, James E. and Ivezi{\'c}, {\v Z}eljko and Schneider, Donald P. and Anderson, Scott F. and Brinkmann, J. and Bahcall, Neta A. and Connolly, Andrew J. and Csabai, Istv{\'a}n and Doi, Mamoru and Fukugita, Masataka and Geballe, Tom and Grebel, Eva K. and Harbeck, Daniel and Hennessy, Gregory and Lamb, Don Q. and Miknaitis, Gajus and Munn, Jeffrey A. and Nichol, Robert and Okamura, Sadanori and Pier, Jeffrey R. and Prada, Francisco and Richards, Gordon T. and Szalay, Alex and York, Donald G.},
  year = {2001},
  month = dec,
  journal = {\aj},
  volume = {122},
  number = {6},
  pages = {2833},
  publisher = {IOP Publishing},
  issn = {1538-3881},
  doi = {10.1086/324111},
  urldate = {2025-01-30},
  langid = {english}
}

@article{kaurovCOSMICREIONIZATIONCOMPUTERS2015,
  title = {{{COSMIC REIONIZATION ON COMPUTERS}}. {{III}}. {{THE CLUMPING FACTOR}}},
  author = {Kaurov, Alexander A. and Gnedin, Nickolay Y.},
  year = {2015},
  month = sep,
  journal = {\apj},
  volume = {810},
  number = {2},
  pages = {154},
  publisher = {The American Astronomical Society},
  issn = {0004-637X},
  doi = {10.1088/0004-637X/810/2/154},
  urldate = {2025-01-30},
  langid = {english}
}

@article{johnsonStellarPopulationInference2021,
  title = {Stellar {{Population Inference}} with {{Prospector}}},
  author = {Johnson, Benjamin D. and Leja, Joel and Conroy, Charlie and Speagle, Joshua S.},
  year = {2021},
  month = may,
  journal = {\apjs},
  volume = {254},
  number = {2},
  pages = {22},
  publisher = {The American Astronomical Society},
  issn = {0067-0049},
  doi = {10.3847/1538-4365/abef67},
  urldate = {2025-01-30},
  langid = {english}
}

@article{johnsonProspectorStellarPopulation2019,
  title = {Prospector: {{Stellar}} Population Inference from Spectra and {{SEDs}}},
  shorttitle = {Prospector},
  author = {Johnson, Benjamin D. and Leja, Joel L. and Conroy, Charlie and Speagle, Joshua S.},
  year = {2019},
  month = may,
  journal = {Astrophysics Source Code Library},
  pages = {ascl:1905.025},
  urldate = {2025-01-30}
}

@article{jaiswarMOSELSurveyUnwrapping2024,
  title = {{{MOSEL}} Survey: {{Unwrapping}} the {{Epoch}} of {{Reionisation}} through Mimic Galaxies at {{Cosmic Noon}}},
  shorttitle = {{{MOSEL}} Survey},
  author = {Jaiswar, Ravi and Gupta, Anshu and da Cunha, Elisabete and Trott, Cathryn M. and Harshan, Anishya and Battisti, Andrew and Forrest, Ben},
  year = {2024},
  month = jan,
  journal = {\pasa},
  volume = {41},
  pages = {e040},
  issn = {1323-3580, 1448-6083},
  doi = {10.1017/pasa.2024.37},
  urldate = {2025-01-29},
  langid = {english}
}

@article{munozReionizationJWSTPhoton2024,
  title = {Reionization after {{JWST}}: A Photon Budget Crisis?},
  shorttitle = {Reionization after {{JWST}}},
  author = {Mu{\~n}oz, Julian B and Mirocha, Jordan and Chisholm, John and Furlanetto, Steven R and Mason, Charlotte},
  year = {2024},
  month = nov,
  journal = {Monthly Notices of the Royal Astronomical Society: Letters},
  volume = {535},
  number = {1},
  pages = {L37-L43},
  issn = {1745-3925},
  doi = {10.1093/mnrasl/slae086},
  urldate = {2025-01-24}
}

@article{simmondsLowmassBurstyGalaxies2024,
  title = {Low-Mass Bursty Galaxies in {{JADES}} Efficiently Produce Ionizing Photons and Could Represent the Main Drivers of Reionization},
  author = {Simmonds, C and Tacchella, S and Hainline, K and Johnson, B D and McClymont, W and Robertson, B and Saxena, A and Sun, F and Witten, C and Baker, W M and Bhatawdekar, R and Boyett, K and Bunker, A J and Charlot, S and {Curtis-Lake}, E and Egami, E and Eisenstein, D J and Hausen, R and Maiolino, R and Maseda, M V and Scholtz, J and Williams, C C and Willott, C and Witstok, J},
  year = {2024},
  month = jan,
  journal = {\mnras},
  volume = {527},
  number = {3},
  pages = {6139--6157},
  issn = {0035-8711},
  doi = {10.1093/mnras/stad3605},
  urldate = {2025-01-29}
}

@article{robertsonNEWCONSTRAINTSCOSMIC2013,
  title = {{{NEW CONSTRAINTS ON COSMIC REIONIZATION FROM THE}} 2012 {{HUBBLE ULTRA DEEP FIELD CAMPAIGN}}},
  author = {Robertson, Brant E. and Furlanetto, Steven R. and Schneider, Evan and Charlot, Stephane and Ellis, Richard S. and Stark, Daniel P. and McLure, Ross J. and Dunlop, James S. and Koekemoer, Anton and Schenker, Matthew A. and Ouchi, Masami and Ono, Yoshiaki and {Curtis-Lake}, Emma and Rogers, Alexander B. and Bowler, Rebecca A. A. and Cirasuolo, Michele},
  year = {2013},
  month = apr,
  journal = {\apj},
  volume = {768},
  number = {1},
  pages = {71},
  publisher = {The American Astronomical Society},
  issn = {0004-637X},
  doi = {10.1088/0004-637X/768/1/71},
  urldate = {2025-01-30},
  langid = {english}
}

@article{lillyCanadaFranceRedshiftSurvey1996a,
  title = {The {{Canada-France Redshift Survey}}: {{The Luminosity Density}} and {{Star Formation History}} of the {{Universe}} to {{Z}} Approximately 1},
  shorttitle = {The {{Canada-France Redshift Survey}}},
  author = {Lilly, S. J. and Le Fevre, O. and Hammer, F. and Crampton, David},
  year = {1996},
  month = mar,
  journal = {\apj},
  volume = {460},
  pages = {L1},
  publisher = {IOP},
  issn = {0004-637X},
  doi = {10.1086/309975},
  urldate = {2025-01-30}
}

@article{aghanimPlanck2018Results2020a,
  title = {Planck 2018 Results - {{I}}. {{Overview}} and the Cosmological Legacy of {{Planck}}},
  author = {{Planck Collaboration}},
  year = {2020},
  month = sep,
  journal = {\aap},
  volume = {641},
  pages = {A1},
  publisher = {EDP Sciences},
  issn = {0004-6361, 1432-0746},
  doi = {10.1051/0004-6361/201833880},
  urldate = {2025-01-29},
  copyright = {{\copyright} ESO 2020},
  langid = {english}
}

@article{aghanimPlanck2018Results2020b,
  title = {Planck 2018 Results - {{VI}}. {{Cosmological}} Parameters},
  author = {{Planck Collaboration}},
  year = {2020},
  month = sep,
  journal = {\aap},
  volume = {641},
  pages = {A6},
  publisher = {EDP Sciences},
  issn = {0004-6361, 1432-0746},
  doi = {10.1051/0004-6361/201833910},
  urldate = {2025-01-29},
  copyright = {{\copyright} ESO 2020},
  langid = {english}
}

@article{kuhlenConcordanceModelsReionization2012,
  title = {Concordance Models of Reionization: Implications for Faint Galaxies and Escape Fraction Evolution},
  shorttitle = {Concordance Models of Reionization},
  author = {Kuhlen, Michael and {Faucher-Gigu{\`e}re}, Claude-Andr{\'e}},
  year = {2012},
  month = jun,
  journal = {\mnras},
  volume = {423},
  number = {1},
  pages = {862--876},
  issn = {0035-8711},
  doi = {10.1111/j.1365-2966.2012.20924.x},
  urldate = {2025-01-29}
}

@article{baluThermalReionizationHistory2023,
  title = {Thermal and Reionization History within a Large-Volume Semi-Analytic Galaxy Formation Simulation},
  author = {Balu, Sreedhar and Greig, Bradley and Qiu, Yisheng and Power, Chris and Qin, Yuxiang and Mutch, Simon and Wyithe, J Stuart B},
  year = {2023},
  month = apr,
  journal = {\mnras},
  volume = {520},
  number = {3},
  pages = {3368--3382},
  issn = {0035-8711},
  doi = {10.1093/mnras/stad281},
  urldate = {2025-01-28}
}

@article{kannanIntroducingThesanProject2022,
  title = {Introducing the Thesan Project: Radiation-Magnetohydrodynamic Simulations of the Epoch of Reionization},
  shorttitle = {Introducing the Thesan Project},
  author = {Kannan, R and Garaldi, E and Smith, A and Pakmor, R and Springel, V and Vogelsberger, M and Hernquist, L},
  year = {2022},
  month = apr,
  journal = {\mnras},
  volume = {511},
  number = {3},
  pages = {4005--4030},
  issn = {0035-8711},
  doi = {10.1093/mnras/stab3710},
  urldate = {2025-01-28}
}

@article{finkelsteinConditionsReionizingUniverse2019,
  title = {Conditions for {{Reionizing}} the {{Universe}} with a {{Low Galaxy Ionizing Photon Escape Fraction}}},
  author = {Finkelstein, Steven L. and D'Aloisio, Anson and Paardekooper, Jan-Pieter and Jr, Russell Ryan and Behroozi, Peter and Finlator, Kristian and Livermore, Rachael and Sanderbeck, Phoebe R. Upton and Vecchia, Claudio Dalla and Khochfar, Sadegh},
  year = {2019},
  month = jul,
  journal = {\apj},
  volume = {879},
  number = {1},
  pages = {36},
  publisher = {The American Astronomical Society},
  issn = {0004-637X},
  doi = {10.3847/1538-4357/ab1ea8},
  urldate = {2025-01-28},
  langid = {english}
}

@article{bouwensREIONIZATIONPLANCKDERIVED2015,
  title = {{{REIONIZATION AFTER PLANCK}}: {{THE DERIVED GROWTH OF THE COSMIC IONIZING EMISSIVITY NOW MATCHES THE GROWTH OF THE GALAXY UV LUMINOSITY DENSITY}}*},
  shorttitle = {{{REIONIZATION AFTER PLANCK}}},
  author = {Bouwens, R. J. and Illingworth, G. D. and Oesch, P. A. and Caruana, J. and Holwerda, B. and Smit, R. and Wilkins, S.},
  year = {2015},
  month = sep,
  journal = {\apj},
  volume = {811},
  number = {2},
  pages = {140},
  publisher = {The American Astronomical Society},
  issn = {0004-637X},
  doi = {10.1088/0004-637X/811/2/140},
  urldate = {2025-01-28},
  langid = {english}
}

@article{madauCosmicReionizationJWST2024,
  title = {Cosmic {{Reionization}} in the {{JWST Era}}: {{Back}} to {{AGNs}}?},
  shorttitle = {Cosmic {{Reionization}} in the {{JWST Era}}},
  author = {Madau, Piero and Giallongo, Emanuele and Grazian, Andrea and Haardt, Francesco},
  year = {2024},
  month = aug,
  journal = {\apj},
  volume = {971},
  pages = {75},
  publisher = {IOP},
  issn = {0004-637X},
  doi = {10.3847/1538-4357/ad5ce8},
  urldate = {2025-01-24}
}

@article{madauCosmicReionizationPlanck2015,
  title = {Cosmic {{Reionization}} after {{Planck}}: {{Could Quasars Do It All}}?},
  shorttitle = {Cosmic {{Reionization}} after {{Planck}}},
  author = {Madau, Piero and Haardt, Francesco},
  year = {2015},
  month = nov,
  journal = {\apj},
  volume = {813},
  pages = {L8},
  publisher = {IOP},
  issn = {0004-637X},
  doi = {10.1088/2041-8205/813/1/L8},
  urldate = {2025-01-24}
}

@article{masonModelindependentConstraintsHydrogenionizing2019,
  title = {Model-Independent Constraints on the Hydrogen-Ionizing Emissivity at z {$>$} 6},
  author = {Mason, Charlotte A and Naidu, Rohan P and Tacchella, Sandro and Leja, Joel},
  year = {2019},
  month = oct,
  journal = {\mnras},
  volume = {489},
  number = {2},
  pages = {2669--2676},
  issn = {0035-8711},
  doi = {10.1093/mnras/stz2291},
  urldate = {2025-01-24}
}

@article{robertsonCosmicReionizationEarly2015a,
  title = {Cosmic {{Reionization}} and {{Early Star-forming Galaxies}}: {{A Joint Analysis}} of {{New Constraints}} from {{Planck}} and the {{Hubble Space Telescope}}},
  shorttitle = {Cosmic {{Reionization}} and {{Early Star-forming Galaxies}}},
  author = {Robertson, Brant E. and Ellis, Richard S. and Furlanetto, Steven R. and Dunlop, James S.},
  year = {2015},
  month = apr,
  journal = {\apj},
  volume = {802},
  pages = {L19},
  publisher = {IOP},
  issn = {0004-637X},
  doi = {10.1088/2041-8205/802/2/L19},
  urldate = {2025-01-24}
}

@article{shullCriticalStarFormation2012,
  title = {Critical {{Star Formation Rates}} for {{Reionization}}: {{Full Reionization Occurs}} at {{Redshift}} z {$\approx$} 7},
  shorttitle = {Critical {{Star Formation Rates}} for {{Reionization}}},
  author = {Shull, J. Michael and Harness, Anthony and Trenti, Michele and Smith, Britton D.},
  year = {2012},
  month = mar,
  journal = {\apj},
  volume = {747},
  pages = {100},
  publisher = {IOP},
  issn = {0004-637X},
  doi = {10.1088/0004-637X/747/2/100},
  urldate = {2025-01-24}
}

@article{boquienCIGALEPythonCode2019b,
  title = {{{CIGALE}}: A Python {{Code Investigating GALaxy Emission}}},
  shorttitle = {{{CIGALE}}},
  author = {Boquien, M. and Burgarella, D. and Roehlly, Y. and Buat, V. and Ciesla, L. and Corre, D. and Inoue, A. K. and Salas, H.},
  year = {2019},
  month = feb,
  journal = {\aap},
  volume = {622},
  pages = {A103},
  publisher = {EDP Sciences},
  issn = {0004-6361, 1432-0746},
  doi = {10.1051/0004-6361/201834156},
  urldate = {2025-01-24},
  copyright = {{\copyright} ESO 2019},
  langid = {english}
}

@article{beckerNewMeasurementsIonizing2013,
  title = {New Measurements of the Ionizing Ultraviolet Background over 2~{$<~$}z~{$<~$}5 and Implications for Hydrogen Reionization},
  author = {Becker, George D. and Bolton, James S.},
  year = {2013},
  month = dec,
  journal = {\mnras},
  volume = {436},
  number = {2},
  pages = {1023--1039},
  issn = {0035-8711},
  doi = {10.1093/mnras/stt1610},
  urldate = {2025-01-24}
}

@article{nakaneLyaEmission7132024a,
  title = {Ly{$\alpha$} {{Emission}} at z = 7--13: {{Clear Evolution}} of {{Ly$\alpha$ Equivalent Width Indicating}} a {{Late Cosmic Reionization History}}},
  shorttitle = {Ly{$\alpha$} {{Emission}} at z = 7--13},
  author = {Nakane, Minami and Ouchi, Masami and Nakajima, Kimihiko and Harikane, Yuichi and Ono, Yoshiaki and Umeda, Hiroya and Isobe, Yuki and Zhang, Yechi and Xu, Yi},
  year = {2024},
  month = may,
  journal = {\apj},
  volume = {967},
  number = {1},
  pages = {28},
  publisher = {The American Astronomical Society},
  issn = {0004-637X},
  doi = {10.3847/1538-4357/ad38c2},
  urldate = {2025-01-24},
  langid = {english}
}

@article{adamPlanckIntermediateResults2016,
  title = {Planck Intermediate Results - {{XLVII}}. {{Planck}} Constraints on Reionization History},
  author = {{Planck Collaboration}},
  year = {2016},
  month = dec,
  journal = {\aap},
  volume = {596},
  pages = {A108},
  publisher = {EDP Sciences},
  issn = {0004-6361, 1432-0746},
  doi = {10.1051/0004-6361/201628897},
  urldate = {2025-01-24},
  copyright = {{\copyright} ESO, 2016},
  langid = {english}
}

@article{daviesQuantitativeConstraintsReionization2018,
  title = {Quantitative {{Constraints}} on the {{Reionization History}} from the {{IGM Damping Wing Signature}} in {{Two Quasars}} at z {$>$} 7},
  author = {Davies, Frederick B. and Hennawi, Joseph F. and Ba{\~n}ados, Eduardo and Luki{\'c}, Zarija and Decarli, Roberto and Fan, Xiaohui and Farina, Emanuele P. and Mazzucchelli, Chiara and Rix, Hans-Walter and Venemans, Bram P. and Walter, Fabian and Wang, Feige and Yang, Jinyi},
  year = {2018},
  month = sep,
  journal = {\apj},
  volume = {864},
  number = {2},
  pages = {142},
  publisher = {The American Astronomical Society},
  issn = {0004-637X},
  doi = {10.3847/1538-4357/aad6dc},
  urldate = {2025-01-24},
  langid = {english}
}

@article{greigConstraintsReionization752019,
  title = {Constraints on Reionization from the z = 7.5 {{QSO ULASJ1342}}+0928},
  author = {Greig, Bradley and Mesinger, Andrei and Ba{\~n}ados, Eduardo},
  year = {2019},
  month = apr,
  journal = {\mnras},
  volume = {484},
  number = {4},
  pages = {5094--5101},
  issn = {0035-8711},
  doi = {10.1093/mnras/stz230},
  urldate = {2025-01-24}
}

@article{hoagConstrainingNeutralFraction2019,
  title = {Constraining the {{Neutral Fraction}} of {{Hydrogen}} in the {{IGM}} at {{Redshift}} 7.5},
  author = {Hoag, A. and Brada{\v c}, M. and Huang, K. and Mason, C. and Treu, T. and Schmidt, K. B. and Trenti, M. and Strait, V. and Lemaux, B. C. and Finney, E. Q. and Paddock, M.},
  year = {2019},
  month = jun,
  journal = {\apj},
  volume = {878},
  number = {1},
  pages = {12},
  publisher = {The American Astronomical Society},
  issn = {0004-637X},
  doi = {10.3847/1538-4357/ab1de7},
  urldate = {2025-01-24},
  langid = {english}
}

@article{masonInferencesTimelineReionization2019,
  title = {Inferences on the Timeline of Reionization at z {$\sim$} 8 from the {{KMOS Lens-Amplified Spectroscopic Survey}}},
  author = {Mason, Charlotte A and Fontana, Adriano and Treu, Tommaso and Schmidt, Kasper B and Hoag, Austin and Abramson, Louis and Amorin, Ricardo and Brada{\v c}, Maru{\v s}a and Guaita, Lucia and Jones, Tucker and Henry, Alaina and Malkan, Matthew A and Pentericci, Laura and Trenti, Michele and Vanzella, Eros},
  year = {2019},
  month = may,
  journal = {\mnras},
  volume = {485},
  number = {3},
  pages = {3947--3969},
  issn = {0035-8711},
  doi = {10.1093/mnras/stz632},
  urldate = {2025-01-24}
}

@article{masonUniverseReionizing72018,
  title = {The {{Universe Is Reionizing}} at z {$\sim$} 7: {{Bayesian Inference}} of the {{IGM Neutral Fraction Using Ly$\alpha$ Emission}} from {{Galaxies}}},
  shorttitle = {The {{Universe Is Reionizing}} at z {$\sim$} 7},
  author = {Mason, Charlotte A. and Treu, Tommaso and Dijkstra, Mark and Mesinger, Andrei and Trenti, Michele and Pentericci, Laura and {de Barros}, Stephane and Vanzella, Eros},
  year = {2018},
  month = mar,
  journal = {\apj},
  volume = {856},
  number = {1},
  pages = {2},
  publisher = {The American Astronomical Society},
  issn = {0004-637X},
  doi = {10.3847/1538-4357/aab0a7},
  urldate = {2025-01-24},
  langid = {english}
}

@article{mcgreerModelindependentEvidenceFavour2015,
  title = {Model-Independent Evidence in Favour of an End to Reionization by z~{$\approx~$}6},
  author = {McGreer, Ian D. and Mesinger, Andrei and D'Odorico, Valentina},
  year = {2015},
  month = feb,
  journal = {\mnras},
  volume = {447},
  number = {1},
  pages = {499--505},
  issn = {0035-8711},
  doi = {10.1093/mnras/stu2449},
  urldate = {2025-01-24}
}

@article{tangJWSTNIRSpecObservations2024,
  title = {{{JWST}}/{{NIRSpec Observations}} of {{Lyman}} {$\alpha$} {{Emission}} in {{Star-forming Galaxies}} at 6.5 {$\lessequivlnt$} z {$\lessequivlnt$} 13},
  author = {Tang, Mengtao and Stark, Daniel P. and Topping, Michael W. and Mason, Charlotte and Ellis, Richard S.},
  year = {2024},
  month = nov,
  journal = {The Astrophysical Journal},
  volume = {975},
  number = {2},
  pages = {208},
  publisher = {The American Astronomical Society},
  issn = {0004-637X},
  doi = {10.3847/1538-4357/ad7eb7},
  urldate = {2025-02-13},
  langid = {english}
}

@ARTICLE{umedaSILVERRUSHXIVLya2024,
       author = {{Umeda}, Hiroya and {Ouchi}, Masami and {Kikuta}, Satoshi and {Harikane}, Yuichi and {Ono}, Yoshiaki and {Shibuya}, Takatoshi and {Inoue}, Akio K. and {Shimasaku}, Kazuhiro and {Liang}, Yongming and {Matsumoto}, Akinori and {Saito}, Shun and {Kusakabe}, Haruka and {Kageura}, Yuta and {Nakane}, Minami},
        title = "{SILVERRUSH. XIV. Lya Luminosity Functions and Angular Correlation Functions from \raisebox{-0.5ex}\textasciitilde20,000 Lya Emitters at z\raisebox{-0.5ex}\textasciitilde2.2-7.3 from upto 24 ${\rm deg}^2$ HSC-SSP and CHORUS Surveys: Linking the Post-Reionization Epoch to the Heart of Reionization}",
      journal = {arXiv e-prints},
         year = 2024,
        month = nov,
          eid = {arXiv:2411.15495},
        pages = {arXiv:2411.15495},
          doi = {10.48550/arXiv.2411.15495},
archivePrefix = {arXiv},
       eprint = {2411.15495},
 primaryClass = {astro-ph.GA},
       adsurl = {https://ui.adsabs.harvard.edu/abs/2024arXiv241115495U}
}

@misc{gnedinModelingCosmicReionization2022,
  title = {Modeling {{Cosmic Reionization}}},
  author = {Gnedin, Nickolay Y. and Madau, Piero},
  year = {2022},
  month = aug,
  journal = {arXiv.org},
  urldate = {2025-01-24},
  howpublished = {https://arxiv.org/abs/2208.02260v2},
  langid = {english}
}

@article{madauRadiativeTransferClumpy1999,
  title = {Radiative {{Transfer}} in a {{Clumpy Universe}}. {{III}}. {{The Nature}} of {{Cosmological Ionizing Sources}}},
  author = {Madau, Piero and Haardt, Francesco and Rees, Martin J.},
  year = {1999},
  month = apr,
  journal = {\aj},
  volume = {514},
  number = {2},
  pages = {648},
  publisher = {IOP Publishing},
  issn = {0004-637X},
  doi = {10.1086/306975},
  urldate = {2025-01-24},
  langid = {english}
}

@ARTICLE{smithLymanContinuumEscape2020,
       author = {{Smith}, Brent M. and {Windhorst}, Rogier A. and {Cohen}, Seth H. and {Koekemoer}, Anton M. and {Jansen}, Rolf A. and {White}, Cameron and {Borthakur}, Sanchayeeta and {Hathi}, Nimish and {Jiang}, Linhua and {Rutkowski}, Michael and {Ryan}, Jr., Russell E. and {Inoue}, Akio K. and {O'Connell}, Robert W. and {MacKenty}, John W. and {Conselice}, Christopher J. and {Silk}, Joseph I.},
        title = "{The Lyman Continuum Escape Fraction of Galaxies and AGN in the GOODS Fields}",
      journal = {\apj},
         year = 2020,
        month = jul,
       volume = {897},
       number = {1},
          eid = {41},
        pages = {41},
          doi = {10.3847/1538-4357/ab8811},
archivePrefix = {arXiv},
       eprint = {2004.04360},
 primaryClass = {astro-ph.GA},
       adsurl = {https://ui.adsabs.harvard.edu/abs/2020ApJ...897...41S}
}

@ARTICLE{smithLymanContinuumEmission2024,
       author = {{Smith}, Brent M. and {Windhorst}, Rogier A. and {Teplitz}, Harry and {Hayes}, Matthew and {Rafelski}, Marc and {Dickinson}, Mark and {Mehta}, Vihang and {Hathi}, Nimish P. and {MacKenty}, John and {Yung}, L.~Y. Aaron and {Koekemoer}, Anton M. and {Soto}, Emmaris and {Conselice}, Christopher J. and {Lucas}, Ray A. and {Wang}, Xin and {Kim}, Keunho J. and {Alavi}, Anahita and {Grogin}, Norman A. and {Sunnquist}, Ben and {Prichard}, Laura and {Jansen}, Rolf A. and {Uvcandels Team}},
        title = "{Lyman Continuum Emission from Active Galactic Nuclei at 2.3 {\ensuremath{\lesssim}} z {\ensuremath{\lesssim}} 3.7 in the UVCANDELS Fields}",
      journal = {\apj},
         year = 2024,
        month = mar,
       volume = {964},
       number = {1},
          eid = {73},
        pages = {73},
          doi = {10.3847/1538-4357/ad1ef0},
archivePrefix = {arXiv},
       eprint = {2401.03094},
 primaryClass = {astro-ph.GA},
       adsurl = {https://ui.adsabs.harvard.edu/abs/2024ApJ...964...73S}
}

@ARTICLE{fermi-latcollaborationGammarayDeterminationUniverses2018,
       author = {{Fermi-LAT Collaboration} and {Abdollahi}, S. and {Ackermann}, M. and {Ajello}, M. and {Atwood}, W.~B. and {Baldini}, L. and {Ballet}, J. and {Barbiellini}, G. and {Bastieri}, D. and {Becerra Gonzalez}, J. and {Bellazzini}, R. and {Bissaldi}, E. and {Blandford}, R.~D. and {Bloom}, E.~D. and {Bonino}, R. and {Bottacini}, E. and {Buson}, S. and {Bregeon}, J. and {Bruel}, P. and {Buehler}, R. and {Cameron}, R.~A. and {Caputo}, R. and {Caraveo}, P.~A. and {Cavazzuti}, E. and {Charles}, E. and {Chen}, S. and {Cheung}, C.~C. and {Chiaro}, G. and {Ciprini}, S. and {Cohen-Tanugi}, J. and {Cominsky}, L.~R. and {Conrad}, J. and {Costantin}, D. and {Cutini}, S. and {D'Ammando}, F. and {de Palma}, F. and {Desai}, A. and {Digel}, S.~W. and {Di Lalla}, N. and {Di Mauro}, M. and {Di Venere}, L. and {Dom{\'\i}nguez}, A. and {Favuzzi}, C. and {Fegan}, S.~J. and {Finke}, J. and {Franckowiak}, A. and {Fukazawa}, Y. and {Funk}, S. and {Fusco}, P. and {Gallardo Romero}, G. and {Gargano}, F. and {Gasparrini}, D. and {Giglietto}, N. and {Giordano}, F. and {Giroletti}, M. and {Green}, D. and {Grenier}, I.~A. and {Guillemot}, L. and {Guiriec}, S. and {Hartmann}, D.~H. and {Hays}, E. and {Helgason}, K. and {Horan}, D. and {J{\'o}hannesson}, G. and {Kocevski}, D. and {Kuss}, M. and {Larsson}, S. and {Latronico}, L. and {Li}, J. and {Longo}, F. and {Loparco}, F. and {Lott}, B. and {Lovellette}, M.~N. and {Lubrano}, P. and {Madejski}, G.~M. and {Magill}, J.~D. and {Maldera}, S. and {Manfreda}, A. and {Marcotulli}, L. and {Mazziotta}, M.~N. and {McEnery}, J.~E. and {Meyer}, M. and {Michelson}, P.~F. and {Mizuno}, T. and {Monzani}, M.~E. and {Morselli}, A. and {Moskalenko}, I.~V. and {Negro}, M. and {Nuss}, E. and {Ojha}, R. and {Omodei}, N. and {Orienti}, M. and {Orlando}, E. and {Ormes}, J.~F. and {Palatiello}, M. and {Paliya}, V.~S. and {Paneque}, D. and {Perkins}, J.~S. and {Persic}, M. and {Pesce-Rollins}, M. and {Petrosian}, V. and {Piron}, F. and {Porter}, T.~A. and {Primack}, J.~R. and {Principe}, G. and {Rain{\`o}}, S. and {Rando}, R. and {Razzano}, M. and {Razzaque}, S. and {Reimer}, A. and {Reimer}, O. and {Saz Parkinson}, P.~M. and {Sgr{\`o}}, C. and {Siskind}, E.~J. and {Spandre}, G. and {Spinelli}, P. and {Suson}, D.~J. and {Tajima}, H. and {Takahashi}, M. and {Thayer}, J.~B. and {Tibaldo}, L. and {Torres}, D.~F. and {Torresi}, E. and {Tosti}, G. and {Tramacere}, A. and {Troja}, E. and {Valverde}, J. and {Vianello}, G. and {Vogel}, M. and {Wood}, K. and {Zaharijas}, G.},
        title = "{A gamma-ray determination of the Universe's star formation history}",
      journal = {Science},
         year = 2018,
        month = nov,
       volume = {362},
       number = {6418},
        pages = {1031-1034},
          doi = {10.1126/science.aat8123},
archivePrefix = {arXiv},
       eprint = {1812.01031},
 primaryClass = {astro-ph.HE},
       adsurl = {https://ui.adsabs.harvard.edu/abs/2018Sci...362.1031F}
}

@ARTICLE{koushanGAMADEVILSConstraining2021,
       author = {{Koushan}, Soheil and {Driver}, Simon P. and {Bellstedt}, Sabine and {Davies}, Luke J. and {Robotham}, Aaron S.~G. and {Lagos}, Claudia del P. and {Hashemizadeh}, Abdolhosein and {Obreschkow}, Danail and {Thorne}, Jessica E. and {Bremer}, Malcolm and {Holwerda}, B.~W. and {Hopkins}, Andrew M. and {Jarvis}, Matt J. and {Siudek}, Malgorzata and {Windhorst}, Rogier A.},
        title = "{GAMA/DEVILS: constraining the cosmic star formation history from improved measurements of the 0.3-2.2 {\ensuremath{\mu}}m extragalactic background light}",
      journal = {\mnras},
         year = 2021,
        month = may,
       volume = {503},
       number = {2},
        pages = {2033-2052},
          doi = {10.1093/mnras/stab540},
archivePrefix = {arXiv},
       eprint = {2102.12323},
 primaryClass = {astro-ph.CO},
       adsurl = {https://ui.adsabs.harvard.edu/abs/2021MNRAS.503.2033K}
}

@ARTICLE{bagleyCEERSEpochNIRCam2023a,
       author = {{Bagley}, Micaela B. and {Finkelstein}, Steven L. and {Koekemoer}, Anton M. and {Ferguson}, Henry C. and {Arrabal Haro}, Pablo and {Dickinson}, Mark and {Kartaltepe}, Jeyhan S. and {Papovich}, Casey and {P{\'e}rez-Gonz{\'a}lez}, Pablo G. and {Pirzkal}, Nor and {Somerville}, Rachel S. and {Willmer}, Christopher N.~A. and {Yang}, Guang and {Yung}, L.~Y. Aaron and {Fontana}, Adriano and {Grazian}, Andrea and {Grogin}, Norman A. and {Hirschmann}, Michaela and {Kewley}, Lisa J. and {Kirkpatrick}, Allison and {Kocevski}, Dale D. and {Lotz}, Jennifer M. and {Medrano}, Aubrey and {Morales}, Alexa M. and {Pentericci}, Laura and {Ravindranath}, Swara and {Trump}, Jonathan R. and {Wilkins}, Stephen M. and {Calabr{\`o}}, Antonello and {Cooper}, M.~C. and {Costantin}, Luca and {de la Vega}, Alexander and {Hilbert}, Bryan and {Hutchison}, Taylor A. and {Larson}, Rebecca L. and {Lucas}, Ray A. and {McGrath}, Elizabeth J. and {Ryan}, Russell and {Wang}, Xin and {Wuyts}, Stijn},
        title = "{CEERS Epoch 1 NIRCam Imaging: Reduction Methods and Simulations Enabling Early JWST Science Results}",
      journal = {\apjl},
         year = 2023,
        month = mar,
       volume = {946},
       number = {1},
          eid = {L12},
        pages = {L12},
          doi = {10.3847/2041-8213/acbb08},
archivePrefix = {arXiv},
       eprint = {2211.02495},
 primaryClass = {astro-ph.IM},
       adsurl = {https://ui.adsabs.harvard.edu/abs/2023ApJ...946L..12B}
}

@ARTICLE{finkelsteinCEERSKeyPaper2023a,
       author = {{Finkelstein}, Steven L. and {Bagley}, Micaela B. and {Ferguson}, Henry C. and {Wilkins}, Stephen M. and {Kartaltepe}, Jeyhan S. and {Papovich}, Casey and {Yung}, L.~Y. Aaron and {Arrabal Haro}, Pablo and {Behroozi}, Peter and {Dickinson}, Mark and {Kocevski}, Dale D. and {Koekemoer}, Anton M. and {Larson}, Rebecca L. and {Le Bail}, Aur{\'e}lien and {Morales}, Alexa M. and {P{\'e}rez-Gonz{\'a}lez}, Pablo G. and {Burgarella}, Denis and {Dav{\'e}}, Romeel and {Hirschmann}, Michaela and {Somerville}, Rachel S. and {Wuyts}, Stijn and {Bromm}, Volker and {Casey}, Caitlin M. and {Fontana}, Adriano and {Fujimoto}, Seiji and {Gardner}, Jonathan P. and {Giavalisco}, Mauro and {Grazian}, Andrea and {Grogin}, Norman A. and {Hathi}, Nimish P. and {Hutchison}, Taylor A. and {Jha}, Saurabh W. and {Jogee}, Shardha and {Kewley}, Lisa J. and {Kirkpatrick}, Allison and {Long}, Arianna S. and {Lotz}, Jennifer M. and {Pentericci}, Laura and {Pierel}, Justin D.~R. and {Pirzkal}, Nor and {Ravindranath}, Swara and {Ryan}, Russell E. and {Trump}, Jonathan R. and {Yang}, Guang and {Bhatawdekar}, Rachana and {Bisigello}, Laura and {Buat}, V{\'e}ronique and {Calabr{\`o}}, Antonello and {Castellano}, Marco and {Cleri}, Nikko J. and {Cooper}, M.~C. and {Croton}, Darren and {Daddi}, Emanuele and {Dekel}, Avishai and {Elbaz}, David and {Franco}, Maximilien and {Gawiser}, Eric and {Holwerda}, Benne W. and {Huertas-Company}, Marc and {Jaskot}, Anne E. and {Leung}, Gene C.~K. and {Lucas}, Ray A. and {Mobasher}, Bahram and {Pandya}, Viraj and {Tacchella}, Sandro and {Weiner}, Benjamin J. and {Zavala}, Jorge A.},
        title = "{CEERS Key Paper. I. An Early Look into the First 500 Myr of Galaxy Formation with JWST}",
      journal = {\apjl},
         year = 2023,
        month = mar,
       volume = {946},
       number = {1},
          eid = {L13},
        pages = {L13},
          doi = {10.3847/2041-8213/acade4},
archivePrefix = {arXiv},
       eprint = {2211.05792},
 primaryClass = {astro-ph.GA},
       adsurl = {https://ui.adsabs.harvard.edu/abs/2023ApJ...946L..13F}
}

@ARTICLE{bagleyNextGenerationDeep2024,
       author = {{Bagley}, Micaela B. and {Pirzkal}, Nor and {Finkelstein}, Steven L. and {Papovich}, Casey and {Berg}, Danielle A. and {Lotz}, Jennifer M. and {Leung}, Gene C.~K. and {Ferguson}, Henry C. and {Koekemoer}, Anton M. and {Dickinson}, Mark and {Kartaltepe}, Jeyhan S. and {Kocevski}, Dale D. and {Somerville}, Rachel S. and {Yung}, L.~Y. Aaron and {Backhaus}, Bren E. and {Casey}, Caitlin M. and {Castellano}, Marco and {Ch{\'a}vez Ortiz}, {\'O}scar A. and {Chworowsky}, Katherine and {Cox}, Isabella G. and {Dav{\'e}}, Romeel and {Davis}, Kelcey and {Estrada-Carpenter}, Vicente and {Fontana}, Adriano and {Fujimoto}, Seiji and {Gardner}, Jonathan P. and {Giavalisco}, Mauro and {Grazian}, Andrea and {Grogin}, Norman A. and {Hathi}, Nimish P. and {Hutchison}, Taylor A. and {Jaskot}, Anne E. and {Jung}, Intae and {Kewley}, Lisa J. and {Kirkpatrick}, Allison and {Larson}, Rebecca L. and {Matharu}, Jasleen and {Natarajan}, Priyamvada and {Pentericci}, Laura and {P{\'e}rez-Gonz{\'a}lez}, Pablo G. and {Ravindranath}, Swara and {Rothberg}, Barry and {Ryan}, Russell and {Shen}, Lu and {Simons}, Raymond C. and {Snyder}, Gregory F. and {Trump}, Jonathan R. and {Wilkins}, Stephen M.},
        title = "{The Next Generation Deep Extragalactic Exploratory Public (NGDEEP) Survey}",
      journal = {\apjl},
         year = 2024,
        month = apr,
       volume = {965},
       number = {1},
          eid = {L6},
        pages = {L6},
          doi = {10.3847/2041-8213/ad2f31},
archivePrefix = {arXiv},
       eprint = {2302.05466},
 primaryClass = {astro-ph.GA},
       adsurl = {https://ui.adsabs.harvard.edu/abs/2024ApJ...965L...6B}
}

@ARTICLE{eisensteinOverviewJWSTAdvanced2023a,
       author = {{Eisenstein}, Daniel J. and {Willott}, Chris and {Alberts}, Stacey and {Arribas}, Santiago and {Bonaventura}, Nina and {Bunker}, Andrew J. and {Cameron}, Alex J. and {Carniani}, Stefano and {Charlot}, Stephane and {Curtis-Lake}, Emma and {D'Eugenio}, Francesco and {Endsley}, Ryan and {Ferruit}, Pierre and {Giardino}, Giovanna and {Hainline}, Kevin and {Hausen}, Ryan and {Jakobsen}, Peter and {Johnson}, Benjamin D. and {Maiolino}, Roberto and {Rieke}, Marcia and {Rieke}, George and {Rix}, Hans-Walter and {Robertson}, Brant and {Stark}, Daniel P. and {Tacchella}, Sandro and {Williams}, Christina C. and {Willmer}, Christopher N.~A. and {Baker}, William M. and {Baum}, Stefi and {Bhatawdekar}, Rachana and {Boyett}, Kristan and {Chen}, Zuyi and {Chevallard}, Jacopo and {Circosta}, Chiara and {Curti}, Mirko and {Danhaive}, A. Lola and {DeCoursey}, Christa and {de Graaff}, Anna and {Dressler}, Alan and {Egami}, Eiichi and {Helton}, Jakob M. and {Hviding}, Raphael E. and {Ji}, Zhiyuan and {Jones}, Gareth C. and {Kumari}, Nimisha and {L{\"u}tzgendorf}, Nora and {Laseter}, Isaac and {Looser}, Tobias J. and {Lyu}, Jianwei and {Maseda}, Michael V. and {Nelson}, Erica and {Parlanti}, Eleonora and {Perna}, Michele and {Pusk{\'a}s}, D{\'a}vid and {Rawle}, Tim and {Rodr{\'\i}guez Del Pino}, Bruno and {Sandles}, Lester and {Saxena}, Aayush and {Scholtz}, Jan and {Sharpe}, Katherine and {Shivaei}, Irene and {Silcock}, Maddie S. and {Simmonds}, Charlotte and {Skarbinski}, Maya and {Smit}, Renske and {Stone}, Meredith and {Suess}, Katherine A. and {Sun}, Fengwu and {Tang}, Mengtao and {Topping}, Michael W. and {{\"U}bler}, Hannah and {Villanueva}, Natalia C. and {Wallace}, Imaan E.~B. and {Whitler}, Lily and {Witstok}, Joris and {Woodrum}, Charity},
        title = "{Overview of the JWST Advanced Deep Extragalactic Survey (JADES)}",
      journal = {arXiv e-prints},
         year = 2023,
        month = jun,
          eid = {arXiv:2306.02465},
        pages = {arXiv:2306.02465},
          doi = {10.48550/arXiv.2306.02465},
archivePrefix = {arXiv},
       eprint = {2306.02465},
 primaryClass = {astro-ph.GA},
       adsurl = {https://ui.adsabs.harvard.edu/abs/2023arXiv230602465E}
}

@ARTICLE{riekeJADESInitialData2023,
       author = {{Rieke}, Marcia J. and {Robertson}, Brant and {Tacchella}, Sandro and {Hainline}, Kevin and {Johnson}, Benjamin D. and {Hausen}, Ryan and {Ji}, Zhiyuan and {Willmer}, Christopher N.~A. and {Eisenstein}, Daniel J. and {Pusk{\'a}s}, D{\'a}vid and {Alberts}, Stacey and {Arribas}, Santiago and {Baker}, William M. and {Baum}, Stefi and {Bhatawdekar}, Rachana and {Bonaventura}, Nina and {Boyett}, Kristan and {Bunker}, Andrew J. and {Cameron}, Alex J. and {Carniani}, Stefano and {Charlot}, Stephane and {Chevallard}, Jacopo and {Chen}, Zuyi and {Curti}, Mirko and {Curtis-Lake}, Emma and {Danhaive}, A. Lola and {DeCoursey}, Christa and {Dressler}, Alan and {Egami}, Eiichi and {Endsley}, Ryan and {Helton}, Jakob M. and {Hviding}, Raphael E. and {Kumari}, Nimisha and {Looser}, Tobias J. and {Lyu}, Jianwei and {Maiolino}, Roberto and {Maseda}, Michael V. and {Nelson}, Erica J. and {Rieke}, George and {Rix}, Hans-Walter and {Sandles}, Lester and {Saxena}, Aayush and {Sharpe}, Katherine and {Shivaei}, Irene and {Skarbinski}, Maya and {Smit}, Renske and {Stark}, Daniel P. and {Stone}, Meredith and {Suess}, Katherine A. and {Sun}, Fengwu and {Topping}, Michael and {{\"U}bler}, Hannah and {Villanueva}, Natalia C. and {Wallace}, Imaan E.~B. and {Williams}, Christina C. and {Willott}, Chris and {Whitler}, Lily and {Witstok}, Joris and {Woodrum}, Charity},
        title = "{JADES Initial Data Release for the Hubble Ultra Deep Field: Revealing the Faint Infrared Sky with Deep JWST NIRCam Imaging}",
      journal = {\apjs},
         year = 2023,
        month = nov,
       volume = {269},
       number = {1},
          eid = {16},
        pages = {16},
          doi = {10.3847/1538-4365/acf44d},
archivePrefix = {arXiv},
       eprint = {2306.02466},
 primaryClass = {astro-ph.GA},
       adsurl = {https://ui.adsabs.harvard.edu/abs/2023ApJS..269...16R}
}

@ARTICLE{conroyModelingPanchromaticSpectral2013a,
       author = {{Conroy}, Charlie},
        title = "{Modeling the Panchromatic Spectral Energy Distributions of Galaxies}",
      journal = {\araa},
         year = 2013,
        month = aug,
       volume = {51},
       number = {1},
        pages = {393-455},
          doi = {10.1146/annurev-astro-082812-141017},
archivePrefix = {arXiv},
       eprint = {1301.7095},
 primaryClass = {astro-ph.CO},
       adsurl = {https://ui.adsabs.harvard.edu/abs/2013ARA&A..51..393C}
}

@ARTICLE{harikaneJWSTNIRSpecFirst2023b,
       author = {{Harikane}, Yuichi and {Zhang}, Yechi and {Nakajima}, Kimihiko and {Ouchi}, Masami and {Isobe}, Yuki and {Ono}, Yoshiaki and {Hatano}, Shun and {Xu}, Yi and {Umeda}, Hiroya},
        title = "{A JWST/NIRSpec First Census of Broad-line AGNs at z = 4-7: Detection of 10 Faint AGNs with M $_{BH}$ {}10$^{6}$-{}10$^{8}$ M $_{{\ensuremath{\odot}}}$ and Their Host Galaxy Properties}",
      journal = {\apj},
         year = 2023,
        month = dec,
       volume = {959},
       number = {1},
          eid = {39},
        pages = {39},
          doi = {10.3847/1538-4357/ad029e},
archivePrefix = {arXiv},
       eprint = {2303.11946},
 primaryClass = {astro-ph.GA},
       adsurl = {https://ui.adsabs.harvard.edu/abs/2023ApJ...959...39H}
}

@ARTICLE{madaudickinson2014,
       author = {{Madau}, Piero and {Dickinson}, Mark},
        title = "{Cosmic Star-Formation History}",
      journal = {\araa},
         year = 2014,
        month = aug,
       volume = {52},
        pages = {415-486},
          doi = {10.1146/annurev-astro-081811-125615},
archivePrefix = {arXiv},
       eprint = {1403.0007},
 primaryClass = {astro-ph.CO},
       adsurl = {https://ui.adsabs.harvard.edu/abs/2014ARA&A..52..415M}
}

@ARTICLE{okeSecondaryStandardStars1983,
       author = {{Oke}, J.~B. and {Gunn}, J.~E.},
        title = "{Secondary standard stars for absolute spectrophotometry.}",
      journal = {\apj},
         year = 1983,
        month = mar,
       volume = {266},
        pages = {713-717},
          doi = {10.1086/160817},
       adsurl = {https://ui.adsabs.harvard.edu/abs/1983ApJ...266..713O}
}

@ARTICLE{lagosFarultravioletFarinfraredGalaxy2019,
       author = {{Lagos}, Claudia del P. and {Robotham}, Aaron S.~G. and {Trayford}, James W. and {Tobar}, Rodrigo and {Bravo}, Mat{\'\i}as and {Bellstedt}, Sabine and {Davies}, Luke J.~M. and {Driver}, Simon P. and {Elahi}, Pascal J. and {Obreschkow}, Danail and {Power}, Chris},
        title = "{From the far-ultraviolet to the far-infrared - galaxy emission at 0 {\ensuremath{\leq}} z {\ensuremath{\leq}} 10 in the SHARK semi-analytic model}",
      journal = {\mnras},
         year = 2019,
        month = nov,
       volume = {489},
       number = {3},
        pages = {4196-4216},
          doi = {10.1093/mnras/stz2427},
archivePrefix = {arXiv},
       eprint = {1908.03423},
 primaryClass = {astro-ph.GA},
       adsurl = {https://ui.adsabs.harvard.edu/abs/2019MNRAS.489.4196L}
}

@ARTICLE{larsonCEERSDiscoveryAccreting2023a,
       author = {{Larson}, Rebecca L. and {Finkelstein}, Steven L. and {Kocevski}, Dale D. and {Hutchison}, Taylor A. and {Trump}, Jonathan R. and {Arrabal Haro}, Pablo and {Bromm}, Volker and {Cleri}, Nikko J. and {Dickinson}, Mark and {Fujimoto}, Seiji and {Kartaltepe}, Jeyhan S. and {Koekemoer}, Anton M. and {Papovich}, Casey and {Pirzkal}, Nor and {Tacchella}, Sandro and {Zavala}, Jorge A. and {Bagley}, Micaela and {Behroozi}, Peter and {Champagne}, Jaclyn B. and {Cole}, Justin W. and {Jung}, Intae and {Morales}, Alexa M. and {Yang}, Guang and {Zhang}, Haowen and {Zitrin}, Adi and {Amor{\'\i}n}, Ricardo O. and {Burgarella}, Denis and {Casey}, Caitlin M. and {Ch{\'a}vez Ortiz}, {\'O}scar A. and {Cox}, Isabella G. and {Chworowsky}, Katherine and {Fontana}, Adriano and {Gawiser}, Eric and {Grazian}, Andrea and {Grogin}, Norman A. and {Harish}, Santosh and {Hathi}, Nimish P. and {Hirschmann}, Michaela and {Holwerda}, Benne W. and {Juneau}, St{\'e}phanie and {Leung}, Gene C.~K. and {Lucas}, Ray A. and {McGrath}, Elizabeth J. and {P{\'e}rez-Gonz{\'a}lez}, Pablo G. and {Rigby}, Jane R. and {Seill{\'e}}, Lise-Marie and {Simons}, Raymond C. and {de La Vega}, Alexander and {Weiner}, Benjamin J. and {Wilkins}, Stephen M. and {Yung}, L.~Y. Aaron and {Ceers Team}},
        title = "{A CEERS Discovery of an Accreting Supermassive Black Hole 570 Myr after the Big Bang: Identifying a Progenitor of Massive z > 6 Quasars}",
      journal = {\apjl},
         year = 2023,
        month = aug,
       volume = {953},
       number = {2},
          eid = {L29},
        pages = {L29},
          doi = {10.3847/2041-8213/ace619},
archivePrefix = {arXiv},
       eprint = {2303.08918},
 primaryClass = {astro-ph.GA},
       adsurl = {https://ui.adsabs.harvard.edu/abs/2023ApJ...953L..29L}
}

@article{shenBolometricQuasarLuminosity2020,
  title = {The Bolometric Quasar Luminosity Function at {\emph{z}} = 0--7},
  author = {Shen, Xuejian and Hopkins, Philip F and {Faucher-Gigu{\`e}re}, Claude-Andr{\'e} and Alexander, D M and Richards, Gordon T and Ross, Nicholas P and Hickox, R C},
  year = {2020},
  month = jul,
  journal = {\mnras},
  volume = {495},
  number = {3},
  pages = {3252--3275},
  issn = {0035-8711, 1365-2966},
  doi = {10.1093/mnras/staa1381},
  urldate = {2022-11-11},
  langid = {english}
}

@article{maiolinoJADESDiversePopulation2023a,
  title = {{{JADES}} - {{The}} Diverse Population of Infant Black Holes at 4 {$<$} z {$<$} 11: {{Merging}}, Tiny, Poor, but Mighty},
  shorttitle = {{{JADES}} - {{The}} Diverse Population of Infant Black Holes at 4 {$<$} z {$<$} 11},
  author = {Maiolino, Roberto and Scholtz, Jan and {Curtis-Lake}, Emma and Carniani, Stefano and Baker, William and de Graaff, Anna and Tacchella, Sandro and {\"U}bler, Hannah and D'Eugenio, Francesco and Witstok, Joris and Curti, Mirko and Arribas, Santiago and Bunker, Andrew J. and Charlot, St{\'e}phane and Chevallard, Jacopo and Eisenstein, Daniel J. and Egami, Eiichi and Ji, Zhiyuan and Jones, Gareth C. and Lyu, Jianwei and Rawle, Tim and Robertson, Brant and Rujopakarn, Wiphu and Perna, Michele and Sun, Fengwu and Venturi, Giacomo and Williams, Christina C. and Willott, Chris},
  year = {2024},
  month = nov,
  journal = {Astronomy \& Astrophysics},
  volume = {691},
  pages = {A145},
  publisher = {EDP Sciences},
  issn = {0004-6361, 1432-0746},
  doi = {10.1051/0004-6361/202347640},
  urldate = {2025-02-21},
  copyright = {{\copyright} The Authors 2024},
  langid = {english}
}

@article{greeneUNCOVERSpectroscopyConfirms2024,
  title = {{{UNCOVER Spectroscopy Confirms}} the {{Surprising Ubiquity}} of {{Active Galactic Nuclei}} in {{Red Sources}} at z {$>$} 5},
  author = {Greene, Jenny E. and Labbe, Ivo and Goulding, Andy D. and Furtak, Lukas J. and Chemerynska, Iryna and Kokorev, Vasily and Dayal, Pratika and Volonteri, Marta and Williams, Christina C. and Wang, Bingjie and Setton, David J. and Burgasser, Adam J. and Bezanson, Rachel and Atek, Hakim and Brammer, Gabriel and Cutler, Sam E. and Feldmann, Robert and Fujimoto, Seiji and Glazebrook, Karl and {de Graaff}, Anna and Khullar, Gourav and Leja, Joel and Marchesini, Danilo and Maseda, Michael V. and Matthee, Jorryt and Miller, Tim B. and Naidu, Rohan P. and Nanayakkara, Themiya and Oesch, Pascal A. and Pan, Richard and Papovich, Casey and Price, Sedona H. and {van Dokkum}, Pieter and Weaver, John R. and Whitaker, Katherine E. and Zitrin, Adi},
  year = {2024},
  month = mar,
  journal = {\apj},
  volume = {964},
  pages = {39},
  publisher = {IOP},
  issn = {0004-637X},
  doi = {10.3847/1538-4357/ad1e5f},
  urldate = {2024-07-30}
}

@article{atekExtremeFaintEnd2018,
  title = {The Extreme Faint End of the {{UV}} Luminosity Function at z {$\sim$} 6 through Gravitational Telescopes: A Comprehensive Assessment of Strong Lensing Uncertainties},
  shorttitle = {The Extreme Faint End of the {{UV}} Luminosity Function at z {$\sim$} 6 through Gravitational Telescopes},
  author = {Atek, Hakim and Richard, Johan and Kneib, Jean-Paul and Schaerer, Daniel},
  year = {2018},
  month = oct,
  journal = {\mnras},
  volume = {479},
  pages = {5184--5195},
  publisher = {OUP},
  issn = {0035-8711},
  doi = {10.1093/mnras/sty1820},
  urldate = {2024-07-25}
}

@article{nakajimaJWSTCensusMassMetallicity2023a,
  title = {{{JWST Census}} for the {{Mass-Metallicity Star Formation Relations}} at z = 4-10 with {{Self-consistent Flux Calibration}} and {{Proper Metallicity Calibrators}}},
  author = {Nakajima, Kimihiko and Ouchi, Masami and Isobe, Yuki and Harikane, Yuichi and Zhang, Yechi and Ono, Yoshiaki and Umeda, Hiroya and Oguri, Masamune},
  year = {2023},
  month = dec,
  journal = {\apjs},
  volume = {269},
  pages = {33},
  publisher = {IOP},
  issn = {0067-0049},
  doi = {10.3847/1538-4365/acd556},
  urldate = {2024-07-12}
}

@article{juodzbalisEPOCHSVIIDiscovery2023,
  title = {{{EPOCHS VII}}: {{Discovery}} of High Redshift (\$6.5 {$<$} z {$<$} 12\$) {{AGN}} Candidates in {{JWST ERO}} and {{PEARLS}} Data},
  shorttitle = {{{EPOCHS VII}}},
  author = {Juod{\v z}balis, Ignas and Conselice, Christopher J. and Singh, Maitrayee and Adams, Nathan and Ormerod, Katherine and Harvey, Thomas and Austin, Duncan and Volonteri, Marta and Cohen, Seth H. and Jansen, Rolf A. and Summers, Jake and Windhorst, Rogier A. and D'Silva, Jordan C. J. and Koekemoer, Anton M. and Coe, Dan and Driver, Simon P. and Frye, Brenda and Grogin, Norman A. and Marshall, Madeline A. and Nonino, Mario and Pirzkal, Nor and Robotham, Aaron and Ryan, Jr and Ortiz III, Rafael and Tompkins, Scott and Willmer, Christopher N. A. and Yan, Haojing},
  year = {2023},
  month = aug,
  journal = {\mnras},
  volume = {525},
  number = {1},
  eprint = {2307.07535},
  primaryclass = {astro-ph},
  pages = {1353--1364},
  issn = {0035-8711, 1365-2966},
  doi = {10.1093/mnras/stad2396},
  urldate = {2023-09-11},
  archiveprefix = {arxiv}
}

@article{mattheeLittleRedDots2024,
  title = {Little {{Red Dots}}: {{An Abundant Population}} of {{Faint Active Galactic Nuclei}} at z {$\sim$} 5 {{Revealed}} by the {{EIGER}} and {{FRESCO JWST Surveys}}},
  shorttitle = {Little {{Red Dots}}},
  author = {Matthee, Jorryt and Naidu, Rohan P. and Brammer, Gabriel and Chisholm, John and Eilers, Anna-Christina and Goulding, Andy and Greene, Jenny and Kashino, Daichi and Labbe, Ivo and Lilly, Simon J. and Mackenzie, Ruari and Oesch, Pascal A. and Weibel, Andrea and Wuyts, Stijn and Xiao, Mengyuan and Bordoloi, Rongmon and Bouwens, Rychard and {van Dokkum}, Pieter and Illingworth, Garth and Kramarenko, Ivan and Maseda, Michael V. and Mason, Charlotte and Meyer, Romain A. and Nelson, Erica J. and Reddy, Naveen A. and Shivaei, Irene and Simcoe, Robert A. and Yue, Minghao},
  year = {2024},
  month = mar,
  journal = {\apj},
  volume = {963},
  pages = {129},
  publisher = {IOP},
  issn = {0004-637X},
  doi = {10.3847/1538-4357/ad2345},
  urldate = {2024-06-18}
}
\bibliographystyle{aasjournalv7}

\end{document}